\documentclass[aps, prb, 10pt, twocolumn, superscriptaddress,floatfix]{revtex4-1}
\usepackage{graphicx}
\usepackage{color}
\usepackage{overpic}
\usepackage{braket}
\usepackage{bm}
\usepackage{tensor}
\usepackage[breaklinks=true,colorlinks,citecolor=blue,linkcolor=blue,urlcolor=blue]{hyperref}
\usepackage{comment}
\usepackage{tensor}
\usepackage{tabularx}  
\usepackage{bbold}
\usepackage{amssymb}
\usepackage{booktabs}  
\usepackage{makecell}  
\usepackage{microtype}
\usepackage{titlesec} 
\AtBeginDocument{
\newcommand{\I}{\mathrm{i}}

}
\usepackage{mathtools}
\usepackage[normalem]{ulem}

\begin{document}

\title{Quantum Electrodynamics of graphene Landau levels in a deep sub-wavelength hyperbolic phonon polariton cavity}

\author{Gian Marcello Andolina}
\affiliation{JEIP, UAR 3573 CNRS, Collège de France, PSL Research University, 11 Place Marcelin Berthelot,  F-75321 Paris, France}

\author{Matteo Ceccanti}
\affiliation{ICFO - Institut de Ciències Fotòniques, The Barcelona Institute of Science and Technology, Castelldefels
(Barcelona), Spain}

\author{Bianca Turini}
\affiliation{ICFO - Institut de Ciències Fotòniques, The Barcelona Institute of Science and Technology, Castelldefels
(Barcelona), Spain}

\author{Riccardo Riolo}
\affiliation{Dipartimento di Fisica dell’Università di Pisa, Largo Bruno Pontecorvo 3, I-56127 Pisa, Italy}

\author{Marco Polini}
\affiliation{Dipartimento di Fisica dell’Università di Pisa, Largo Bruno Pontecorvo 3, I-56127 Pisa, Italy}
\affiliation{ICFO - Institut de Ciències Fotòniques, The Barcelona Institute of Science and Technology, Castelldefels
(Barcelona), Spain}

\author{Marco Schir{\`o}}
\affiliation{JEIP, UAR 3573 CNRS, Collège de France, PSL Research University, 11 Place Marcelin Berthelot,  F-75321 Paris, France}

\author{Frank H.L. Koppens}
\affiliation{ICFO - Institut de Ciències Fotòniques, The Barcelona Institute of Science and Technology, Castelldefels
(Barcelona), Spain}
\affiliation{ICREA-Institució Catalana de Recerca i Estudis Avançats, Barcelona, Spain}

\begin{abstract}
The confinement of electromagnetic radiation within extremely small volumes offers an effective means to significantly enhance light-matter interactions, to the extent that zero-point quantum vacuum fluctuations can influence and control the properties of materials.
Here, we develop a theoretical framework for the quantum electrodynamics of graphene Landau levels embedded in a deep sub-wavelength hyperbolic cavity, where light is confined into ultrasmall mode volumes. By studying the spectrum, we discuss the emergence of polaritons, and disentangle the contributions of resonant quantum vacuum effects from those of purely electrostatic interactions. Finally, we study the hybridization between magnetoplasmons and the cavity modes.
\end{abstract}

\maketitle

\section{Introduction}

Electromagnetic radiation is among the most versatile probes for investigating the properties of matter, and technological advancements enable the exploitation of various spectral regions to interrogate material characteristics.
For instance, X-ray diffraction techniques elucidate the atomic arrangement within crystals \cite{Compton}, Raman spectroscopy characterizes the vibrational modes of molecules \cite{Raman}, time-resolved pump-probe spectroscopy reveals ultrafast electron dynamics, and both terahertz and angle-resolved photoemission spectroscopy (ARPES) provide insights into low-energy excitations and electronic band structures \cite{Damascelli}. After the seminal work of Purcell \cite{Purcell1946}, it is now well established that light can be tailored by means of resonant structures - or \textit{cavities} - to enhance its interaction with matter. Indeed, the emergence of cavity quantum electrodynamics (C-QED) \cite{Raimond01,Haroche13}, where mirrors confine the electromagnetic field, has enabled the measurement of quantum properties of a single atom coupled to a single photon.

Recently, a new paradigm has emerged: light can also be used to control matter \cite{cavityqm,BlochReview}, altering electronic structures and inducing novel states of matter. As a paradigmatic example, cold atoms in {\it driven} optical cavities exhibit altered many-body out-of-equilibrium properties, leading to emergent phases \cite{Ritsch2013,Baumann2010}.  Remarkably, due to vacuum fluctuations, even an empty, non-driven cavity \cite{VacuumReview} can change the properties of a material. Although only a few pioneering experiments have confirmed this concept \cite{Jarc2023,appugliese2022,enkner2024}, a growing body of theoretical predictions suggests that vacuum fluctuations could be harnessed to control the properties of matter \cite{Lu2024,schlawin2019,nguyen2023,wang2019,gao2020,Mazza2019,Sentef2018} leading to new cavity-induced mesoscopic phenomena, such as ferroelectricity\cite{Latini2021} and enhanced superconductivity\cite{Kozin2024}. 
These methods generally necessitate advanced control over solid-state systems, such as two-dimensional (2D) electron gases subject to a quantizing magnetic field \cite{appugliese2022,enkner2024}.

Light-matter hybridization manifests itself as an avoided crossing between energy levels, known as the vacuum Rabi splitting $\Omega$\cite{Sanchez-Mondragon1983, agarwal1984, agarwal1985,Brune1996,Raimond01}, which in solid-state systems can be probed in transmission and reflection experiments\cite{Khitrova2006,Todorov_PhysRevLett_2010,Todorov_PhysRevLett_2009,Scalari_Science_2012,Bitton}.
This phenomenon can be understood within both classical and quantum settings\cite{DDB_arxiv_2023}  but, in the context of C-QED, the quantum nature of the electric field $\hat{E}$ unveils a general correspondence between the vacuum Rabi splitting and quantum vacuum fluctuations of the electric field, $\Omega \sim  \sqrt{ \langle\hat{E}^2\rangle} $ \cite{dalibard1982,ciuti2005,DDB_arxiv_2023}. 
A promising strategy to enhance vacuum fluctuations (and consequently the coupling) is to confine the electromagnetic field to a small volume $V$\cite{chang2006, Kockum19, tame2013}, as the amplitude of fluctuations scales as $1/\sqrt{V}$. While Fabry-P\'{e}rot cavities are often used\cite{Raimond01,Haroche13}, the minimal achievable volume in these objects is limited by the so-called {\it diffraction limit}, $V \geq \lambda_{\rm d}^3$, where $\lambda_{\rm d}\equiv \lambda_{0}/2 $ and $\lambda_0$ is the wavelength in free space. This limit imposes a significant obstacle to the compression and impact of quantum vacuum fluctuations.

Conversely, sub-wavelength cavities, made from an optically active electromagnetic environment, differ from optical cavities as they can confine light in a volume $V \ll \lambda_{\rm d}^3$, well below the {\it diffraction limit} \cite{chang2006,koppens2011,tame2013,bozhevolnyi2006, oulton2009, baumberg2019, tame2013}. Paradigmatic examples of sub-wavelength cavities are plasmonic systems and metamaterials \cite{Maier2007}. Plasmonic nanocavities exploit the excitation of surface plasmons at metal-dielectric interfaces to confine electromagnetic fields to volumes significantly smaller than the diffraction limit \cite{Schuller2010,Ciraci2012, Chikkaraddy2016}. 
While these systems can achieve extremely high compression factors\cite{Alcaraz2018,Barra-Burillo2021}, they suffer from a low quality factor ($Q\approx 10$) stemming from the highly
dissipative nature of metallic systems. Among these, split-ring resonators operating in the terahertz regime\cite{Maissen2014} have been extensively used in solid-state cavities experiments, due to their integrability with high-quality two-dimensional electron gases. More generally, polaritonic systems serve as promising platforms for solid-state C-QED, as they can host deep sub-wavelength optical modes and offer great tunability \cite{dai2014}. 
While this approach is experimentally promising, most current theories are developed for Fabry-P\'{e}rot cavities, where excitations are described by a transverse vector potential $\hat{\bm{A}}$. As the influence of these cavities on material is predicted to be extremely small for equilibrium properties \cite{andolina2024,Amelio2021}, these theories are thus applied to sub-wavelength cavities by using a phenomenological substitution to enhance the Rabi frequency with a compression factor $A = V / \lambda_{\rm d}^3$. The compression factor quantifies how much a sub-wavelength cavity surpasses the diffraction limit \cite{Keller2017,arwas2023}, reaching values of $A \approx 10^{-10}$.
Given a Rabi frequency $\Omega$ derived from optical cavity theory, which scales as $\sim 1/\sqrt{V}$, the same quantity for a sub-wavelength cavity is often obtained\cite{gao2020,schlawin2019,wang2019,nguyen2023,zezyulin2022,lenk2022,arwas2023,pavosevic2022,Dag2024} by substituting $\Omega \to \Omega /\sqrt{A}$, thus enhancing light-matter interactions by a factor up to $10^{5}$. This phenomenological procedure underlies striking predictions in sub-wavelength cavities, such as cavity-mediated superconductivity \cite{schlawin2019}, breakdown of topological integer quantum Hall effect \cite{arwas2023}, and cavity control of topological properties \cite{wang2019,nguyen2023}. 
\\
However, it has been shown \cite{andolina2024} that, in the generalized Coulomb gauge, sub-wavelength cavities characterized by a strong compression $A\ll 1$, cannot be described by a transverse vector potential $\hat{\bm{A}}$. Instead, they are more accurately described by a scalar potential $\hat{\phi}$, reflecting their electrostatic character\cite{andolina2024,pantazopoulos2024,Muniain2024}. This underscores the need for a comprehensive theory of sub-wavelength cavity QED based on the coupling with a scalar potential $\hat{\phi}$ solving the Poisson equation. Developing this theory---and applying it to the specific problem of cavity QED of Landau levels in graphene---is the main goal of the present article.

The excitations of a sub-wavelength cavity are not transverse photons but longitudinal hybrid light-matter excitations such as plasmons\cite{GPlasmonic} and polaritons\cite{PCavities}. Recently, sub-wavelength cavities have been shown~\cite{riolo2024} to alter the quasiparticle properties of Fermi liquids realized in weakly correlated 2D electron systems.
Among sub-wavelength platforms, hyperbolic phonon-polariton (HPP) cavities \cite{dai2014,caldwell2014,HerzigSheinfux,li2015, Autore} based on hexagonal boron nitride (hBN) offer an unparalleled balance of high compression factor ($A \approx 10^{-6}$) and impressive quality factors ($Q\approx 400$) \cite{HerzigSheinfux}. 
Due to the strong anisotropy in its optical phonon spectrum\cite{caldwell2014}, hBN exhibits an upper ($\sim 168-200$~meV) and a lower ($\sim 94-103$~meV) Reststrahlen band, where the in-plane and out-of-plane permittivity components are opposite in sign, leading to a hyperbolic polaritonic dispersion.  The resulting ray-like nature of hBN mid-infrared phonon polaritons allows for extreme wavelength compression of light, enabling electromagnetic waves to be confined to deeply sub-wavelength scales\cite{li2015}. On the other hand, the bulk nature of hBN phonon polaritons protects them from Ohmic loss mechanisms typical of metal-dielectric interfaces, enhancing the lifetime and making HPP cavities highly effective for mid-IR photonic applications\cite{caldwell2014}.

In this paper, we present a fully non-local quantum theory to describe the C-QED of 2D electron systems coupled to sub-wavelength cavities, with the goal of deriving vacuum Rabi frequencies from a fully microscopic framework without relying on any phenomenological assumptions. To achieve this, we develop a quantization scheme for sub-wavelength cavities that allows the calculation of quantum vacuum fluctuations of the cavity electric field. This approach enables the use of standard methods and techniques characteristic of quantum optics and C-QED. 
While our theory is general for any sub-wavelength cavity, we apply it to an experimentally relevant scenario.
Specifically, we investigate electrons in graphene subject to a strong perpendicular magnetic field ${\bm B}$ and coupled to the sub-wavelength modes of HPP cavity.  This ${\bm B}$ field induces circular electron motion (cyclotron motion), leading to the emergence of quantized energy levels known as Landau levels (LLs)\cite{Giuliani_and_Vignale,PrangeGirvin1987}. The excitations of HPP cavities made from hBN are typically around $0.1~{\rm eV}$ \cite{HerzigSheinfux}, hence LL transitions can be tuned to this energy range by applying a magnetic field of a few Tesla.  Therefore, if the coupling is higher than the losses, LLs transitions hybridize with resonant cavity modes, forming Landau-phonon-polaritons (LPhPs) \cite{hagenmuller2010,hagenmuller2012,Chirolli2012,Pellegrino2014,ashida2023}.

To describe LPhPs, we develop a comprehensive microscopic model of quantum vacuum fluctuations in deep sub-wavelength HPP cavities. Our microscopic theory can predict the optimal cavity size to maximize light-matter coupling, measured by the vacuum Rabi frequencies. Importantly, the deep sub-wavelength regime allows cavity and matter excitations to hybridize at very high wavevectors, necessitating a fully {\it non-local} theoretical framework beyond the {\it dipole approximation} \cite{Svendsen2021}. Indeed, the maximal vacuum Rabi frequencies are obtained for finite momenta. Scattering-type near-field optical microscopy (s-SNOM) \cite{basov2016}, capable of probing collective modes at such high wavevectors, provides a promising platform to measure our predictions. The fully microscopic theory developed here can have profound implications for the control of materials in sub-wavelength cavities. As a major example, it will be interesting in the future to use our theory to study quantum transport in these 2D electron systems embedded in HPP cavities. Several experiments showed that sub-wavelength split-ring cavities affect the integer \cite{appugliese2022,enkner2024} and fractional \cite{enkner2024b} quantum Hall, generating significant theoretical interest \cite{ciuti2021,arwas2023,bacciconi2023,rokaj2023,rokaj2022}. In principle, our framework can be applied to study how HPP cavities modify magnetotransport \cite{paravicini2019} and topology in general \cite{arwas2023,nguyen2023,wang2019,Zeno, Dmytruk2022,Dmytruk2024, Dag2024}. In our case, as the collective magnetic excitations are strongly coupled to the cavity, a fully many-body treatment of quantum transport would be needed, rather than a single-particle approach. 
Since in sub-wavelength cavities the light-matter coupling occurs via the density operator, rather than via the current, these cavities circumvent the so-called ``no-go theorems"\cite{andolina_prb_2019,andolina_epjp_2022,rzazewski_prl_1975,nataf_naturecommun_2010} and thus can, in principle, realize a superradiant phase transition. However, since the excitations of the cavity are encoded in the scalar potential $\hat{\phi}$, the nature of the phase transition would be ferroelectric\cite{debernardis2018,rouse2023}, and would not lead to photon condensation.

This paper is organized as follows. In Sect.~\ref{Sec:summ}, we summarize the main concepts and key results of this work, which are presented in greater detail in the subsequent sections. In Section~\ref{sec:HHP_cavities}, we develop a quantum theory of HPP cavities. Sect.~\ref{sec:LLs} provides an overview of the physics of LLs in graphene. In Sect.~\ref{sec:Polaritons}, we delve into the details of light-matter interactions and analyze the resulting polaritonic spectrum.
Additional technical details regarding the electrostatic properties of hexagonal boron nitride are provided in Appendix \ref{app0}, while the equivalence between the present quantization scheme and standard approaches is discussed in Appendix~\ref{app_alt}. The density–density response function of graphene Landau levels is reviewed in Appendix~\ref{app1}.

\section{Summary of main results}
\label{Sec:summ}
In this article, we present our main findings as described in what follows.
First, in Sect.~\ref{HPP_cavities_summary_results}, we solve the Poisson equation with a frequency-dependent permittivity tensor within a sub-wavelength cavity to determine the scalar Green’s function. This solution enables us to quantize the scalar potential $\hat{\phi}$, expressing it in terms of the quantum modes of the cavity. This provides a theoretical framework for describing electromagnetic and quantum vacuum fluctuations within a sub-wavelength cavity.
In Sect.~\ref{LLs_summary}, we review the physics of LLs emerging in graphene subjected to a strong perpendicular magnetic field. Crucially, we highlight that, in sub-wavelength cavities, electrons couple through the density as they are coupled to a scalar potential $\hat{\phi}$. This is in stark contrast with traditional Fabry-Pérot cavities, where the light-matter interaction is implemented through the minimal coupling mechanism, which couples the electronic transverse current to the transverse vector potential. In Sect.~\ref{subsec:pol} we develop the C-QED of a 2D material embedded in a sub-wavelength cavity. By formulating and diagonalizing the Hopfield Hamiltonian, we study polaritonic states that arise from the hybridization of LL transitions with cavity modes. This also allows us to directly calculate, without invoking any phenomenological procedure, the vacuum Rabi frequencies of the system, showing the system achieves ultra-strong and super-strong coupling. 
While our theory is specifically applied to graphene embedded in an hBN cavity, the same theory can be readily applied to other 2D materials within sub-wavelength cavities.
Finally, in Sect.~\ref{sec:comparison} we comment on the relation of theory with the previous literature.

\subsection{Theory of the Green's function of HPP cavities}
\label{HPP_cavities_summary_results}

Generally, electromagnetic cavities are fully characterized by the solutions of Maxwell's equations\cite{Jackson}. In sub-wavelength cavities the most important contribution to the description of the electromagnetic properties is provided by the solution of the frequency-dependent Poisson equation in the presence of an external charge\cite{andolina2024,riolo2024}:
\begin{equation}
\label{eqn:scalar_Poisson}
-\sum_{i,j}\partial_i [\epsilon_{ij}(\bm{r}, \omega)\partial_j \phi(\bm{r}, \omega)] = 4\pi\rho_{\rm ext}(\bm{r}, \omega)~,
\end{equation}
where $\bm{r}$ is the position, $\partial_i$ is a shorthand for the spatial derivative $\partial/\partial r_i$, $\omega$ is the frequency, $\epsilon_{ij}(\bm{r}, \omega)$ is the dielectric tensor of the cavity embedding, $\phi(\bm{r}, \omega)$ is the potential, and $\rho_{\rm ext}(\bm{r}, \omega)$ is the free charge. In the case of sub-wavelength cavities, the speed of light $c$ can be considered infinite, as $\lambda \ll \lambda_{0}$, being $\lambda$ the characteristic wavelength of the cavity. Under this condition, we restrict our analysis to the electrostatic component of the total electromagnetic field.  Consequently, the electromagnetic behavior is described by the electrostatic Poisson equation (Eq.~\eqref{eqn:scalar_Poisson}), which incorporates the non-trivial frequency dependence of the response of the metallo-dielectric environment through the dielectric tensor $\epsilon_{ij}(\bm{r}, \omega)$.

\begin{figure}[t]
  \vspace{1.em}
   \begin{overpic}[width=1\columnwidth]{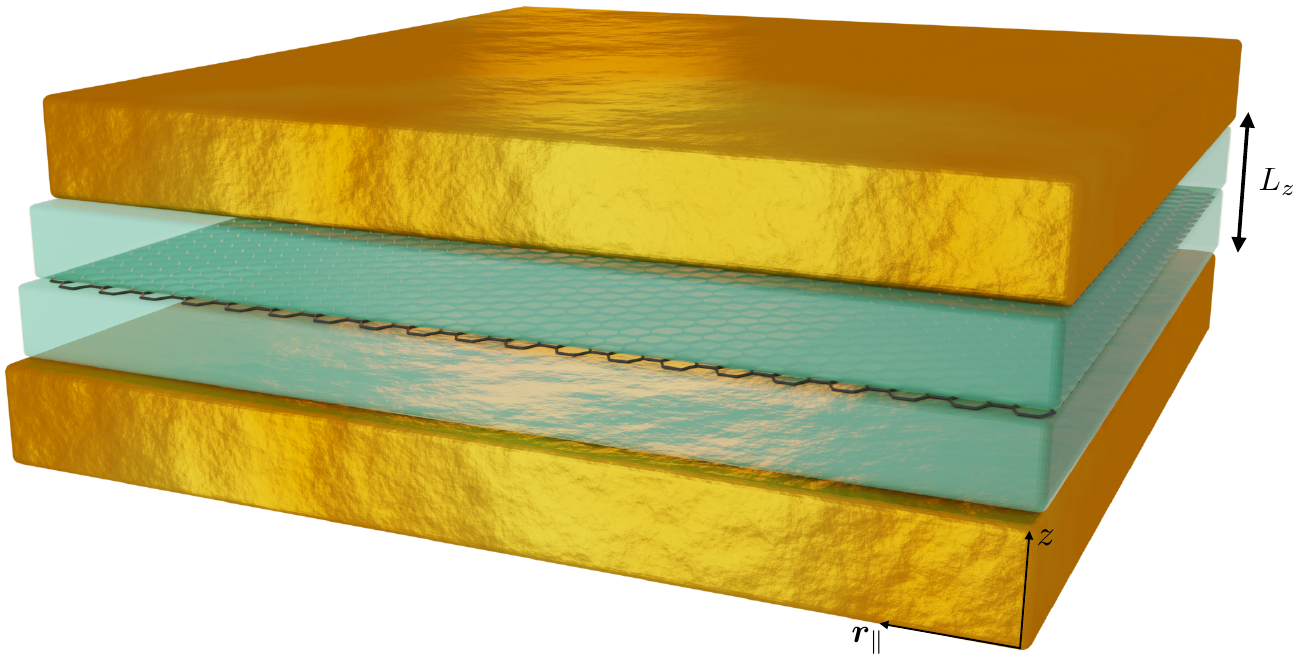} \put(0,80){\normalsize }\end{overpic}
\caption{(Color online) A schematic representation of the system under study. The HPP cavity is formed by two metallic mirrors (depicted in gold) located at $|z| > L_z/2$. The region between the mirrors is filled with hBN, shown in light blue. A graphene sheet is positioned at $z=0$ and subjected to an external perpendicular magnetic field. \label{fig_sketch}}
\end{figure}

The cavity geometry considered is shown in Fig.~\ref{fig_sketch}. We assume translational invariance along the $\hat{\bm{e}}_x$ and $\hat{\bm{e}}_y$ directions and denote the in-plane coordinates by the vector \( \bm{r}_\parallel = x \hat{\bm{e}}_x + y \hat{\bm{e}}_y \). Along the direction perpendicular to this plane, i.e., the \( \hat{\bm{e}}_z \) direction, and within the region \( |z| \leq L_z/2 \), space is assumed to be filled by an hBN layer of thickness \( L_z \) \cite{dai2014,caldwell2014,HerzigSheinfux,li2015,Autore}.  Notice that, as we focus on energies $\hbar\omega \sim 0.1~\mathrm{eV}$, corresponding to a free-space wavelength $\lambda_0 \equiv (2\pi c/\omega) \approx 1.2\times 10^{4}~\mathrm{nm}$, and choose a cavity thickness $L_z \sim 10\!-\!100~\mathrm{nm}$, we are in the deep subwavelength regime $L_z\ll \lambda_0$, where the speed of light can be effectively regarded as infinite. hBN is characterized by a uniaxial permittivity tensor\cite{tomadin2015,riolo2024} $\bm{\epsilon}(\bm{r}, \omega) = \text{diag}(\epsilon_x(\bm{r}, \omega), \epsilon_x(\bm{r}, \omega), \epsilon_z(\bm{r}, \omega))$. This dielectric tensor is initially assumed to be real, with its imaginary part introduced later through the substitution $\omega \to \omega + \I \eta$. While this approach is not generally required, it is particularly advantageous in this context, as it allows us to quantize the field first and incorporate losses subsequently.
Further details on this tensor are reported in Appendix \ref{app0}. For $|z| > L_z/2$, we assume the presence of two perfect conductors, acting as mirrors.  Consequently, the boundary conditions for the potential inside hBN ( $|z| \leq L_z/2$) are given by \cite{Jackson}: $\phi(\bm{r}_{\parallel}, z = \pm L_z/2) = 0$.  
Realistic metallic mirrors possess a frequency-dependent dielectric response, characterized by a plasma resonance at frequency $\Omega_{\rm Mir}$: they strongly screen quasi-static fields for $\omega \ll \Omega_{\rm Mir}$, while becoming transparent to electromagnetic radiation for $\omega \gg \Omega_{\rm Mir}$. For gold, $\hbar\Omega_{\rm Mir} \approx 9~\mathrm{eV}$, so in the frequency range of interest ($\hbar\omega \sim 0.1~\mathrm{eV}$) the perfect-conductor approximation is well justified.

Instead of directly solving Eq.~\eqref{eqn:scalar_Poisson} for an arbitrary external charge distribution, it is convenient to solve for the associated scalar bare Green's function\cite{andolina2024} $g^{(0)}(\bm{r},\bm{r}', \omega)$ corresponding to the solution of Eq.~\eqref{eqn:scalar_Poisson} in the presence of a test charge $\rho^{({\rm test},\bm r')}_{\rm ext}(\bm r,\omega)=\delta(\bm r-\bm r')$ at a given position $\bm{r}'$. Note that the bare Green's function $g^{(0)}(\bm{r},\bm{r}', \omega)$ describes how an \textit{empty} HPP cavity responds to a test charge, while the effect of an embedded material will be treated later in Section \ref{subsec:pol}. 
Once the Green's function has been obtained, the potential $\phi(\bm r, \omega)$ can be obtained directly as a convolution between $g^{(0)}(\bm{r},\bm{r}', \omega)$ and the external density $\rho_{\rm ext}(\bm r',\omega)$. 
The scalar Green's function of an HPP cavity can be obtained from the following Fourier transform with respect to the 2D coordinate ${\bm r}_{\|}$ and expansion in normal modes:

\begin{equation}
\begin{split}
\label{eqn:green_function_hBN_expansion_Main}
& g^{(0)}(\bm q_{\parallel},z,z', \omega)= \\
&\frac{8\pi}{ L_z}\sum_{n_z=0}^{\infty}\frac{\sin\left(q_{z,n_z} [z+\frac{L_z}{2}]\right) \sin\left(q_{z,n_z} [z'+\frac{L_z}{2}]\right)}{{\epsilon}_z(\omega)q_{z,n_z}^2+{\epsilon}_x(\omega)q_\parallel^2} ~,
\end{split}
\end{equation}
where $\bm{q}_{\parallel}$ is the wavevector in the $\hat{\bm e}_x-\hat{\bm e}_y$ plane, ${q}_{\parallel}$ is its modulus, and  $q_{z,n_z} \equiv n_z\pi/L_z$ represents the quantized components of the wavevector along the out-of-plane $\hat{\bm e}_z$ direction. The derivation of this equation is detailed in Sec.~\ref{Subsec:modes}. Eq.~\eqref{eqn:green_function_hBN_expansion_Main} represents a key theoretical result, as it shows that the scalar Green's function can be expanded into distinct modes, which can then be quantized systematically. Notice that the Green's function depends only on the modulus ${q}_{\parallel}$ as the system is invariant under rotations in the $\hat{\bm e}_x-\hat{\bm e}_y$ plane.

\begin{figure}[t]
  \vspace{1.em}
   \begin{overpic}[width=0.9\columnwidth]{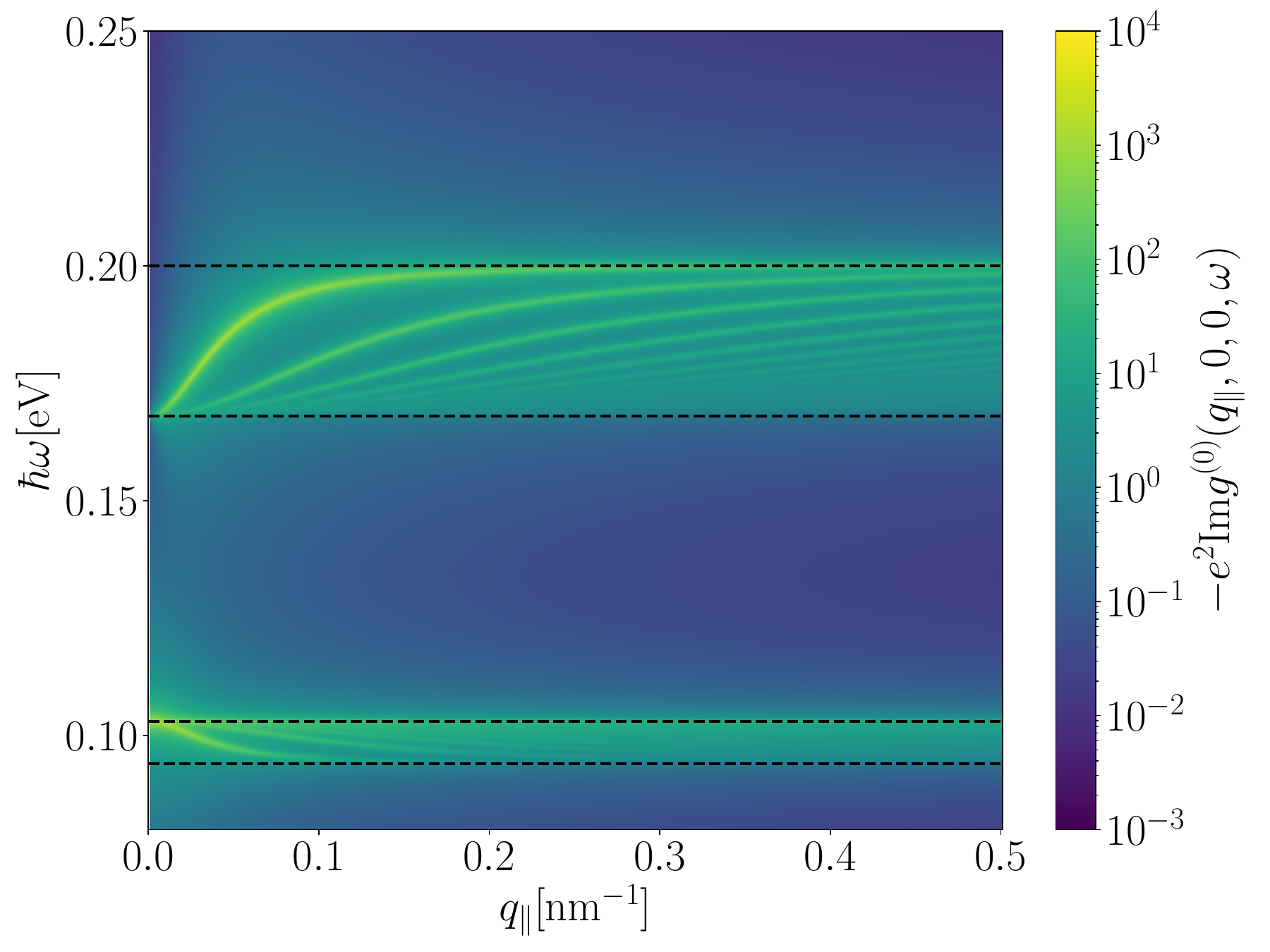} \put(0,80){\normalsize (a)}\end{overpic}
   \begin{overpic}[width=0.9\columnwidth]{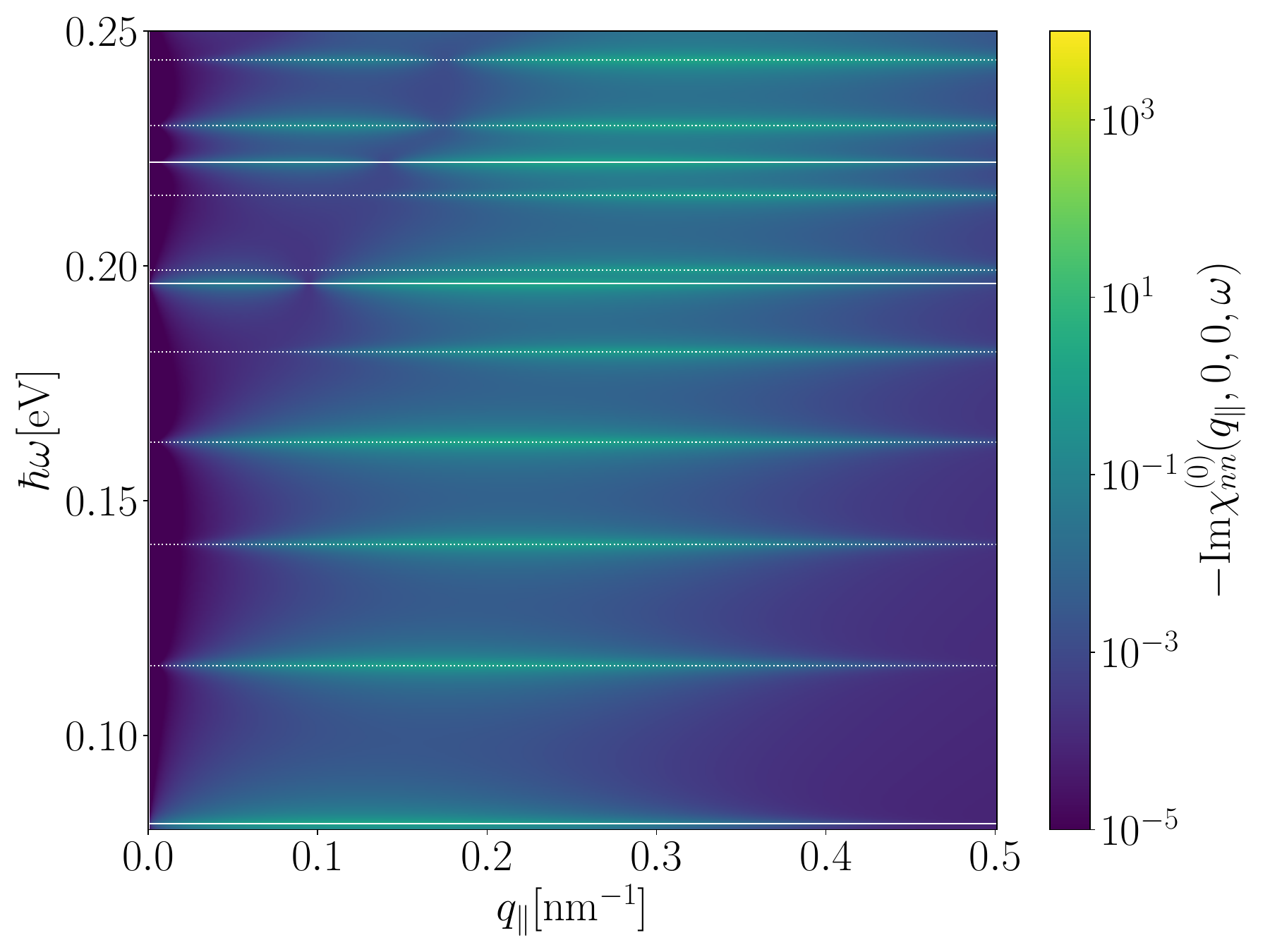}\put(0,80){\normalsize (b)}\end{overpic}
 \caption{(Color online) Different spectra are plotted as functions of $q_\parallel$ (expressed in ${\rm nm}^{-1}$) and $\hbar\omega$ (expressed in $\rm{eV}$). Panel (a) shows the imaginary part of the bare Green's function of the HPP cavity, i.e.~$-e^2{\rm Im}~g^{(0)}(q_{\parallel},0,0, \omega)$, for $L_z=50~{\rm nm}$. All modes are taken into account via resummation of the series.  Panel (b) displays the imaginary part of the bare density-density response function for LLs in graphene, $-{\rm Im}~\chi^{(0)}_{\rm nn} ({q}_{\parallel}, \omega)$, taking $B=5~{\rm T}$ and a cutoff for the LLs $N_{\rm c}=30$. In both panels we choose $\hbar \eta= 5\times 10^{-4}{\rm eV}$.  \label{fig1}}
\end{figure}

Fig.~\ref{fig1}(a) displays the spectrum $-{\rm Im}~g^{(0)}({q}_{\parallel}, 0, 0, \omega)$ of the HPP cavity, which quantifies the absorption of the cavity at the position $z=z^\prime=0$, where graphene will be placed.  The scalar Green's function $g^{(0)}({q}_{\parallel}, z, z', \omega)$ exhibits poles when the condition ${\epsilon}_z(\omega) q_{z, n_z}^2 + {\epsilon}_x(\omega) q_{\parallel}^2 = 0$ is satisfied. This occurs only in the {\it lower} ($\omega_{z}^{\rm T} < \omega < \omega_{z}^{\rm L}$) and {\it upper} ($\omega_{x}^{\rm T} < \omega < \omega_{x}^{\rm L}$) {\it Reststrahlen bands}.
The longitudinal and transverse phonon frequencies, $\omega_{i}^{\rm L}$ and $\omega_{i}^{\rm T}$, are marked by black dotted lines in Fig.~\ref{fig1}(a).  
We denote the poles of the bare scalar Green's function with the symbol $\omega_{{q}_{\parallel}, n_z, s}$, where $s = +1$ ($s= -1$) refers to the upper (lower) Reststrahlen band. 
There are important differences and similarities between the scalar Green's function in Eq.~\eqref{eqn:green_function_hBN_expansion_Main} for HPP cavities and the full dyadic Green's function for conventional Fabry-P\'{e}rot cavities \cite{andolina2024,Charlie_Thesis}. 
The latter can be obtained in a similar geometry, consisting of two infinite metallic mirrors separated by a distance $L_z$, but without any hBN material between them.
Both Green's functions share key features, such as poles corresponding to mode frequencies and a prefactor of $1/L_z$ (see Eq.~\eqref{eqn:green_function_hBN_expansion_Main}), which encodes the compression mechanism. However, in both cases, the dependence on $L_z$ is non-trivial, as the mode frequencies $\omega_{{q}_{\parallel}, n_z, s}$ themselves depend on the cavity length $L_z$. 
In Fabry-P\'{e}rot cavities, the poles, which are associated with the energies of cavity photons, appear only in the vectorial part of the dyadic Green's function \cite{andolina2024}, while the scalar part only represents the Coulomb potential modified by the image charges induced in the mirrors. Conversely, in HPP cavities, poles corresponding to the phonon-polariton energies are present in both the scalar and vectorial components of the dyadic Green's function. Nevertheless, due to the sub-wavelength nature of HPP cavity modes ($\lambda_{\rm p} = 2\pi/q_{\parallel} \ll \lambda_0$) the scalar Green's function $g^{(0)}({q}_{\parallel}, z, z', \omega)$ is the dominant contribution to the system's electromagnetic properties \cite{andolina2024}.

Note that, in the static limit ($\omega = 0$), the scalar Green's function reduces to the Coulomb potential, as screened by the metallic mirrors at $z = \pm L_z/2$ and the static polarization charges inside hBN. Taking also the limit $L_z \to \infty$, $g^{(0)}({q}_{\parallel}, 0, 0, 0)$ reduces to the 2D Coulomb potential, as renormalized by the static dielectric properties of hBN, i.e.~$\lim_{L_z \to \infty} g^{(0)}({q}_{\parallel}, 0, 0, 0) = 2\pi / (q_{\parallel} \sqrt{\epsilon_x(0) \epsilon_z(0)})$. We conclude that Eq.~\eqref{eqn:green_function_hBN_expansion_Main} accurately describes both the static screening ($\omega = 0$) induced by hBN polarization charges and metallic mirrors, and the resonant structure at finite $\omega$, occurring at the location of the poles $ \omega_{{q}_{\parallel}, n_z, s}$.

In free space, the Poisson equation in Eq.~\eqref{eqn:scalar_Poisson} lacks a time-dependent component, making $\phi(\bm{r})$ a completely classical field.
However, in HPP cavities, the non-trivial frequency dependence of the dielectric permittivity tensor causes the emergence of resonant modes\cite{Gubbin_JChemPhys_2022}, which can be described as quantum mechanical modes. 
Consequently, the poles of Eq.~\eqref{eqn:green_function_hBN_expansion_Main} correspond to the quantum mechanical modes of the cavity Hamiltonian, $\hat{\mathcal{H}}_{\rm cav}=\sum_{\bm{q}_\parallel,n_z,s}\hbar \omega_{{q}_\parallel,n_z,s}\hat{a}^\dagger_{\bm{q}_\parallel,n_z,s}\hat{a}_{\bm{q}_\parallel,n_z,s}$, where the operator $\hat{a}^\dagger_{\bm{q}_\parallel,n_z,s}$ ($\hat{a}_{\bm{q}_\parallel,n_z,s}$) creates (destroys) a cavity excitation.
Since the cavity is deeply sub-wavelength, the electromagnetic field is well described by a scalar potential
\begin{align}
\label{eqn:potential}
    \hat{\phi}(\bm{r}_{\parallel},z)&=\sum_{\bm{q}_{\parallel},n_z,s}{\phi_{{q}_\parallel,n_z,s}(z)}\left(\frac{e^{-\I\bm{q}_\parallel\cdot\bm{r}_{\parallel}}}{{\sqrt{S}}}\hat{a}^\dagger_{\bm{q}_\parallel,n_z,s}+{\rm h.c.}\right)~, 
\end{align}

where $S$ is the surface area of the cavity, and $ \phi_{{q}_\parallel,n_z,s}(z)\equiv\sin\left(q_{z,n_z} [z+\frac{L_z}{2}]\right)\sqrt{{2}/({N_{{q}_\parallel,n_z,s}L_z})}$ are the amplitude coefficients of the modes, with $N_{{q}_\parallel,n_z,s}$ being a normalization factor. The procedure to determine this normalization factor will be discussed momentarily and detailed in Subsection~\ref{Subsec:quantization}.
Eq.~\eqref{eqn:potential} is one of our key results, as it enables the solution of the Poisson equation (Eq.~\eqref{eqn:scalar_Poisson}) to be expressed in terms of creation and annihilation operators. This formulation allows for a direct quantification of quantum vacuum fluctuations of the field. The corresponding physical electric field associated with the scalar potential is obtained by taking the gradient of Eq.~\eqref{eqn:potential}.

The normalization factors are determined by matching the imaginary part of the quantum scalar Green’s function, \( g_{\rm q}^{(0)}( q_{\parallel}, z, z', \omega) \), to its classical counterpart, \( g^{(0)}( q_{\parallel}, z, z', \omega) \). The quantum Green’s function is computed from the scalar potential $\hat{\phi}(\bm{r}_{\parallel}, z)$ given in Eq.~\eqref{eqn:potential} using the standard Kubo formula\cite{Giuliani_and_Vignale}, whereas the classical Green’s function follows directly from the Poisson equation. Crucially, despite this matching of imaginary parts, their real parts differ significantly: the quantum Green’s function vanishes at high frequencies, whereas the classical Green’s function approaches a finite limit, defined as $g_\infty^{(0)}( q_{\parallel}, z, z') \equiv \lim_{\omega \to \infty} g^{(0)}( q_{\parallel}, z, z', \omega)$. As a consequence, \( g_{\rm q}^{(0)}( q_{\parallel}, z, z', \omega) \) directly fulfills the Kramers-Kronig relations, whereas \( g^{(0)}( q_{\parallel}, z, z', \omega) \) requires subtraction of the high-frequency limit \( g_\infty^{(0)}( q_{\parallel}, z, z') \) to satisfy them. Explicitly, this yields the decomposition 
\begin{equation}
    g^{(0)}( q_{\parallel}, z, z', \omega) = g_{\rm q}^{(0)}( q_{\parallel}, z, z', \omega) + g_\infty^{(0)}( q_{\parallel}, z, z')~,
\end{equation} 
clearly separating resonant quantum mechanical contributions \( g_{\rm q}^{(0)} \)—associated with excitation modes \( \hat{a}_{\bm{q}_\parallel,n_z,s},\hat{a}^\dagger_{\bm{q}_\parallel,n_z,s} \)—from frequency-independent Coulomb interaction renormalizations encapsulated within \( g_\infty^{(0)} \). Further details on this procedure are discussed in Sec.~\ref{sec:HHP_cavities}.

\subsection{Theory of Landau levels in Graphene and their coupling to a sub-wavelength cavity}
\label{LLs_summary}
A single-layer graphene is now considered, positioned at the center of the HPP cavity (i.e.~at $z = 0$), subject to a strong, uniform, and static perpendicular magnetic field $B$. In this regime, electronic excitations are inter-Landau-level transitions \cite{Goerbig11,ando1982,roldan2009,roldan2010}. In the Landau gauge, the energy spectrum of electrons roaming in this crystal is given by massless Dirac fermion Landau levels (LLs), $\varepsilon_{\lambda, n} = \lambda (\hbar v_{\rm F} / \ell_B) \sqrt{2n}$, where $\ell_B = \sqrt{\hbar c / (eB)}$ is the magnetic length, $e$ is the electron charge, $v_{\rm F}$ is the Fermi velocity in graphene, $k_x$ is the electronic wavevector along the $x$-axis, which accounts for the degeneracy of the spectrum, $n$ is a non-negative integer representing the LL index, and $\lambda=\pm$ denotes conduction/valence band states. We denote by the symbol $N_{\rm F}$ the index of the last occupied LL. Here, we focus on the charge neutrality point of graphene, choosing $N_{\rm F}=0$.
In Fig.~\ref{fig1}(b), we plot the non-interacting spectrum of LLs, $-{\rm Im} \chi^{(0)}_{\rm nn} ({q}_{\parallel}, \omega)$, where $\chi^{(0)}_{\rm nn} ({q}_{\parallel}, \omega)$ is the bare density-density response function calculated using the Kubo formula \cite{roldan2009,Giuliani_and_Vignale}. 
This quantity is fully non-local and valid for all in-plane wavevectors ${q}_\parallel$. This feature is particularly relevant, as the deep sub-wavelength nature of hBN enables the exploration of high wave-vector regions. This limit of rapidly varying matter fields ($q_\parallel\sim 0.1~{\rm nm}^{-1}$) has been investigated in SNOM experiments\cite{dai2014} also in presence of a magnetic field\cite{wehmeier2024}.

The LLs spectrum shows electron-hole pair excitations associated with transitions $\lambda, k_x, n \to \lambda', k_x', n'$ and corresponding transition energies $\varepsilon_{\lambda', n'} - \varepsilon_{\lambda, n}$.
Due to the $k_x$ degeneracy, bare LL transitions do not depend on the momentum transfer ${q}_\parallel = k_x' - k_x$ and thus appear as flat lines in the ${q}_\parallel - \hbar\omega$ plane. The magnitude of the bare density-density current response function $\chi^{(0)}_{\rm nn} ({q}_{\parallel}, \omega)$ critically depends on the matrix elements of the electronic density operator

\begin{equation}
\begin{split}
\label{eq:dipole}
  \mathcal{M}_{n^\prime,\lambda^\prime,n,\lambda}({q}_\parallel)\equiv \sum_{k^\prime_x,k_x}
\lvert\braket{k_x^\prime,n',\lambda'|\hat n_{\bm q_\parallel}|k_x,n,\lambda}\rvert^2~,\\
\end{split}
\end{equation}

where $|k_x,n,\lambda\rangle$ are the electronic state associated with the quantum number $\lambda, k_x, n$ and  $\hat n_{\bm q_\parallel}$ is the momentum-resolved electronic density.
In Fig.~\ref{fig1}(b) the transitions between valence-band states $\lambda=-,n$ and conduction-band states $\lambda^\prime=+,n^\prime$ are shown.    
In the long-wavelength limit \(q_\parallel \ell_B \ll 1\), only \textit{dipole-allowed} transitions ($n^\prime = n \pm 1$), represented by solid white lines, couple effectively to light. However, due to the deeply sub-wavelength nature of our cavity, we consider significantly larger in-plane wavevectors \(q_\parallel \approx 0.1~{\rm nm}^{-1}\), corresponding to \(q_\parallel \ell_B \sim 1\) as we consider a magnetic field of \(B = 5~{\rm T}\) (\(\ell_B \approx 11.5~{\rm nm}\)). In this regime, also {\it dipole-forbidden} transitions (indicated by dashed white lines) are relevant \cite{Rivera2016}. Indeed, as illustrated in Fig.~\ref{fig1}(b), all the discrete spectrum of Landau-level transitions is optically active.

We now move on to discuss the coupling between matter degrees of freedom and cavity modes.
In the Coulomb gauge, the LLs and the cavity modes interact through the electrostatic interaction Hamiltonian \cite{andolina2024,Giuliani_and_Vignale}:
\begin{equation}
\label{eq:H_int}
\hat{\mathcal{H}}_{\rm int} = -e\int d\bm{r}_\parallel \, \hat{\phi}(\bm{r}_{\parallel},z=0) \hat{n}(\bm{r}_\parallel)~,
\end{equation}
where $\hat{\phi}(\bm{r}_{\parallel},z=0)$ is the scalar potential of the cavity reported in Eq.~\eqref{eqn:potential} and  evaluated at the vertical location of graphene and $\hat{n}(\bm{r}_\parallel)$ is the electron density operator in graphene.
However, the interaction Hamiltonian in Eq.~\eqref{eq:H_int} includes only the dynamical field component \( g_{\rm q}^{(0)}(q_{\parallel}, z, z', \omega) \). The frequency-independent contribution \( g_\infty^{(0)}(q_{\parallel}, z, z') \) must also be included as a standard two-body screened Coulomb interaction:
\begin{equation}
\label{eq:H_Coulomb}
\hat{\mathcal{H}}_{\rm C} = \frac{1}{2S}\sum_{\bm{q}_\parallel} \hat{n}_{-\bm{q}_\parallel}e^2g_{\infty}^{(0)}(\bm q_{\parallel},0,0)\hat{n}_{\bm{q}_\parallel}~,
\end{equation}

where $\hat{n}_{\bm{q}_\parallel}$ is the momentum-resolved density operator. In the previous summation, the ${\bm q}_{\parallel} = {0}$ contribution is canceled by the inclusion of a positive background (Jellium) to ensure charge neutrality \cite{Giuliani_and_Vignale}. Since our analysis focuses on finite $q_\parallel$, this procedure does not affect the results.
Eq.~\eqref{eq:H_Coulomb} represents the Coulomb interaction unaffected by the phononic resonances of hBN, $\omega_i^{\rm T}, \omega_i^{\rm L}$. 
In contrast, the term \( g_{\infty}^{(0)}(\bm q_{\parallel},0,0) \) accounts for screening effects arising from high-frequency modes, such as those in hBN, encoded in \( \epsilon_i(\infty) \), as well as the influence of the metallic mirrors, which is incorporated through the boundary conditions of Eq.~\eqref{eqn:scalar_Poisson}.

An important difference between sub-wavelength cavities and conventional Fabry P\'erot cavities lies in the nature of the light-matter coupling. In sub-wavelength cavities, the interaction Hamiltonian involves a fundamentally different coupling compared to the traditional minimal coupling $\hat{\bm{p}} \to \hat{\bm{p}}+(e/c) \hat{\bm{A}}$ interaction in Fabry P\'erot cavities, where $\hat{\bm{A}}$ is the transverse vector potential and $\hat{\bm{p}}$ is the momentum operator. The minimal coupling results in a diamagnetic term in the continuum-model Hamiltonian~\cite{andolina_prb_2019,Ruggenthaler2023,Schaefer2020} or introduces a non-linear Peierl's phase in truncated lattice models~\cite{DeBernardis18,savasta2021gauge,distefano2019,dmytruk2021}. In contrast, Eq.~\eqref{eq:H_int} represents a purely linear coupling, which is a distinctive feature of interactions within sub-wavelength cavities. 
The details of this section are provided in Sec.~\ref{sec:LLs}.
\subsection{The polaritonic spectrum and the vacuum Rabi frequency}
\label{subsec:pol}

The coupling between graphene LLs and cavity modes leads to the formation of hybrid light-matter states known as polaritons \cite{hopfield1958,Kavokin11}. 
Within the Random Phase Approximation (RPA)\cite{Giuliani_and_Vignale}, polaritons are the zeroes of the dielectric function 
\begin{equation}
\begin{split}
\label{eqn:epsilon}
\epsilon_{\rm RPA}({q}_{\parallel},\omega) \equiv 1 - e^2 g^{(0)}({q}_\parallel, 0, 0, \omega) \chi^{(0)}_{\rm nn}({q}_{\parallel}, \omega)~.
\end{split}
\end{equation}
Once \( \epsilon_{\rm RPA}({q}_{\parallel},\omega) \) is known, the dressed electronic spectrum can be evaluated as  
\(S_{\rm el}({q}_{\parallel},\omega) = -({\hbar}/{\pi}) \, {\rm Im}~\chi_{\rm nn}({q}_{\parallel},\omega)\) with \( \chi_{\rm nn}({q}_{\parallel},\omega) = \chi^{(0)}_{\rm nn}({q}_{\parallel},\omega)/\epsilon_{\rm RPA}({q}_{\parallel},\omega) \).
The poles of \( \epsilon_{\rm RPA}({q}_{\parallel},\omega) \) manifest as resonance peaks in the spectrum. Although we focus here on the electronic spectrum, the cavity spectrum \( S_{\rm cav}({q}_{\parallel},\omega) \) exhibits analogous features, reflecting the hybrid nature of the light-matter modes.
Alternatively, one can diagonalize the total Hamiltonian $\hat{\cal{H}}_{\rm tot}=\hat{\cal{H}}_{\rm LLs}+\hat{\cal{H}}_{\rm C}+\hat{\cal{H}}_{\rm cav}+\hat{\cal{H}}_{\rm int}$. In the single excitation manifold, electron-hole excitations between LLs can be treated as bosonic quasiparticles~\cite{ciuti2005}. In this limit, the total Hamiltonian reduces to a Hopfield Hamiltonian\cite{hopfield1958}, a quadratic bosonic Hamiltonian that describes light-matter excitations and can be exactly diagonalized in terms of polaritonic modes. The polariton energies obtained through this procedure coincide with the zeros of the RPA dielectric function \( \epsilon_{\rm RPA}({q}_{\parallel}, \omega) \).

\begin{figure}[t]
  \centering
  \vspace{1.em}
  \begin{overpic}[width=0.9\columnwidth]{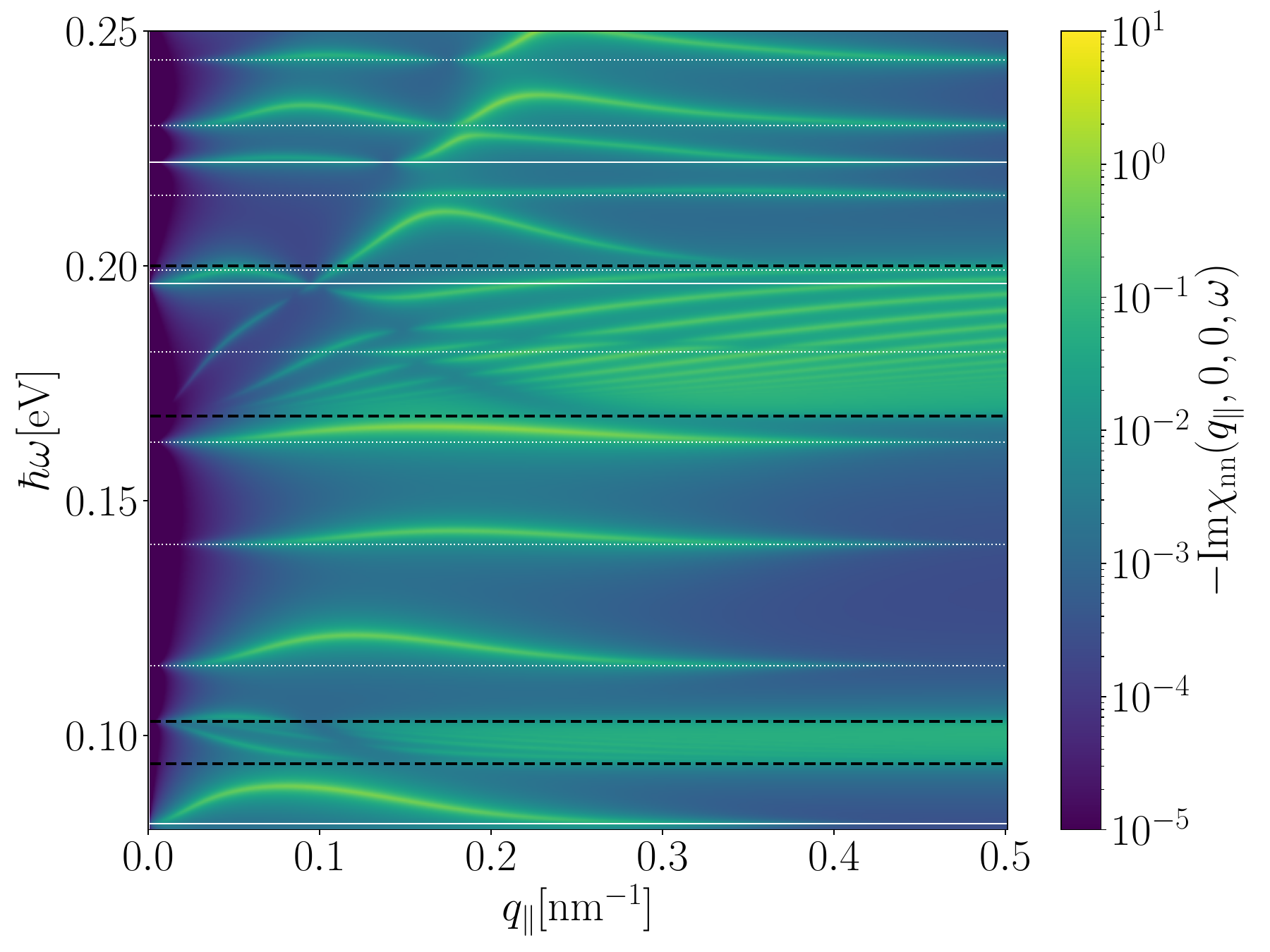}\put(0,80){\normalsize (a)}\end{overpic}
   \begin{overpic}[width=0.9\columnwidth]{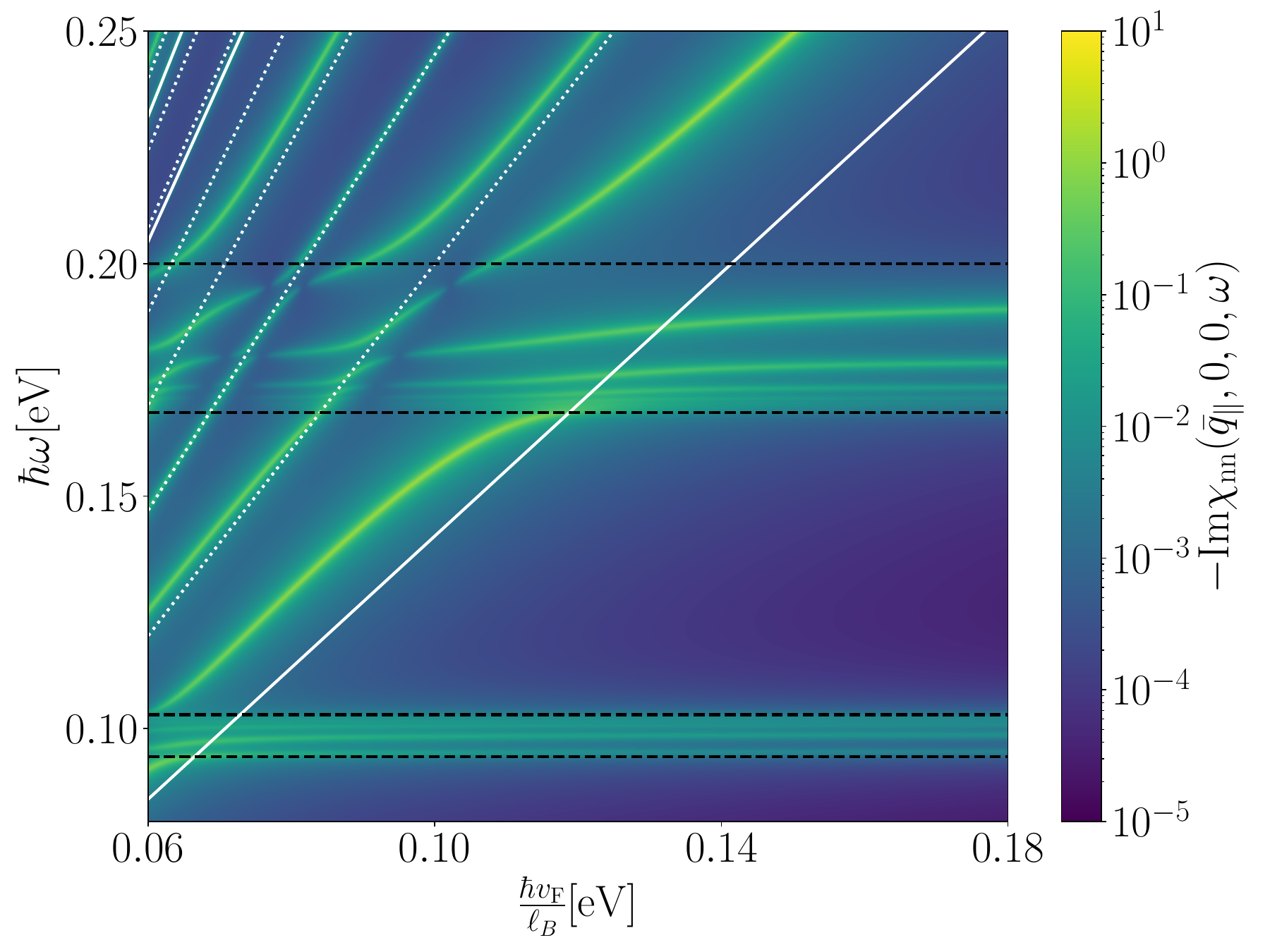}\put(0,80){\normalsize (b)}\end{overpic}
  \caption{(Color online)
The RPA density-density response of the coupled system (HPP cavity plus graphene), $-{\rm Im}~{\chi}_{\rm nn} ({q}_{\parallel}, \omega)$. Panel (a) shows $-{\rm Im}~{\chi}_{\rm nn} ({q}_{\parallel}, \omega)$ as a function of the wavevector $q_\parallel$ for a given magnetic field, $B=5~{\rm T}$ (corresponding to $\ell_B\approx 11.5~{\rm nm}$). Panel (b) shows the same quantity as a function of $\hbar v_{\rm F}/\ell_B$, for fixed momentum $\bar{q}_\parallel=0.1~{\rm nm}^{-1}$. Other parameters are $L_z=50~{\rm nm}$, $N_{\rm c}=30$, and $\hbar \eta= 5\times 10^{-4}~{\rm eV}$. For $g^{(0)}( q_{\parallel}, z, z', \omega)$, the expression resummed over all modes was employed. \label{fig2}}
\end{figure}

Fig.~\ref{fig2} shows the electronic spectrum at the neutrality point $N_{\rm F}=0$. Panel (a) presents this spectrum as a function of the wavevector $q_\parallel$, while panel (b) displays the spectrum as a function of $(\hbar v_{\rm F}/\ell_B)$. The latter spectrum can be experimentally obtained by varying the external magnetic field $B$ \cite{zhang2005}.
Inside the Reststrahlen bands (defined by $\omega_{{\rm T},i} < \omega < \omega_{{\rm L},i}$, with these frequencies indicated by black dashed lines), the LL transitions hybridize with HPP cavity modes. This hybridization gives rise to an energy splitting quantified by the vacuum Rabi frequency. These vacuum Rabi frequencies are explicitly calculated below, where the strength of this coupling it is found to reach up to 10\% of the cavity frequency, placing the system in the {\it ultra-strong coupling} regime \cite{Kockum19,geiser2012}. Moreover, due to the highly structured mode spectrum of the HPP cavity, a single LL transition simultaneously couples to multiple cavity modes—a scenario referred to as {\it super-strong coupling} \cite{kuzmin2019}. This regime gives rise to a rich polaritonic spectrum that goes beyond the simple picture of a two-level system interacting with a single mode.
This region of the spectrum inside the Reststrahlen bands is mainly controlled by the LL transitions encoded in ${\chi}^{(0)}_{\rm nn}({q}_{\parallel},\omega)$ and by the quantum Green's function associated with the cavity modes ${g}^{(0)}_{\rm q }({q}_\parallel, 0, 0, \omega)$, corresponding to the expansion in Eq.~\eqref{eqn:potential}. 
Outside the Reststrahlen bands, LL transitions exhibit a $q_\parallel$-dependent positive frequency shift, the \emph{depolarization shift}, \cite{Giuliani_and_Vignale,shtrichman2001}, which yields a finite dispersion as seen in Fig.~\ref{fig2}(a).
The depolarization shift is a collective Coulombic blue shift of electronic transitions, originating from the self-consistent longitudinal electric field generated by charge-density oscillations. In confined geometries, this effect is enhanced by electrostatic field confinement (e.g., due to metallic screening) and therefore persists even off resonance. In particular, it is not associated with resonant polaritonic poles but with the non-resonant part of the scalar Green's function \cite{Giuliani_and_Vignale,shtrichman2001}.
This separation between an instantaneous (Coulombic) renormalization of the matter resonance and resonant photonic poles is also emphasized in dipole-gauge treatments of confined intersubband polaritons, where the depolarization shift is encoded in the Coulomb/self-polarization sector of the Hamiltonian rather than in cavity-mode resonances \cite{todorov2012}.
Consistently, panel (b) shows that, even off-resonance, the LL transitions do not coincide with the non-interacting case (indicated by white solid and dashed lines). This shift originates from the frequency-independent Green's function $g_\infty^{(0)}({q}_{\parallel}, 0, 0)$ rather than from the HPP poles, as it occurs outside the Reststrahlen bands, where $\omega \neq \omega_{{q}_{\parallel}, n_z, s}$.
As detailed in Sect.~\ref{sec:Polaritons}, and in particular in Fig.~\ref{figS4}(b), this effect is present even in the absence of metallic confinement, i.e., in the limit $L_z\to\infty$, which corresponds to removing the metallic mirrors and filling the entire space with hBN. By contrast, strong cavity confinement reduces the effective Coulomb repulsion and therefore suppresses the depolarization shift.

As further detailed in Section~\ref{sec:Polaritons}, our theory allows us to compute the vacuum Rabi frequency $\Omega_{q_\parallel,{n}_z,{s},{\lambda},{n},{\lambda}',{n}'}$, defined as the coupling between the cavity mode $q_\parallel,{n}_z,{s}$ and the inter-LLs transition $n,\lambda \to n^\prime,\lambda^\prime$ given by:

\begin{align}
\label{eq:RabiMain}
& \hbar\Omega_{q_\parallel,{n}_z,{s},{\lambda},{n},{\lambda}',{n}'} = e\phi_{{q}_\parallel,n_z,s}(0)\sqrt{{ \mathcal{M}_{n^\prime,\lambda^\prime,n,\lambda}({q}_\parallel)}} ~,
\end{align}
where $\phi_{{q}_\parallel,n_z,s}$ are the mode's amplitude coefficients defined by Eq.~\eqref{eqn:potential} while ${{ \mathcal{M}_{n^\prime,\lambda^\prime,n,\lambda}({q}_\parallel)}}$ are the matrix elements of the electronic density operator in Eq.~\eqref{eq:dipole}. 
In standard  C-QED \cite{dalibard1982,DDB_arxiv_2023,ciuti2005}, it is well known that the Rabi frequency is proportional to the product of the vacuum fluctuations of the electric field $ \sqrt{ \langle\hat{E}^2\rangle}$ and the dipole moment of the atom in the cavity, $d$, such that $\Omega \sim d  \sqrt{ \langle\hat{E}^2\rangle}$. Eq.~\eqref{eq:RabiMain} is the most important result of this article as it generalizes this concept to non-local ($q_\parallel \neq 0$), sub-wavelength cavities interacting with a 2D material. This equation shows how the coupling depends on the vacuum fluctuations of the cavity field and of the electronic density.

Indeed, Eq.~\eqref{eq:RabiMain} shows that vacuum Rabi frequency is determined by the product of two independent factors: the vacuum fluctuation amplitude, quantified by $\phi_{{q}_\parallel,n_z,s}(0)$, and the matrix elements of the density operator, quantified by $\sqrt{\mathcal{M}_{n', \lambda', n, \lambda}(q_\parallel)}$. To maximize this coupling, each factor can be optimized separately.  As detailed in Sect.~\ref{sec:LLs} (see in particular Fig.~\ref{figS2}(b) and the related discussion), the response of the LLs wavefunctions depends only on product $q_\parallel\ell_B$ and the matrix element $\sqrt{{\mathcal{M}_{n^\prime, \lambda^\prime, n, \lambda}({q}_\parallel)}}$ reaches its maximum when the wavevector is comparable to the inverse of the magnetic length, 

\begin{equation}\label{eq:sim}
    q_\parallel \ell_B \sim 1~.
\end{equation}
Given the strong magnetic fields considered here ($B \approx 5~{\rm T}$, $\ell_B \approx 11.5~{\rm nm}$), this condition implies selecting large in-plane wavevectors, $q_\parallel \sim 0.1~{\rm nm}^{-1}$. In this regime, all LL transitions, including those conventionally considered dipole-forbidden, become optically active and significantly contribute to the overall coupling strength.  
However, as explained in Sect.~\ref{sec:HHP_cavities} (see in particular Fig.~\ref{figS1}(c) and related discussion), the vacuum fluctuations $\phi_{q_\parallel, n_z, s}(0)$ decrease significantly when the cavity dimension $L_z$ is large compared to $q^{-1}_\parallel$, i.e., for $q_\parallel L_z \gg 1$. Hence, to ensure optimal coupling strength, the cavity must satisfy both the LLs condition ($q_\parallel\ell_B\sim1$) and the confinement condition ($q_\parallel L_z \lesssim 1$).

\begin{figure}[t]
\centering
  \vspace{1.em}
  \begin{overpic}[width=0.9\columnwidth]{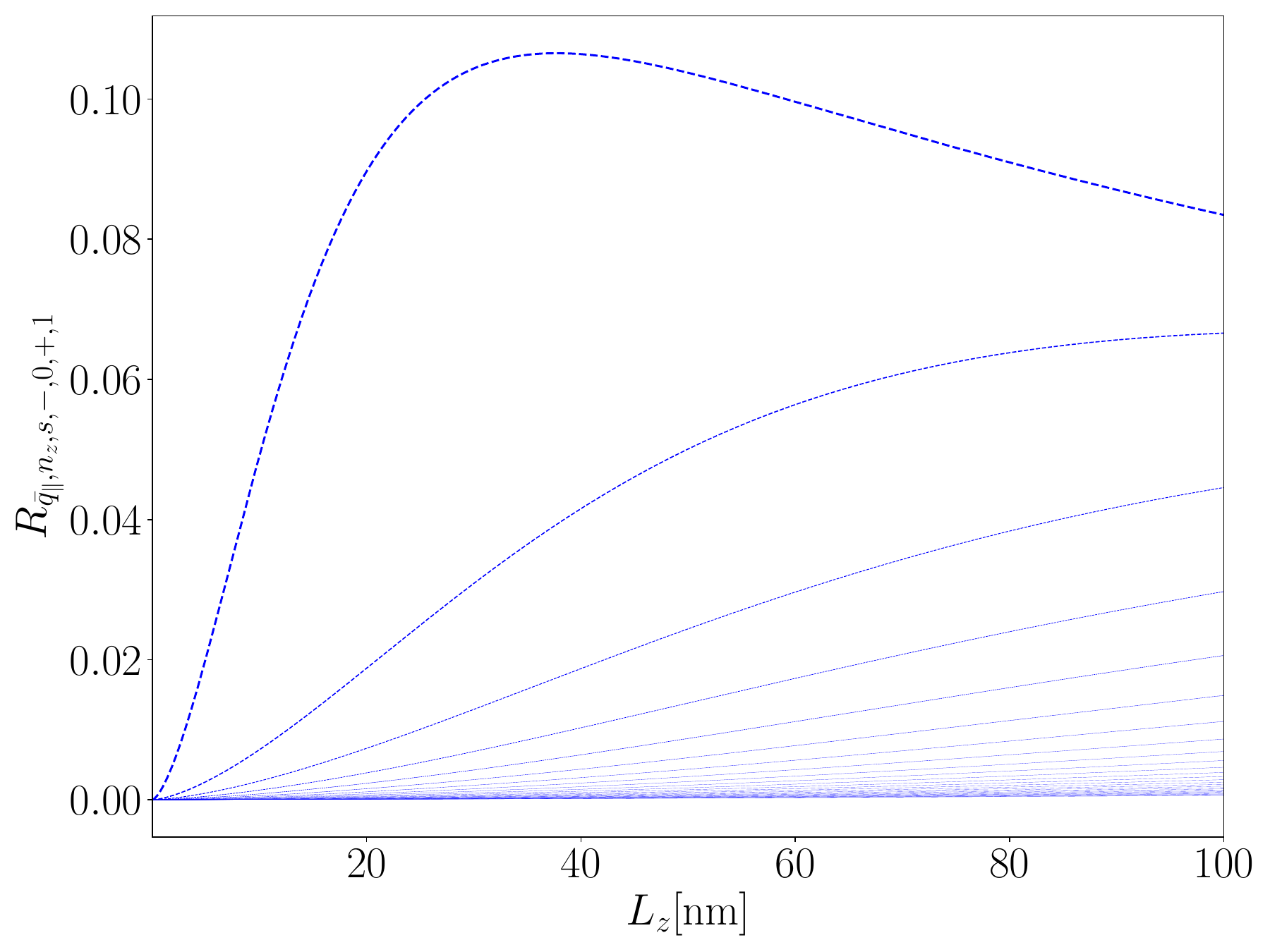}\put(0,80){\normalsize (a)}\end{overpic}
  \vspace*{0.2cm}
  \begin{overpic}[width=0.9\columnwidth]{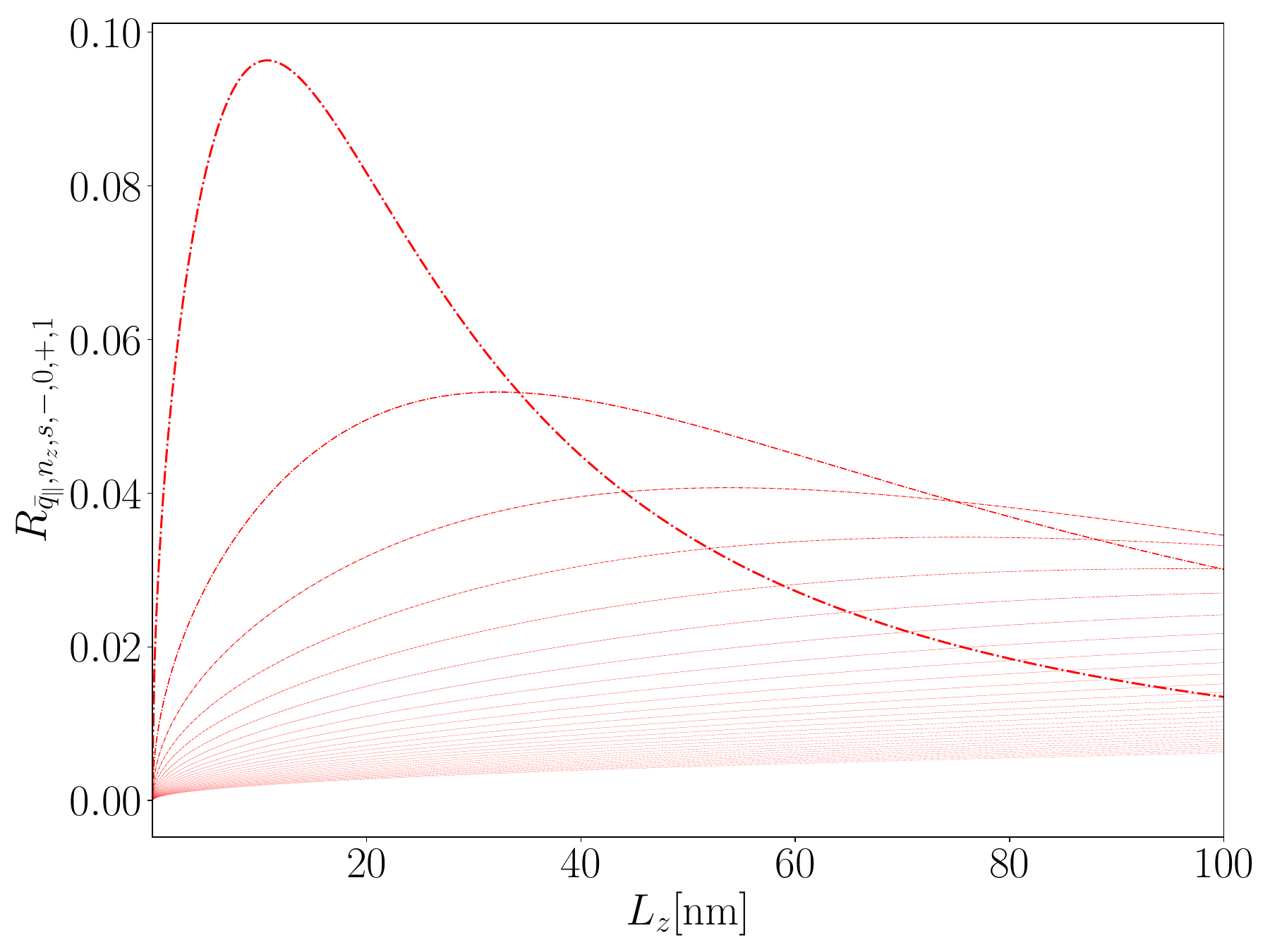}\put(0,80){\normalsize (b)}\end{overpic} 
 \caption{(Color online) Panels (a,b) display the ratio $R_{q_\parallel,n_z,s,-,0,+, 1} = (\Omega_{q_\parallel,n_z,s,-,0,+, 1} / \omega_{q_\parallel,n_z,s})$, where $\Omega_{q_\parallel,n_z,s,-,0,+, 1}$ denotes the Rabi frequency for the Landau level (LL) transition $(0, +1)$, and $\omega_{q_\parallel,n_z,s}$ is the cavity mode frequency. This ratio is shown as a function of $L_z$ (in ${\rm nm}$) for different $n_z$ for a fixed $\bar{q}_\parallel=0.1 {\rm nm^{-1}}$ and $B=10 {\rm T}$. Panels (a) show these quantities for the upper Reststrahlen band using blue dotted curves, while panels (b) present the lower Reststrahlen band with red dashed curves. The line thickness decreases and the color of the curves darkens as the quantum number $n_z$ increases.
  \label{fig_Rabi}}
\end{figure}

In Fig.~\ref{fig_Rabi}, we present the ratio  $R_{q_\parallel,n_z,s,-,0,+, 1}\equiv(\Omega_{\bar{q}_\parallel,n_z,s,-,0,+, 1}/\omega_{\bar{q}_\parallel,n_z,s})$ as a function of $L_z$ (in ${\rm nm}$) for different $n_z$ while $\bar{q}_\parallel=0.1 {\rm nm}^{-1}$. Notably, this quantity exhibits a nonmonotonic dependence on $L_z$, reaching a maximum for a specific $L_z$ while vanishing in both limits $L_z \to 0$ and $L_z \to \infty $. This behavior stands in stark contrast to the naive expectation that the Rabi frequency should scale as $1/\sqrt{L_z}$, an assumption that underlies the phenomenological use of the compression factor \cite{gao2020,schlawin2019,wang2019,nguyen2023,zezyulin2022,lenk2022,arwas2023,pavosevic2022}. While this scaling holds in the regime $\bar{q}_\parallel L_z \gg 1$, it fails for small $L_z$, where the coupling instead vanishes. 
As apparent from Fig.~\ref{fig_Rabi}, the maximum ratio $R_{\bar{q}_\parallel,n_z,s,-,0,+, 1}$, achieved for $n_z = 1$, exceeds $10\%$ for the upper Reststrahlen band and approaches $10\%$ for the lower band. This ratio places the system firmly within the ultra-strong coupling regime, where the coupling strength is a significant fraction of the cavity frequency \cite{Kockum19}. Notably, as further discussed in Section~\ref{sec:Polaritons}, the ultra-strong coupling regime is achieved at a finite wavevector $\bar{q}_\parallel$, emphasizing the necessity of employing a non-local theory. This result underscores the limitations of local approximations and highlights the critical role of spatial dispersion in accurately describing light-matter interactions in sub-wavelength cavities.
Additionally, the simultaneous hybridization of a single LL transition with multiple cavity modes—due to several non-negligible Rabi couplings—places the system in the super-strong coupling regime \cite{kuzmin2019}.

\begin{figure}[t!]
  \centering
  \vspace{1.em}
    \begin{overpic}[width=0.9\columnwidth]{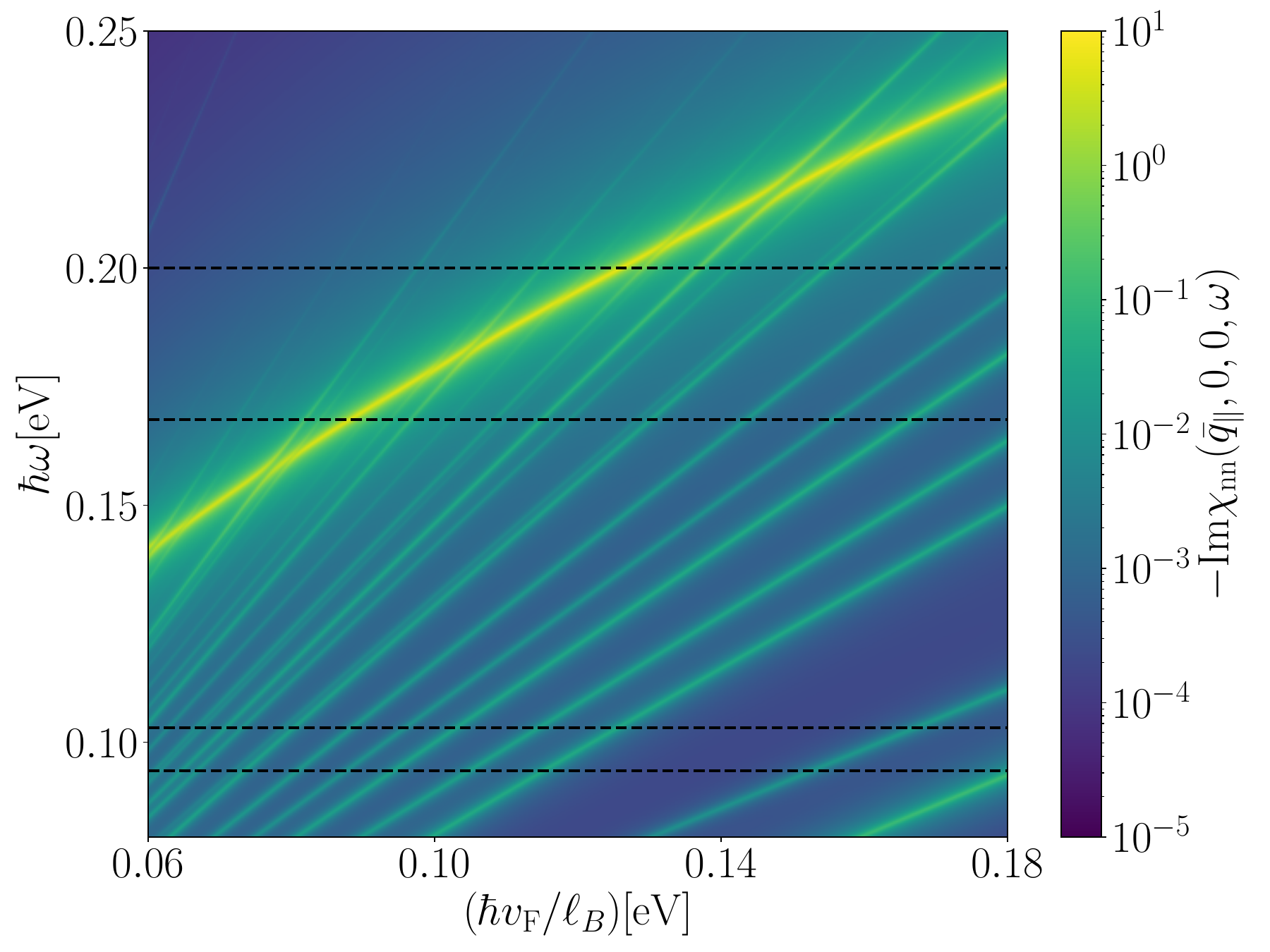}\put(0,80){\normalsize (a)}\end{overpic}
   \begin{overpic}[width=0.9\columnwidth]{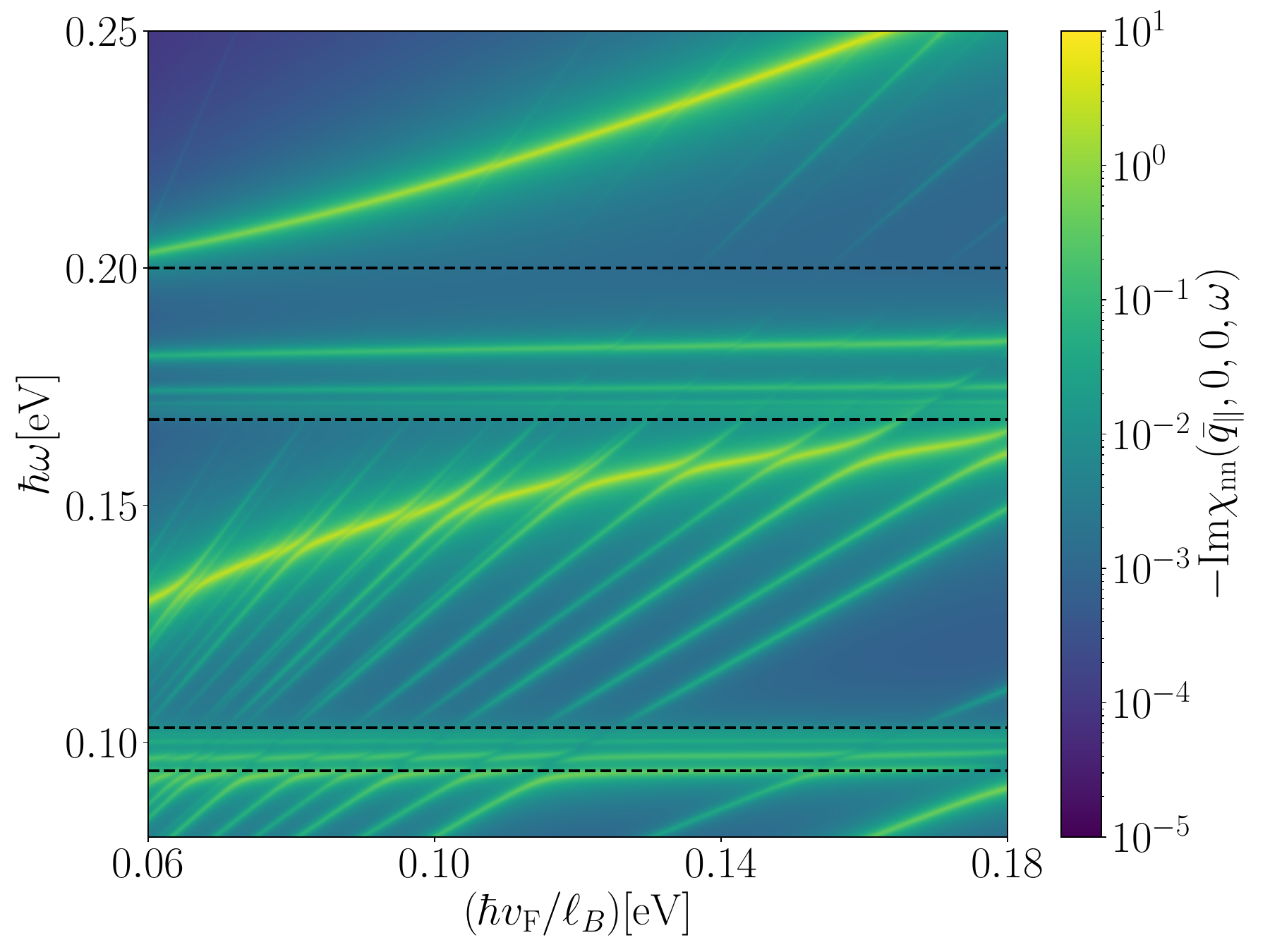}\put(0,80){\normalsize (b)}\end{overpic}
   \begin{overpic}[width=0.9\columnwidth]{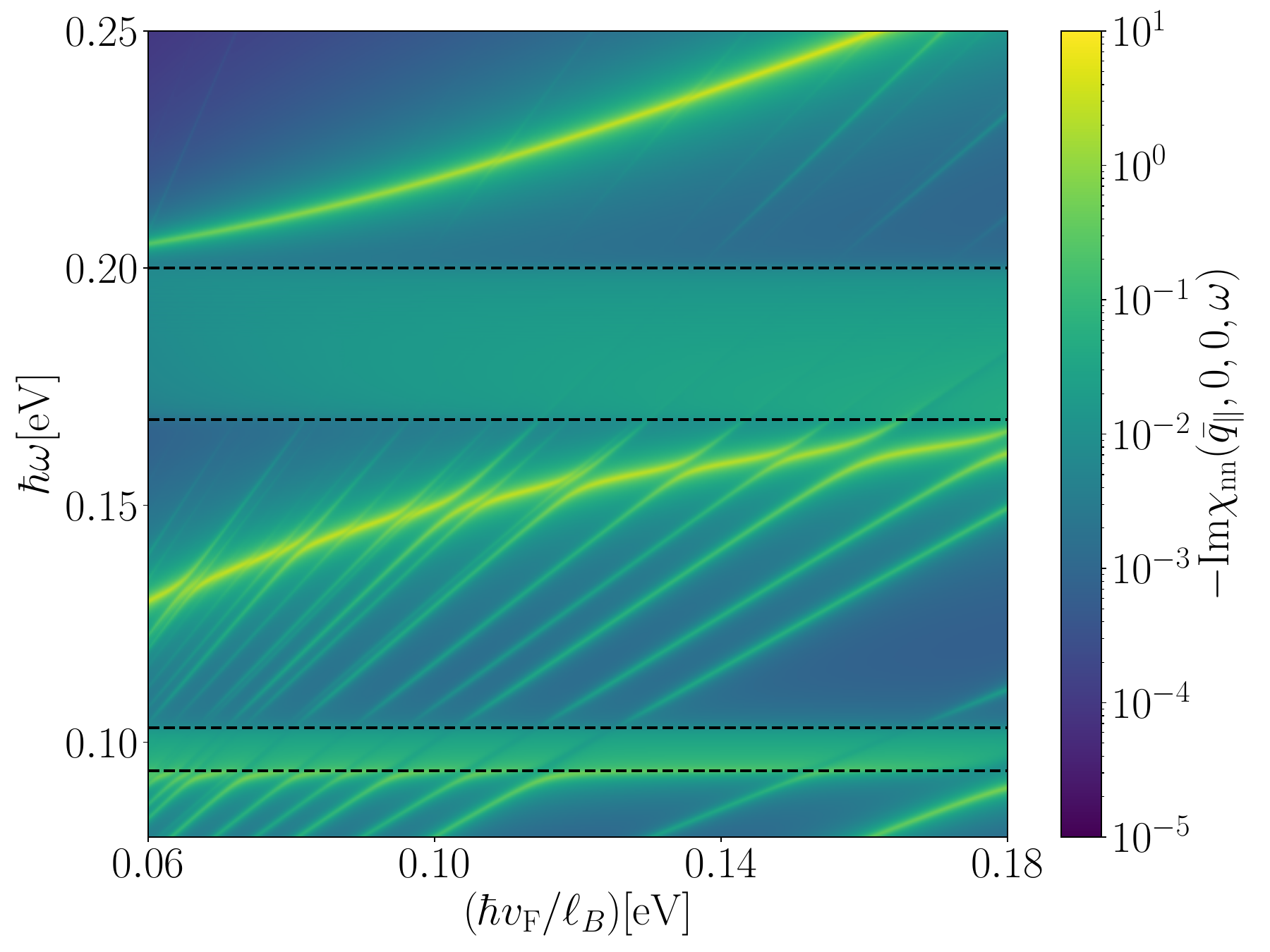}\put(0,80){\normalsize (c)}\end{overpic}
  \caption{(Color online) The RPA proper density-density response of the coupled system (HPP cavity plus graphene), $-{\rm Im}~{\chi}_{\rm nn} (\bar{q}_{\parallel}, \omega)$ as a function of $\hbar v_{\rm F}/\ell_B$ for a doped system ($N_{\rm F}=5$), for a fixed wavevector ($\bar{q}_{\parallel}=0.1{\rm nm}^{-1}$).
Panel (a) illustrates this quantity $-{\rm Im}~{\chi}_{\rm nn} (\bar{q}_{\parallel}, \omega)$ in the case where quantum mechanical modes are not included, $g_{\rm q}^{(0)}( q_{\parallel}, z, z', \omega)=0$. Panel (b) shows the same quantity in the presence of these modes. Other parameters are $L_z=50{\rm nm}$, $N_{\rm c}=30$, $\hbar \eta= 5\times 10^{-4}{\rm eV}$. Panel (c) is the same of panel (b), with $L_z=10^3{\rm nm} $. For $g^{(0)}( q_{\parallel}, z, z', \omega)$, the expression resummed over all modes was employed.\label{fig3} }
\end{figure}

We now examine the spectrum for finite doping. Fig.~\ref{fig3} shows the electronic spectrum for a doped system with $N_{\rm F}=5$, corresponding to a filling factor $\nu=22$, as $\nu=4N_{\rm F}+2$.  At finite charge density $N_{\rm F}\neq 0$ and strong magnetic fields, a plasmonic excitation appears in the system, usually referred to as {\it magnetoplasmon}. Magnetoplasmons \cite{Giuliani_and_Vignale, roldan2009,muravev2016} arise in doped semiconductors under a magnetic field due to the coupling between plasma oscillations of free carriers and their cyclotron resonance. In Fig.~\ref{fig3}(a) we show the spectrum without the contribution of the quantum mechanical modes (which is obtained by setting $g_{\rm q}^{(0)}({q}_{\parallel}, 0, 0, \omega)=0$), revealing the expected magnetoplasmon dispersion\cite{roldan2009}. Fig.~\ref{fig3}(b) is obtained by taking into account the full cavity contribution. Here, the magnetoplasmon strongly hybridizes with the cavity modes achieving a Rabi splitting of $\approx 25\%$ of the cavity frequency. This significant coupling arises from the well-known fact that most of the oscillator strength, which quantifies the ability of a system to absorb or emit electromagnetic radiation, is associated with the magnetoplasmon resonance\cite{Giuliani_and_Vignale}.  Finally, we note that in the $L_z \to \infty$ limit the Rabi splitting shown in Fig.~\ref{fig3}(c) remains of a similar order. Although the coupling between LLs decreases with reduced compression, the corresponding increase in the density of cavity modes compensates, resulting in a splitting\cite{Dai2015, Kumar2015} comparable to the case in which $L_z$ is finite. 
The details are provided in Sect.~\ref{sec:Polaritons}.

\subsection{Relation to prior works}
\label{sec:comparison}

As discussed in the Introduction, many theoretical works on cavity quantum materials rely either on a phenomenological uniform vector potential or on the vector potential derived for Fabry-Pérot cavities, which host modes that fulfill the {\it diffraction limit}. These approaches are often applied to sub-wavelength cavities using a phenomenological compression factor \( A \) \cite{gao2020,schlawin2019,wang2019,nguyen2023,zezyulin2022,lenk2022,arwas2023,pavosevic2022,Dag2024} or by fitting the vacuum field to experimental values.  

Ref.~\onlinecite{andolina2024} demonstrated that, in the generalized Coulomb gauge, sub-wavelength cavities — characterized by a large compression factor \( A=(V/\lambda^3_{\rm d}) \ll 1 \) — exhibit an electric field with a predominantly electrostatic nature, described by the scalar potential \( \hat{\phi} \), while the magnetic field remains negligible.
Consequently, in such cavities, the effective electron-electron interactions, obtained by tracing the degrees of freedom of the cavity, are primarily density-density interactions governed by the dielectric tensor \( \epsilon_{i,j}(\bm{r},\omega) \).
In contrast, transverse current--current interactions (analogous to the magnetostatic Ampère force), which dominate in Fabry--Pérot cavities, are not significantly enhanced in sub-wavelength cavities, since they depend only on the magnetic permeability $\mu(\bm r,\omega)$ rather than on the dielectric tensor. However, in sub-wavelength cavities the magnetic permeability remains approximately that of free space, $\mu(\bm r,\omega)\simeq 1$, thus preventing any substantial enhancement of these interactions. This point is further discussed in Ref.~\onlinecite{andolina2024}, where it is shown that Amperean superconductivity—arising from transverse current--current interactions—cannot be realized in sub-wavelength cavities.
This discussion underscores the necessity of developing a new theoretical framework capable of microscopically describing the vacuum electromagnetic field in sub-wavelength cavities and its coupling with matter.

Here, we developed this framework by quantizing sub-wavelength cavities. For any generic sub-wavelength cavity, our theory enables the quantization of the electric field and thus the calculation of the mode amplitude coefficients, \( \phi_{{q}_\parallel,n_z,s}(z) \), given in Eq.~\eqref{eqn:potential}, which represent the vacuum fluctuations of the electric field, as well as the vacuum Rabi frequencies \( \Omega_{q_\parallel,{n}_z,{s},{\lambda},{n},{\lambda}',{n}'} \) in Eq.~\eqref{eq:RabiMain}. Additionally, our calculations reveal the renormalization of the electrostatic Coulomb interaction, represented by \( g^{(0)}_\infty({q}_\parallel, 0, 0) \).  Our quantization scheme fundamentally departs from the macroscopic QED formalism developed for dispersive and absorbing media \cite{Knoll00,Buhmann08}, which relies on the dyadic Green tensor of Maxwell’s equations and introduces a continuum of bosonic reservoir fields to account for material losses.
In contrast, our approach is fully unitary and specifically formulated for deep-sub-wavelength cavities, where the electromagnetic response is dominated by longitudinal, quasi-electrostatic modes. In this regime, the field quantization is performed directly on the scalar potential $\hat\phi$ associated with the confined modes, which are described by the scalar Green function $g^{(0)}$, without invoking any auxiliary reservoir or dissipative degrees of freedom.

Our scheme also differs from approaches that start from Maxwell’s equations in matter together with explicit polarization dynamics, construct a full Hamiltonian and then quantize the resulting normal modes \cite{Gubbin_JChemPhys_2022,Li18}. As discussed in Sect.~\ref{sec:HHP_cavities}, we bypass the explicit Hamiltonian construction and obtain the correct mode normalization by matching the quantum scalar Green function with its classical counterpart solving the Poisson equation. The formal equivalence between the two procedures is demonstrated in Appendix~\ref{app_alt}.

We applied our theory to electrons in graphene subject to a strong perpendicular magnetic field \( B \) and coupled to the sub-wavelength modes of an HPP cavity. Ref.~\onlinecite{ashida2023} investigated graphene in a magnetic field embedded in a cavity where the mirrors are made of hBN. Despite the apparent similarity, that study focused solely on transverse cavity modes, considering only the excitations of the vector potential \( \hat{\bm A} \), while neglecting the longitudinal cavity modes \( \hat{\phi} \) and the Coulomb interaction \( g^{(0)}_\infty({q}_\parallel, 0, 0) \). However, as demonstrated in Ref.~\onlinecite{andolina2024}, when the field is confined below the diffraction limit, $V\ll \lambda_{\rm d}^3$, the electrostatic contributions encoded in \( g^{(0)}({q}_\parallel, 0, 0,\omega) \) are dominant. Thus, our analysis is valid in the regime of strong confinement and highly non-local wavevectors considered in this article, whereas the theory in Ref.~\onlinecite{ashida2023} applies to the complementary regime of long wavelength where {\it retardation} effects due to the finite speed of light are crucial. 

Numerous previous theoretical works \cite{ciuti2021,rokaj2022,rokaj2023} considered 2DEGs in sub-wavelength split-ring cavities, motivated by pioneering experiments on the quantum Hall effect in such systems \cite{appugliese2022,enkner2024,enkner2024b}. However,  these works \cite{ciuti2021,rokaj2022,rokaj2023} modeled the cavity as a single uniform transverse mode, $\hat{\bm{A}}={{A}}_0\bm{\epsilon}\left(\hat{a}+\hat{a}^\dagger\right)$, where $\hat{a},\hat{a}^\dagger$ are the photonic operators, $\bm{\epsilon}$ is the polarization and ${A}_0=\sqrt{2\pi \hbar c^2/V_{\rm eff} \omega_c} $ is the vacuum cavity field in a Fabry-P\'{e}rot cavity, $\omega_c$ is the cavity frequency and $V_{\rm eff}=A\lambda^3_{\rm d}$ is the effective cavity volume rescaled by the use of the phenomenological compression factor $A$. While this approach may capture the main phenomenology, it neglects the inhomogeneity of the cavity mode \cite{enkner2024b}, lacks a microscopic prediction of the vacuum field and omits frequency-independent electrostatic effects (such as screening due to metal gates), which in our theory are encoded in \( g_\infty(\bm{r},\bm{r}^\prime) \). Therefore, a fully microscopic treatment of the vacuum field in sub-wavelength cavities is necessary to describe these experiments accurately.  In the future, it would be interesting to apply our general theory also to the specific setup of a 2DEG in a split-ring cavity, fully accounting for both the inhomogeneous cavity field and the renormalized Coulomb potential.

\section{Details on ``Theory of the Green's function of HPP cavities"}
\label{sec:HHP_cavities}

In this Section, we outline the key steps and methods used to derive the scalar Green's function for HPP cavities.
Furthermore, we develop a quantization scheme for the electric field in a sub-wavelength cavity by employing the cavity scalar Green's function. This approach enables us to relate the vacuum fluctuations of the scalar potential to the amplitude coefficients of each mode. 
Finally, we decompose the scalar Green's function into its static and dynamical components, allowing us to distinguish between resonant effects and off-resonant contributions associated with screening.

\subsection{Mode decomposition for the scalar Green's function}
\label{Subsec:modes}

We now calculate the scalar Green's function $g^{(0)}(\bm{r}, \bm{r}', \omega)$ for the HPP cavity described above.
This Green's function is determined by the Poisson equation:
\begin{equation}\label{eqn:scalar_green_function}
-\sum_{i,j}\partial_j' [\epsilon_{ji}(\bm{r}', \omega)\partial_i' g^{(0)}(\bm{r}, \bm{r}', \omega)] = 4\pi\delta(\bm{r}-\bm{r}')~,
\end{equation}
where $\epsilon_{ji}(\bm{r}', \omega)$ is the permittivity tensor given in Eq.~\eqref{eq:Drude-Lorentz}, and $\partial_j'$ denotes the partial derivative with respect to the coordinates of $\bm{r}'$.

The electric potential $\phi(\bm{r},\omega)$ in the presence of an external charge $\rho_{\rm ext}(\bm{r}',\omega)$ is given by the convolution of the Green's function with the charge density:

\begin{equation}\label{eqn:electric_potential}
\phi(\bm{r},\omega)=\int d\bm{r}' g^{(0)}(\bm r, \bm r', \omega) \rho_{\rm ext}(\bm{r}',\omega)~.
\end{equation}
We now lay down a generic mode expansion technique to solve Eq.~\eqref{eqn:scalar_green_function}. The Poisson equation can be rewritten as

\begin{equation}\label{eqn:operator_on_green}
\hat{O}_{\bm{r}'}(\omega) g^{(0)}(\bm r, \bm r', \omega) = 4\pi\delta(\bm r-\bm r')~,
\end{equation}
where $\hat{O}_{\bm{r}'}(\omega)f(\bm{r}', \omega) \equiv - \sum_{i,j}\partial_j' [\epsilon_{ji}(\bm{r}', \omega)\partial_i' f(\bm{r}', \omega)]$ and $f(\bm{r}', \omega)$ is an arbitrary function.

We introduce an orthonormal basis set $f_{\mu}(\bm{r},\omega)$ such that:
\begin{equation}\label{eqn:basis_function}
O_{\bm{r}}(\omega)f_\mu(\bm{r},\omega)=\lambda_\mu(\omega)f_\mu(\bm{r},\omega)~,
\end{equation}
where $\lambda_\mu(\omega)$ are the eigenvalues associated with these basis functions. The completeness and orthonormality of the basis set are given by:

\begin{align}\label{eqn:completeness}
  \sum_\mu f_\mu(\bm{r},\omega)f^*_\mu(\bm{r}',\omega)=\delta(\bm{r}-\bm{r}')~, \\
  \label{eqn:orthogonality}
  \int d\bm{r} f^*_\mu(\bm{r},\omega)f_\nu(\bm{r},\omega)=\delta_{\mu,\nu}~.
\end{align}
Using Eq.~\eqref{eqn:completeness} ,we can expand the external density $\rho_{\rm ext}(\bm{r},\omega)$ in terms of the basis set:

\begin{align}
 \rho_{\rm ext}(\bm{r},\omega)=\sum_\mu a_\mu f_\mu(\bm{r},\omega)~, \label{eqn:charge_density_expansion} \\
 a_\mu=\int d\bm{r}' f_\mu(\bm{r}',\omega) \rho_{\rm ext}(\bm{r}',\omega)~. \label{eqn:expansion_coefficients}
\end{align}
The coefficients $a_\mu$ are projections of the charge density onto the basis functions. It can be seen by inspection that the potential:

\begin{equation}\label{eqn:potential_expansion}
 \phi(\bm{r},\omega)=\sum_\mu \frac{4\pi a_\mu f_\mu(\bm r,\omega)}{\lambda_\mu(\omega)}~,
\end{equation}
solves Eq.~\eqref{eqn:scalar_Poisson}. By comparing Eq.~\eqref{eqn:charge_density_expansion} and Eq.~\eqref{eqn:potential_expansion} with Eq.~\eqref{eqn:electric_potential}, we find the following mode expansion for the Green's function:

\begin{equation}\label{eqn:green_function_expansion}
{\rm Re}~g^{(0)}(\bm r, \bm r', \omega)= 4\pi\sum_\mu\frac{f_\mu(\bm r,\omega)f^*_\mu(\bm r',\omega)}{\lambda_\mu(\omega)}~.
\end{equation}
The poles of the Green's function correspond to the condition $\lambda_\mu(\omega) = 0$ and are determined by the frequencies $\omega_{\mu, s}$, where $s$ indexes different solutions of $\lambda_\mu(\omega) = 0$. Assuming the mode function is frequency independent, $f_\mu(\bm{r}, \omega) = f_\mu(\bm{r})$, we calculate the imaginary part of the Green's function ${\rm Im}~g^{(0)}(\bm{r}, \bm{r}', \omega)$ by substituting $\lambda_\mu(\omega) \to \lambda_\mu(\omega + i\eta)$ and taking the limit $\eta \to 0$. In calculating  ${\rm Im}~g^{(0)}(\bm{r}, \bm{r}', \omega)$, we are allowed to expand the denominator around positive (negative) poles $\omega_{\mu, s}$ ($-\omega_{\mu, s}$) as:

\begin{equation}
\begin{split}
  \lambda_\mu(\omega+i\eta)\approx \partial_\omega \lambda_\mu(\omega)|_{\omega=\omega_{\mu, s}}\left(\omega+\I\eta-\omega_{\mu,s}\right)~,\\
  \lambda_\mu(\omega+i\eta)\approx -\partial_\omega \lambda_\mu(\omega)|_{\omega=\omega_{\mu, s}}\left(\omega+\I\eta+\omega_{\mu,s}\right)~,
\end{split}
\end{equation}
where the first equation holds in the vicinity of the positive pole, and the second equation applies near the negative pole. We assumed that $\lambda_\mu(\omega)$ is an even function of $\omega$, making $ \partial_\omega \lambda_\mu(\omega)$ an odd function of $\omega$.
In the limit $\eta \to 0$, this yields:

\begin{equation}
\begin{split}
    \label{eqn:green_function_IM}
&{\rm Im}~g^{(0)}(\bm r, \bm r', \omega)= \\ 
&=-4\pi^2\sum_{\mu,s}\frac{f_\mu(\bm r)f^*_\mu(\bm r')}{\partial_{\omega}\lambda_\mu(\omega)|_{\omega=\omega_{\mu,s} }} \left[\delta(\omega-\omega_{\mu,s})-\delta(\omega+\omega_{\mu,s})\right]~.
\end{split}
\end{equation}
Being a causal response function, the Green's function is expected to fulfill the Kramers-Kronig relations \cite{Giuliani_and_Vignale} and therefore it should vanish in the limit $\omega \to \infty$. However, the term

\begin{equation}
\begin{split}
&g_{\infty}^{(0)}(\bm r,\bm{r}^\prime) \equiv\lim_{\omega\to\infty}  g^{(0)}(\bm r,\bm{r}^\prime, \omega)~.\\
\end{split}
\end{equation}
is generally finite. In the high-frequency limit, far from resonant poles, the function $g_{\infty}^{(0)}(\bm r,\bm{r}^\prime)$ can be assumed to be a real function. However, the real and imaginary parts of the difference $g^{(0)}(\bm{r}, \bm{r}^\prime, \omega) - g_{\infty}^{(0)}(\bm r,\bm{r}^\prime)$, being causal response function that goes to zero for $\omega\to\infty$, satisfy the Kramers-Kronig relations \cite{Giuliani_and_Vignale}:

\begin{align}
\label{eq:KK1}
  & {\rm Re}~g^{(0)}(\bm{r}, \bm{r}^\prime, \omega)= \nonumber\\&= g_{\infty}^{(0)}(\bm r,\bm{r}^\prime)+ \frac{2}{\pi}\int_{0}^{\infty} \frac{\omega'  {\rm Im} ~ g^{(0)}(\bm{r}, \bm{r}^\prime, \omega)}{\omega'^2 - \omega^2} \, d\omega'~,\\ 
  & {\rm Im} ~g^{(0)}(\bm{r}, \bm{r}^\prime, \omega) = \nonumber\\&=-\frac{2 \omega}{\pi} \int_{0}^{\infty} \frac{{\rm Re} \, g^{(0)}(\bm{r}, \bm{r}^\prime, \omega)-g_{\infty}^{(0)}(\bm r,\bm{r}^\prime)}{\omega'^2 - \omega^2} \, d\omega'~. 
\end{align}
We now discuss the specific case of our hPP cavity. Details on the dielectric tensor in hBN are reported in Appendix \ref{app0}.  Here, the mode functions solve the following eigenvalue equation: 

\begin{equation}\label{eq:green_function_hBN}
-\big[\epsilon_z(\omega)\partial^{2}_z+ \epsilon_x(\omega)\nabla^{2}_\parallel\big] f_\mu(\bm r, \omega)=\lambda_\mu(\omega) f_\mu(\bm r,\omega)~,
\end{equation}
with the boundary condition $f_\mu(\bm{r},\omega)|_{z=L_z/2}=f_\mu(\bm{r},\omega)|_{z=-L_z/2}=0$.

\begin{figure*}[t]
\centering
  \vspace{1.em}
  \begin{overpic}[width=0.68\columnwidth]{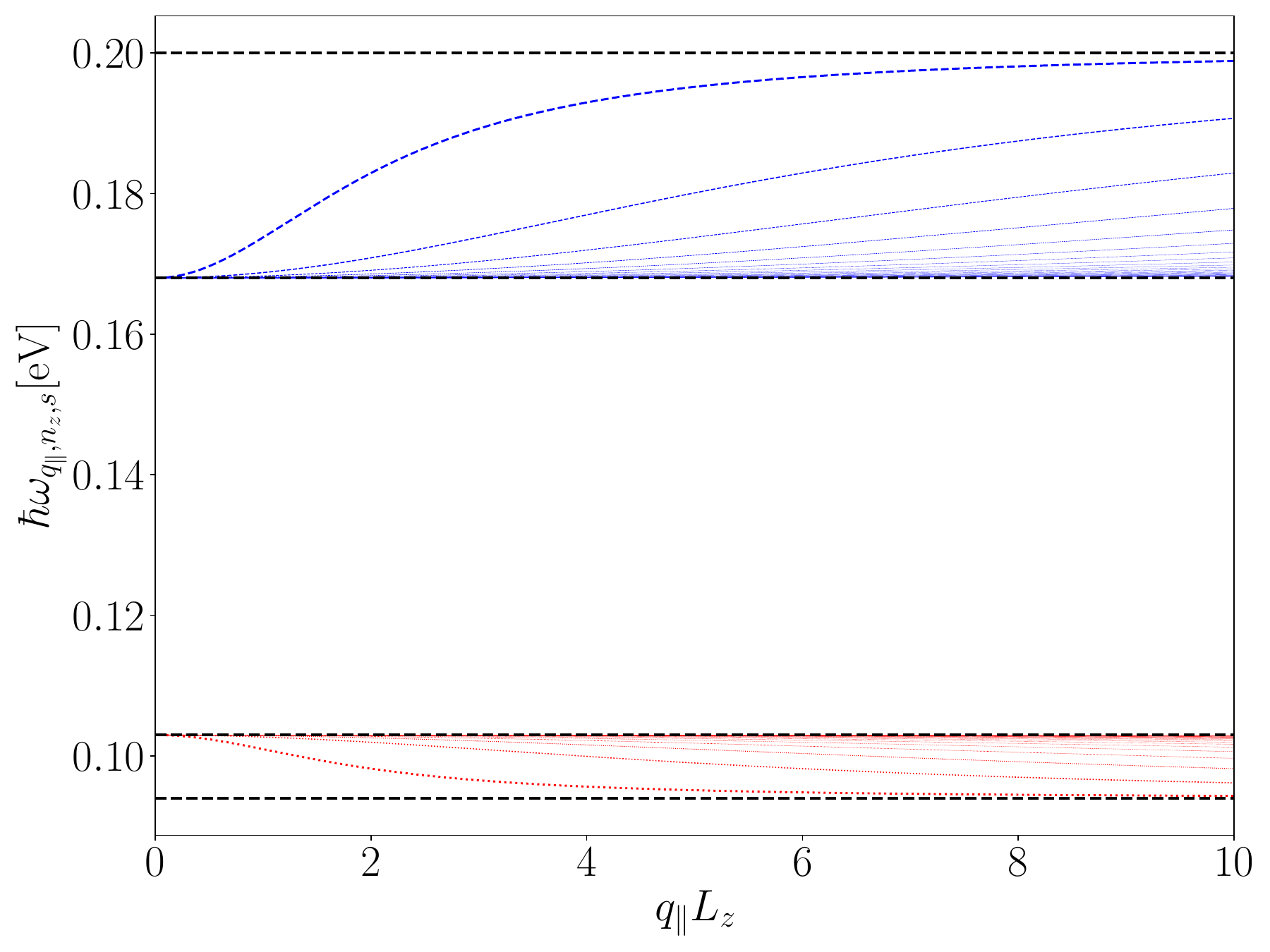}\put(0,80){\normalsize (a)}\end{overpic} 
  \begin{overpic}[width=0.68\columnwidth]{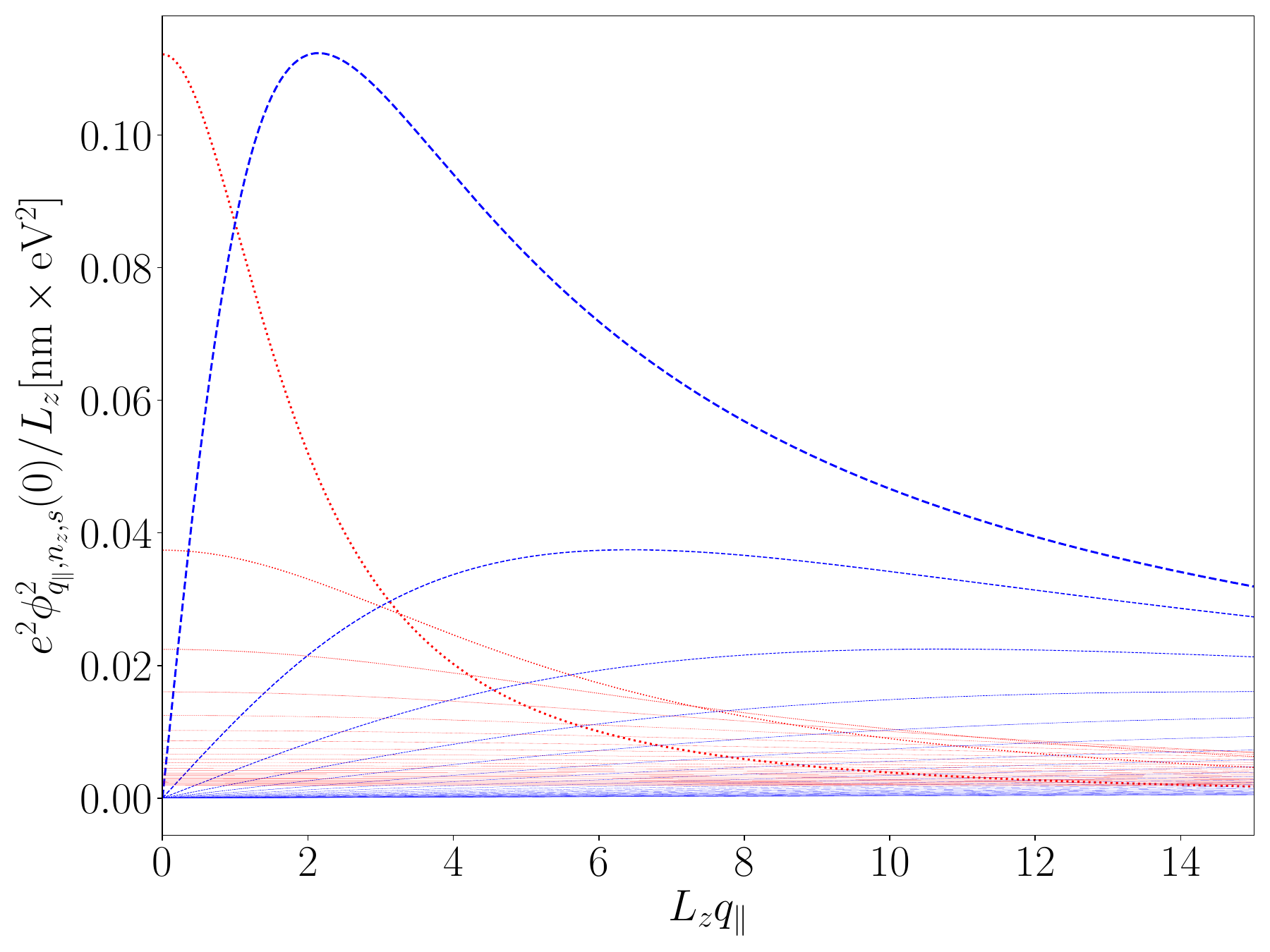}\put(0,80){\normalsize (b)}\end{overpic} 
  \begin{overpic}[width=0.68\columnwidth]{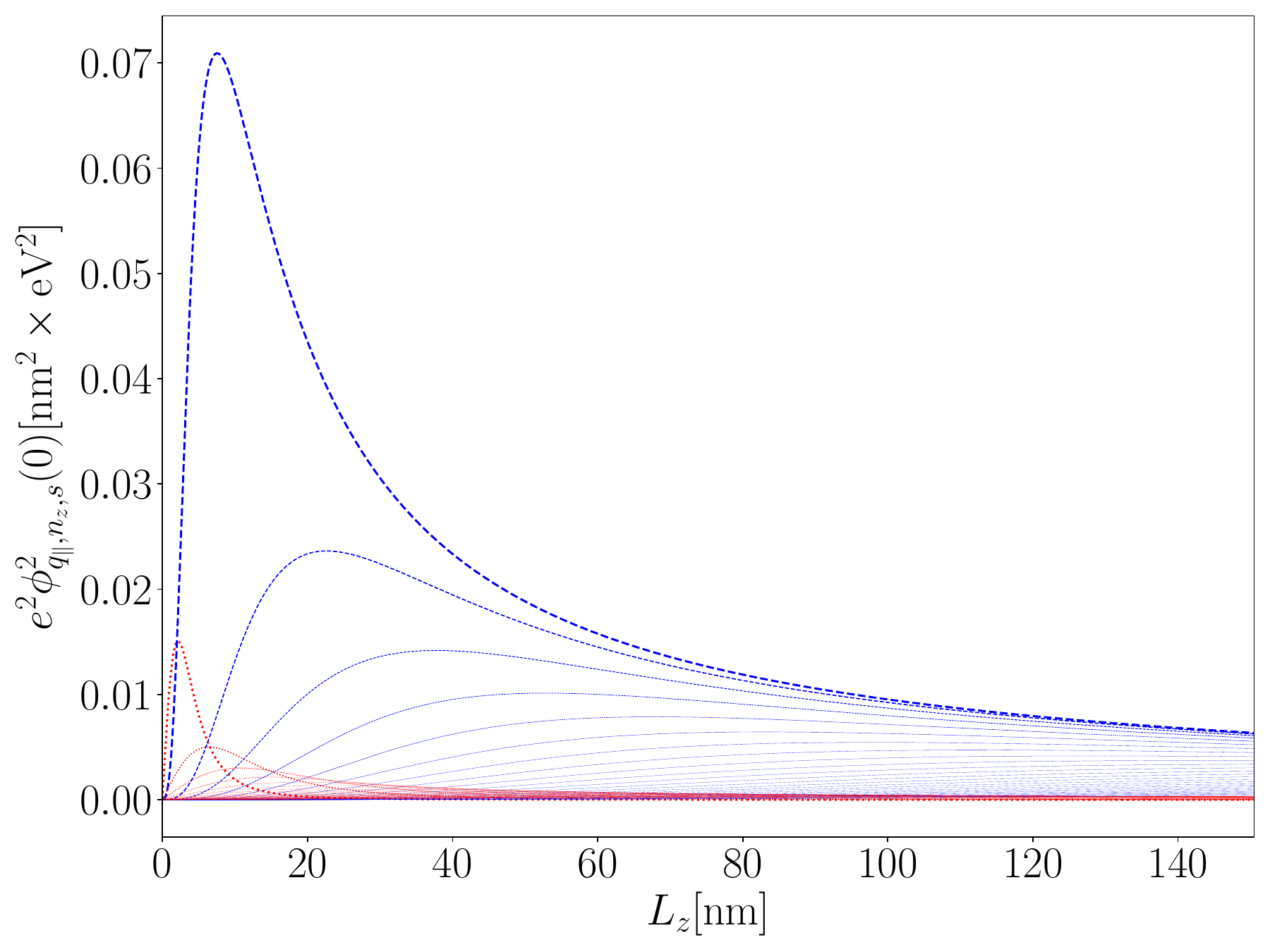}\put(0,80){\normalsize (c)}\end{overpic} 
 \caption{(Color online) Panel (a) shows the cavity frequencies $\hbar\omega_{q_\parallel,n_z,s}$ (in ${\rm eV}$) as a function of $q_\parallel L_z$ for different $n_z$. Each of the two regions corresponding to the Reststrahlen bands is enclosed between two black dashed lines. Panel (b) displays the bare cavity vacuum fluctuations $e^2\phi^2_{{q}_\parallel,n_z,s}(0)$ (normalized by $L_z$ and expressed in ${\rm nm\times eV}^2$) as a function of $q_\parallel L_z$ for different $n_z$. Panel (c) illustrates $e^2\phi^2_{{q}_\parallel,n_z,s}(0)$ (expressed in ${\rm nm}^2\times {\rm eV}^2$) as a function of $L_z$ (in $\text{nm}$) for different $n_z$. In all panels, the upper (lower) Reststrahlen band is shown in blue (red). The thickness of the lines decreases as the quantum number $n_z$ increases. \label{figS1}}
\end{figure*}

The mode functions and eigenvalues for this system are:

\begin{align}\label{eqn:mode_functions_hBN1}
 f_{\bm{q}_{\parallel},n_z}(\bm{r})=\sqrt{\frac{2}{L_zS}}e^{\I\bm{q}_\parallel\cdot\bm{r}_\parallel}\sin\left(q_{z,n_z}\left[z+\frac{L_z}{2}\right]\right)~, \\
 \lambda_{\bm{q}_{\parallel},n_z}(\omega)=\epsilon_z(\omega)q_{z,n_z}^2+\epsilon_x(\omega)q_\parallel^2~,
\end{align}
where $\mu=\bm{q}_\parallel,n_z$ indicate the modes, $\bm{q}_\parallel$ is the in-plane wavenumber, and $q_{z,n_z}=n_z\pi/{L_z}$, $n_z$ being positive integers.
By means of Eq.~\eqref{eqn:green_function_expansion} we can obtain the mode expansions of the scalar Green's function:

\begin{equation}
\begin{split}
\label{eqn:green_function_hBN_expansion}
&g^{(0)}(\bm r_\parallel, \bm r'_\parallel,z,z', \omega)=\\
=&\frac{8\pi}{L_zS}\sum_{\bm{q}_{\parallel},n_z}\frac{e^{\I\bm{q}_\parallel\cdot(\bm{r}_\parallel-\bm{r}'_\parallel)}\sin\left(q_{z,n_z} [z+\frac{L_z}{2}]\right) \sin\left(q_{z,n_z} [z^\prime+\frac{L_z}{2}]\right)}{\epsilon_z(\omega)q_{z,n_z}^2+\epsilon_x(\omega)q_\parallel^2} ~.
\end{split}
\end{equation}
Poles of Eq.~\eqref{eqn:green_function_hBN_expansion} are determined by $\lambda_{\bm{q}_{\parallel},n_z}(\omega)=0$, or, equivalently:
\begin{align}\label{eqn:mode_functions_hBN}
 \frac{\epsilon_z(\omega)}{\epsilon_x(\omega)}=-\left(\frac{q_\parallel}{q_{z,n_z}}\right)^2~.
\end{align}
This equation is solved for $\omega=\omega_{{q}_\parallel,q_{z,n_z},s}$ where $s=1$ ($s=-1$) denotes the upper (lower) Reststrahlen band. 
From this equation, it is clear that: {\it i}) the poles are only present in the hyperbolic region where $\epsilon_z(\omega)\epsilon_x(\omega)<0$ ; {\it ii}) the position of the poles depends only on the product $q_\parallel L_z$ rather than on the two parameters individually.
Fig.~\ref{figS1}(a) shows the cavity energies $\hbar\omega_{{q}_\parallel,q_{z,n_z},s}$ associated with the poles of Eq.~\eqref{eqn:green_function_hBN_expansion} as a function of $q_\parallel L_z$ for different values of $n_z$.

By performing the Fourier transform of Eq.~\eqref{eqn:green_function_hBN_expansion} we obtain Eq.~\eqref{eqn:green_function_hBN_expansion_Main}.
Along the plane $z=0$ (and thus also $z'=0$) the sum over $n_z$ can be performed analytically.  In the case of the scalar Green's function, it results in

\begin{equation}
\begin{split}
\label{eqn:green_function_hBN_classical}
& g^{(0)}(q_{\parallel},0,0, \omega)= \frac{2\pi}{q_\parallel\sqrt{{\epsilon}_x(\omega){\epsilon}_z(\omega)}}\tanh\left[\sqrt{\frac{{\epsilon}_x(\omega)}{{\epsilon}_z(\omega)}}\frac{L_zq_\parallel}{2}\right] ~.
\end{split}
\end{equation}
This expression describes the Coulomb interaction within a cavity, modified by the dielectric environment characterized by the permittivities  ${\epsilon}_x(\omega)$ and ${\epsilon}_z(\omega)$, along the $x$ and $z$ directions, respectively. The presence of the mirrors at the cavity boundaries plays a critical role by effectively introducing image charges that modify the electron-electron interactions. This alteration leads to the appearance of the hyperbolic tangent term, which reflects the confinement effect of the cavity along the $z$-axis.
It is also useful to recast this equation in the following form:

\begin{equation}
\begin{split}
\label{eqn:green_function_hBN_classical1}
& g^{(0)}( q_{\parallel},0,0, \omega)= \frac{2\pi L_z}{q_\parallel L_z\sqrt{{\epsilon}_x(\omega){\epsilon}_z(\omega)}}\tanh\left[\sqrt{\frac{{\epsilon}_x(\omega)}{{\epsilon}_z(\omega)}}\frac{L_zq_\parallel}{2}\right] ~,
\end{split}
\end{equation}
where it is apparent that, for a fixed product $q_\parallel L_z$, the Green's function scales as $L_z$.
In the limit $L_z \to \infty$, the Green's function correctly reduces to the free-space Coulomb potential renormalized by the frequency-dependent dielectric properties of hBN:

\begin{equation}
\begin{split}
\label{eqn:green_function_hBN_coulomb}
\underset{ L_z \to \infty}{\lim}\;g^{(0)}({q}_{\parallel}, 0, 0, \omega) =\frac{ 2\pi }{q_{\parallel} \sqrt{\epsilon_x(\omega) \epsilon_z(\omega)} }~.
\end{split}
\end{equation}
In the limit $\omega\to \infty$, the scalar Green's function reads
\begin{equation}
\begin{split}
\label{eqn:green_function_hBN_classical_inf}
& g_\infty^{(0)}( q_{\parallel},0,0)= \frac{2\pi }{q_\parallel \sqrt{{\epsilon}_{x,\infty}{\epsilon}_{z,\infty}}}\tanh\left[\sqrt{\frac{{\epsilon}_{x,\infty}}{{\epsilon}_{z,\infty}}}\frac{L_zq_\parallel}{2}\right] ~,
\end{split}
\end{equation}

where ${\epsilon}_{x,\infty},{\epsilon}_{z,\infty}$ are the high-frequency limit of the hBN dielectric functions and the dimensionless parameter
\begin{equation}
X=\sqrt{\frac{\epsilon_{x,\infty}}{\epsilon_{z,\infty}}}\,\frac{L_z q_\parallel}{2} ~,
\end{equation}
quantifies the effect of cavity confinement. Since $\tanh[X]\leq 1$, the interaction at finite $L_z$ is always reduced with respect to the bulk case, reflecting electrostatic screening induced by the metallic boundary conditions.

In the opposite limit $X\gg 1$, corresponding to weak confinement (large $L_z$) or short wavelengths, one has $\tanh[X]\simeq 1$. In this regime, the cavity becomes electrostatically irrelevant and the Green’s function reduces to its bulk expression in the absence of metallic boundaries,
\begin{equation}
\underset{q_\parallel L_z \to \infty}{\lim}\;
g_\infty^{(0)}(q_{\parallel},0,0)
= \frac{2\pi}{q_\parallel\sqrt{\epsilon_{x,\infty}\epsilon_{z,\infty}}}\,.
\end{equation}

\subsection{Quantization scheme for sub-wavelength cavities}
\label{Subsec:quantization}

Here, we derive a quantization scheme to associate the Green's function in Eq.~\eqref{eqn:green_function_hBN_expansion_Main} with the corresponding quantum mechanical modes associated with the resonances $\hbar \omega_{{q}_\parallel,n_z,s}$. Before addressing the specific case of hBN sub-wavelength cavities, we consider a generic scenario with a resonance $\hbar\omega_{\mu,s}$, where $\mu$ is a generic index encompassing different quantum numbers (e.g. $\mu=\bm{q}_\parallel,n_z$ in the hBN case) and $s$ denotes different solutions with the same mode function.

The cavity Hamiltonian, associated with these poles, is given by:

\begin{align}
\label{eq:H_cav}
\hat{\mathcal{H}}_{\rm cav} = \sum_{\mu,s} \hbar \omega_{\mu,s} \hat{a}^\dagger_{\mu,s} \hat{a}_{\mu,s}~.
\end{align}
We assume that the scalar potential can be expanded in terms of the modes $\hat{a}^\dagger_{\mu,s}$ and $\hat{a}_{\mu,s}$ as:

\begin{align}
\label{eq:phi_SM}
\hat{\phi}(\bm{r})= \sum_{\mu,s} \frac{1}{\sqrt{N_{\mu,s}}}\left(f_{\mu}(\bm{r}) \hat{a}_{\mu,s} + f^*_{\mu}(\bm{r}) \hat{a}^\dagger_{\mu,s} \right)~,  
\end{align}
where $N_{\mu,s}$ are normalization constants that will be determined later. Eq.~\eqref{eq:phi_SM} allows us to compute the quantum mechanical fluctuations of the field:
\begin{align}
\label{eq:fluct}
  \tensor[_{\rm cav}]{\langle}{} 0| \hat{\phi}(\bm{r})\hat{\phi}(\bm{r}^\prime)|0\tensor[]{\rangle}{_{\rm cav}} &= \sum_{\mu,s} \frac{f_\mu(\bm{r})f^*_\mu(\bm{r}^\prime)}{N_{\mu,s}}~,
\end{align}
where $|0\tensor[]{\rangle}{_{\rm cav}}$ is the vacuum of the cavity Hamiltonian $\hat{\mathcal{H}}_{\rm cav}$.
Assuming the mode expansion in Eq.~\eqref{eq:phi_SM} and the cavity Hamiltonian $\hat{\mathcal{H}}_{\rm cav}$ in Eq.~\eqref{eq:H_cav}, we can use the Kubo formula \cite{Giuliani_and_Vignale} to calculate the scalar Green function associated with the modes, $g_{\rm q}^{(0)}(\bm{r}, \bm{r}^\prime, \tau)$:

\begin{equation}
\begin{split}
 \label{eq:chi}
g_{\rm q}^{(0)}(\bm{r}, \bm{r}^\prime, \tau) \equiv -\frac{\I}{\hbar} \Theta(\tau)  \tensor[_{\rm cav}]{\langle}{} 0|  \big[ \hat{\phi}^{\rm I}(\bm{r}, \tau), \hat{\phi}(\bm{r}^\prime) \big] |0\tensor[]{\rangle}{_{\rm cav}}~,  
\end{split}
\end{equation}
where $\Theta(\tau)$ is the Heaviside function, and $[\hat{X}, \hat{Y}]$ denotes the commutator between the operators $\hat{X}$ and $\hat{Y}$. The term $\hat{\phi}^{\rm I}(\bm{r}, t)$ refers to the electric potential in the interaction picture, specifically defined as $\hat{\phi}^{\rm I}(\bm{r}, t) \equiv U_0^\dagger(t) \hat{\phi}(\bm{r}) U_0(t)$, where $U_0(t) \equiv \exp(-\I \hat{\mathcal{H}}_{\rm cav} t / \hbar)$.

By taking the time Fourier transform of the previous equation,

\begin{equation}
g_{\rm q}^{(0)}(\bm{r}, \bm{r}^\prime, \omega) = \lim\limits_{\eta \to 0} \int_{-\infty}^\infty d\tau~ g_{\rm q}^{(0)}(\bm{r}, \bm{r}^\prime, \tau) e^{\I(\omega + \I\eta)\tau},
\end{equation}
and utilizing the Hermitian property, $f_\mu(\bm{r}) f^*_\mu(\bm{r}^\prime) = f_\mu(\bm{r}^\prime) f^*_\mu(\bm{r})$, we obtain:

\begin{equation}
\begin{split}
\label{eq:green_q}
&g_{\rm q}^{(0)}(\bm{r},\bm{r}^\prime,\omega)=\sum_{\mu,s} \frac{f_\mu(\bm r)f^*_\mu(\bm r^\prime)}{{N}_{\mu,s}}\times\\
&\times\left(\frac{1}{\hbar\omega-\hbar\omega_{\mu,s}+\I\hbar\eta}-\frac{1}{\hbar\omega+\hbar\omega_{\mu,s}+\I\hbar\eta}\right)~.
\end{split}
\end{equation}
This expression reveals how the quantum Green's function is linked to the vacuum virtual excitation of the quantum field $\hat{\phi}(\bm{r}^\prime)$. In fact, 
\begin{equation}
    \tensor[_{\rm cav}]{\langle}{} 0| \hat{\phi}(\bm{r}^\prime)|\mu,s\tensor[]{\rangle}{_{\rm cav}}=f_\mu(\bm r)/\sqrt{{N}_{\mu,s}}
\end{equation}
where $|\mu,s\tensor[]{\rangle}{_{\rm cav}}=a_{\mu,s}^\dagger|0\tensor[]{\rangle}{_{\rm cav}}$ are single-photon states.

This can be interpreted within the general framework of the fluctuation-dissipation theorem \cite{Giuliani_and_Vignale}, which connects ${\rm Im} \, g_{\rm q}^{(0)}(\bm{r}, \bm{r}^\prime, \omega)$ to the spectrum of the fluctuations $S^{(0)}_{\rm cav}(\bm{r},\bm{r}^\prime,\omega)$,

\begin{equation}
\begin{split}
\label{eq:green_FD}
&{\rm Im}~g_{\rm q}^{(0)}(\bm{r},\bm{r}^\prime,\omega)=-\frac{\pi}{\hbar}S^{(0)}_{\rm cav}(\bm{r},\bm{r}^\prime,\omega)~.
\end{split}
\end{equation}
where
\begin{equation}
\begin{split}
\label{eq:green_FD1}
&S^{(0)}_{\rm cav}(\bm{r},\bm{r}^\prime,\omega)\equiv \frac{1}{2\pi}\int_{-\infty}^\infty d\tau  e^{\I\omega \tau}\tensor[_{\rm cav}]{\langle}{} 0|   \hat{\phi}^{\rm I}(\bm{r}, \tau) \hat{\phi}(\bm{r}^\prime)  |0\tensor[]{\rangle}{_{\rm cav}}~.
\end{split}
\end{equation}
Note that $S^{(0)}_{\rm cav}(\bm{r},\bm{r}^\prime,\omega)$ encodes the fluctuation of the field in Eq.~\eqref{eq:fluct} as 
$\int_{-\infty}^\infty d\omega S^{(0)}_{\rm cav}(\bm{r},\bm{r}^\prime,\omega)=  \tensor[_{\rm cav}]{\langle}{} 0| \hat{\phi}(\bm{r})\hat{\phi}(\bm{r}^\prime)|0\tensor[]{\rangle}{_{\rm cav}}$.

By taking the real and imaginary part of Eq.~\eqref{eq:green_q}, and the limit $\eta \to 0$, we have:

\begin{equation}
    \begin{split}
        &{\rm Re} \, g_{\rm q}^{(0)}(\bm{r}, \bm{r}^\prime, \omega) = \sum_\mu \frac{f_\mu(\bm{r}) f^*_\mu(\bm{r}^\prime)}{N_\mu }   \left[ \frac{2 \hbar \omega_\mu}{ (\hbar \omega)^2 - (\hbar \omega_\mu)^2 } \right]~,
    \end{split}
\end{equation}

\begin{equation}
    \begin{split}
    \label{eq:GM_im}
       & {\rm Im} \, g_{\rm q}^{(0)}(\bm{r}, \bm{r}^\prime, \omega) = \\& =-\pi \sum_\mu  \frac{f_\mu(\bm{r}) f^*_\mu(\bm{r}^\prime)}{\hbar N_\mu }  \left[ \delta( \omega - \omega_\mu)-\delta( \omega +  \omega_\mu) \right]~,   
    \end{split}
\end{equation}
By comparing Eq.~\eqref{eq:GM_im} with Eq.~\eqref{eqn:green_function_IM}, by choosing

\begin{equation}
    \begin{split}
      N_{\mu,s} =\frac{1}{4\pi \hbar}\partial_{\omega}\lambda_\mu(\omega)|_{\omega=\omega_{\mu,s}}~,   
    \end{split}
\end{equation}
we can identify the imaginary part of the Green's function ${\rm Im}~g^{(0)}(\bm{r}, \bm{r}^\prime, \omega)$ with the imaginary part of the mode Green's function ${\rm Im}~g_{\rm q}^{(0)}(\bm{r}, \bm{r}^\prime, \omega)$:

\begin{equation}
    \begin{split}
    \label{eqn:Im_g}
      {\rm Im} \, g^{(0)}(\bm{r}, \bm{r}^\prime, \omega)={\rm Im} \, g_{\rm q}^{(0)}(\bm{r}, \bm{r}^\prime, \omega)~.  
    \end{split}
\end{equation}

As $g_{\rm q}^{(0)}(\bm{r}, \bm{r}^\prime, \omega)$ fulfills the Kramers-Kronig relations \cite{Giuliani_and_Vignale}:

\begin{align}
\label{eq:KK2}
  & {\rm Re} \, g_{\rm q}^{(0)}(\bm{r}, \bm{r}^\prime, \omega)=  \frac{2}{\pi}\int_{0}^{\infty} \frac{\omega'  {\rm Im} \, g_{\rm q}^{(0)}(\bm{r}, \bm{r}^\prime, \omega)}{\omega'^2 - \omega^2} \, d\omega'~,\\ 
  & {\rm Im} \, g_{\rm q}^{(0)}(\bm{r}, \bm{r}^\prime, \omega) =-\frac{2 \omega}{\pi} \int_{0}^{\infty} \frac{{\rm Re} \, g_{\rm q}^{(0)}(\bm{r}, \bm{r}^\prime, \omega)}{\omega'^2 - \omega^2} \, d\omega'~. 
\end{align}

By means of Eq.~\eqref{eqn:Im_g} and Eqs.~(\ref{eq:KK1},\ref{eq:KK2}) we find the following decomposition:
\begin{equation}
\begin{split}
\label{eqn:green_function_hBN_expansion_R}
& g^{(0)}(\bm{r}, \bm{r}^\prime, \omega)=  g_{\rm q}^{(0)}(\bm{r}, \bm{r}^\prime, \omega)+g_{\infty}^{(0)}(\bm{r}, \bm{r}^\prime) ~.
\end{split}
\end{equation}
The total Green's function can be decomposed as a sum of the dynamical quantum Green's function $g_{\rm q}^{(0)}(\bm{r}, \bm{r}^\prime, \omega)$ and the static potential $g_{\infty}^{(0)}(\bm{r}, \bm{r}^\prime)$.

We now specialize these results for the case of hBN. The scalar field can be written as:
\begin{equation}
\label{eq:phi_in_modes}
\hat{\phi}(\bm{r})= \sum_{q_\parallel, n_z, s} \frac{1}{\sqrt{N_{q_\parallel, n_z, s}}}\left(f_{{q_\parallel, n_z, s}}(\bm{r}) \hat{a}_{{q_\parallel, n_z, s}} + {\rm h.c.}\right)~,  
\end{equation}

where the normalization factors $N_{q_\parallel, n_z, s}$ can be expressed explicitly in terms of the dielectric function of hBN as follows:

\begin{equation}
\begin{split}
\label{eq:N_q}
&N_{q_\parallel, n_z, s} =\\
 &= \frac{1}{4\pi \hbar} \left[ \partial_{\omega} \epsilon_z(\omega) q_{z, n_z}^2 + \partial_{\omega} \epsilon_x(\omega) q_\parallel^2 \right]_{\omega = \omega_{q_\parallel, n_z, s}}~.
 \end{split}
\end{equation}
By applying Eq.~\eqref{eq:phi_SM} to the hBN cavity, we obtain Eq.~\eqref{eqn:potential} while the explicit expression of the normalization factors $N_{q_\parallel, n_z, s}$ is given in Eq.~\eqref{eq:N_q}.

The Green's function can be expressed in terms of the amplitude coefficients $\phi_{q_\parallel, n_z, s}(z)$ as:

\begin{equation}
\begin{split}
\label{eqn:green_function_hBN_expansion_quantum}
 &g_{\rm q}^{(0)}( q_{\parallel},z,z', \omega)=\sum_{n_z,s} \phi_{{q}_\parallel,n_z,s}(z) \phi_{{q}_\parallel,n_z,s}(z^\prime) \times \\& \times \left(\frac{1}{\hbar\omega-\hbar\omega_{q_\parallel,n_z,s}+\I\hbar\eta}-\frac{1}{\hbar\omega+\hbar\omega_{q_\parallel,n_z,s}+\I\hbar\eta}\right) ~. \\
\end{split}
\end{equation}  
Additionally, the vacuum fluctuations of the field are related to the amplitude coefficients $\phi_{q_\parallel, n_z, s}(z)$:

\begin{align}
\label{eqn:potential_SM}
 \tensor[_{\rm cav}]{\langle}{} 0| \hat{\phi}^2(\bm{r}_\parallel,0)|0\tensor[]{\rangle}{_{\rm cav}}&=\sum_{\bm{q}_\parallel,n_z,s}\frac{1}{S}  \phi^2_{{q}_\parallel,n_z,s}({0})~.
\end{align}
Hence, the amplitude coefficients $\phi_{q_\parallel, n_z, s}(z)$ can be interpreted as a measure of the density of vacuum fluctuations. Moreover, these coefficients also appear in the Green's function, as shown in Eq.~\eqref{eqn:green_function_hBN_expansion_quantum}, indicating that the intensity of a scalar field generated by placing a test charge depends on the vacuum fluctuations of the field in the absence of any charge, represented by $\phi_{q_\parallel, n_z, s}(z)$. It thus is useful to express $\phi_{q_\parallel, n_z, s}(z)$ in the following form:

\begin{widetext}
\begin{align}
\label{eqn:potential_SM1}
 \phi^2_{{q}_\parallel,n_z,s}({0})=\frac{8\pi \hbar L_z}{\left[\partial_{\omega}\epsilon_z(\omega)[(2n_z+1)\pi]^2+\partial_{\omega}\epsilon_x(\omega)(q_\parallel L_z)^2\right]_{\omega=\omega_{q_\parallel,n_z,s}}}~, 
\end{align}

\end{widetext}
where we used $q_{z, n_z} L_z = (2n_z+1)\pi$,  corresponding to the $z$-even modes that remain finite at $z=0$. Note that $\phi^2_{q_\parallel, n_z, s}(0) / L_z$ depends only on the product $q_\parallel L_z$ and not on these variables individually.
Fig.~\ref{figS1}(b,c) shows the vacuum fluctuation coefficients $\phi_{q_\parallel, n_z, s}(z)$ as a function of $q_\parallel L_z$ in panel (b) and as a function of $L_z$ in panel (c). As seen in panel (b), the maximum of $\phi_{q_\parallel, n_z, s}(z)$ occurs at $q_\parallel L_z = 0$ for the lower Reststrahlen band, while for the upper Reststrahlen band, the maximum is reached when $q_\parallel L_z$ is on the order of unity. Given that the vacuum fluctuation coefficients quantify the intensity of the field's response to external perturbations, this implies that the field response is maximized under the same conditions: at $q_\parallel L_z = 0$ for the lower Reststrahlen band and at $q_\parallel L_z \sim 1$ for the upper band.

Finally, in the HPP cavity under study, Eq.~\eqref{eqn:green_function_hBN_expansion_R} reads:

\begin{equation}
\begin{split}
\label{eqn:green_function_hBN_expansion_identity}
& g^{(0)}( q_{\parallel},z,z', \omega)=  g_{\rm q}^{(0)}(q_{\parallel},z,z', \omega)+g_{\infty}^{(0)}(q_{\parallel},z,z') ~,
\end{split}
\end{equation}
where $g_{\infty}^{(0)}( q_{\parallel},z,z')$ represents the high-frequency limit of the Green's function (also given in Eq.~\eqref{eqn:green_function_hBN_classical_inf} where the sum over $n_z$ is carried out):

\begin{equation}
\begin{split}
&g_{\infty}^{(0)}(q_{\parallel},z,z')=\\
&\frac{8\pi}{ L_z}\sum_{n_z=0}^{\infty}\frac{\sin\left(q_{z,n_z} [z+\frac{L_z}{2}]\right) \sin\left(q_{z,n_z} [z^\prime+\frac{L_z}{2}]\right)}{{\epsilon}_{z,\infty}q_{z,n_z}^2+{\epsilon}_{x,\infty}q_\parallel^2}~.
\end{split}
\end{equation}
This means that the quantum expansion of the scalar potential \eqref{eqn:potential} reproduces the correct Green's function $g^{(0)}( q_{\parallel}, z, z', \omega)$, in Eq.~\eqref{eqn:green_function_hBN_expansion}, provided the static potential $g_{\infty}^{(0)}( q_{\parallel}, z, z')$ is added to the Hamiltonian as the standard static Coulomb interaction:

\begin{align}
\label{eq:H_static}
\hat{\mathcal{H}}_{\rm C} = \frac{1}{2S}\sum_{\bm{q}_\parallel} \hat{n}_{-\bm{q}_\parallel}e^2g_{\infty}^{(0)}( q_{\parallel},0,0)\hat{n}_{\bm{q}_\parallel}~,
\end{align}
where $\hat{n}_{\bm{q}_\parallel}$ is the wavevector-resolved density operator
\begin{align}
\hat{n}_{\bm{q}_\parallel}&=\int d\bm{r}_\parallel e^{-i\bm{q}_{\parallel}\cdot\bm{r}_\parallel}\hat{n}(\bm{r}_\parallel)~.
\end{align}

 This concludes the discussion of the quantization scheme. It is worth noting that, unlike approaches that start from Maxwell’s equations in matter, construct a Hamiltonian, and then quantize it, our approach is based entirely on the scalar Green’s function $g^{(0)}$, without explicitly constructing  the Hamiltonian. In Appendix~\ref{app_alt} we demonstrate the formal equivalence between the two methods.

\section{Details on ``Theory of Landau levels in Graphene and their coupling to a sub-wavelength cavity"}
\label{sec:LLs}

\begin{figure*}[t]
\centering
  \vspace{1.em}
  \begin{overpic}[width=0.68\columnwidth]{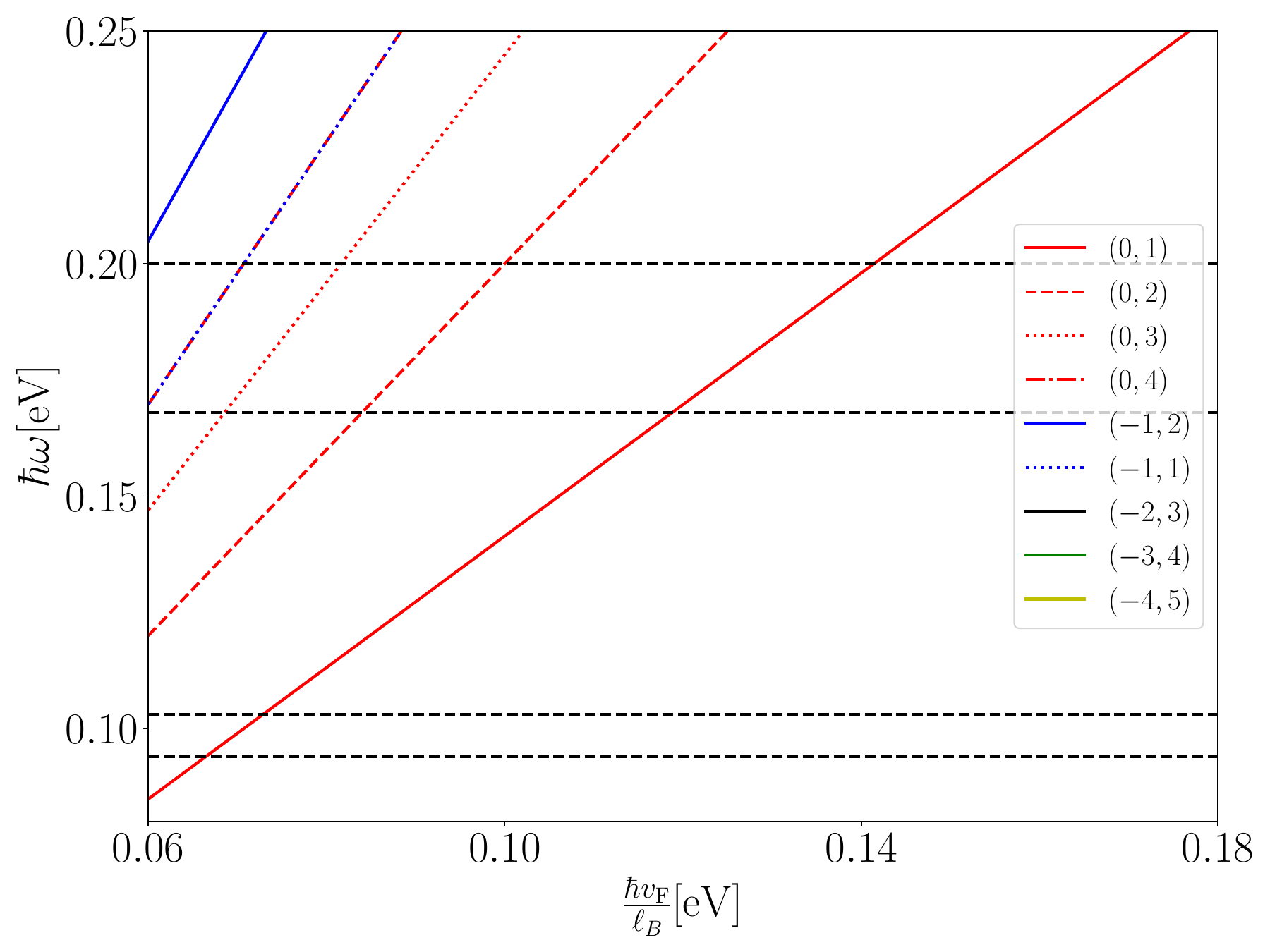}\put(0,80){\normalsize (a)}\end{overpic} 
  \begin{overpic}[width=0.68\columnwidth]{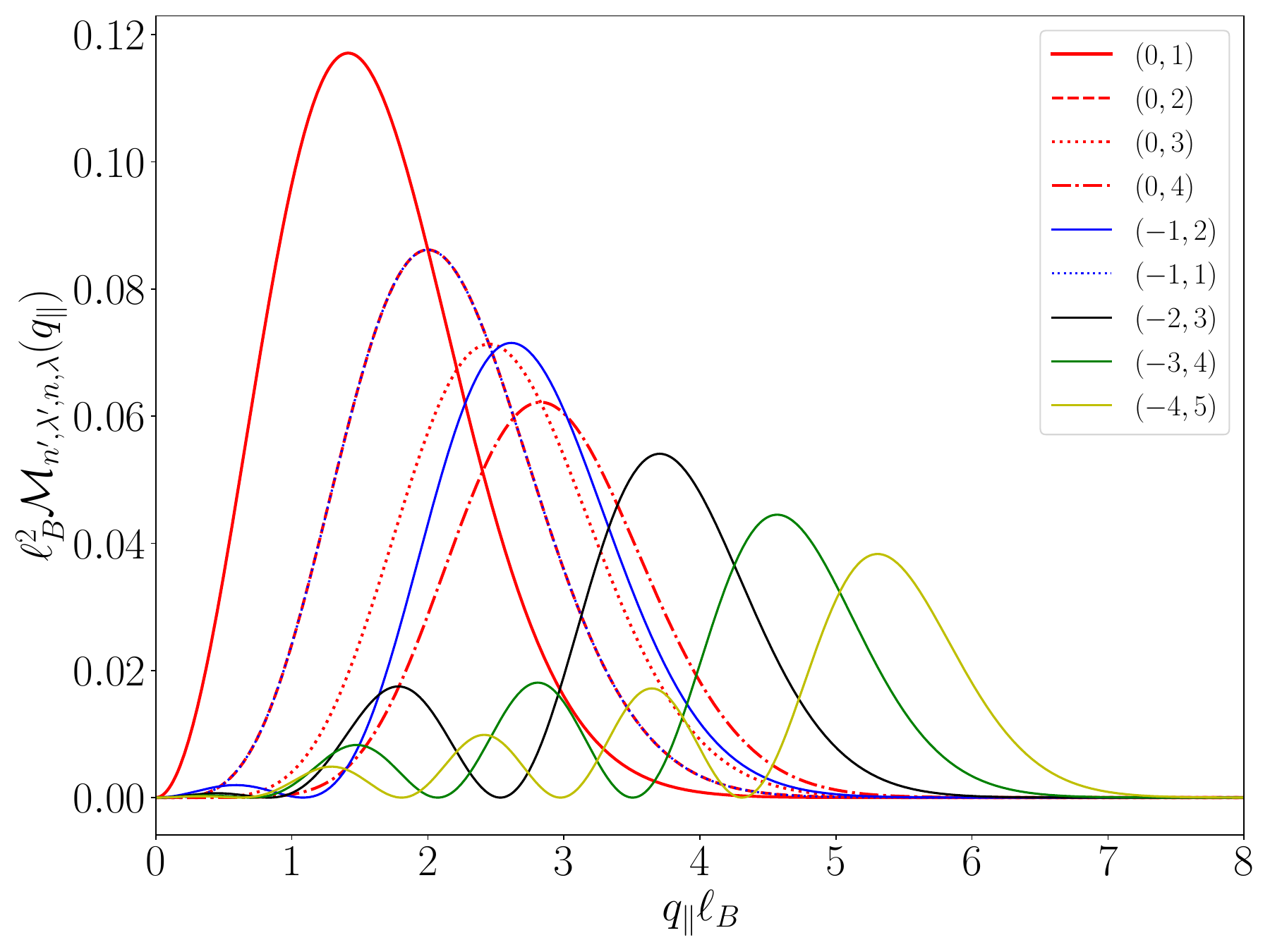}\put(0,80){\normalsize (b)}\end{overpic} 
  \begin{overpic}[width=0.68\columnwidth]{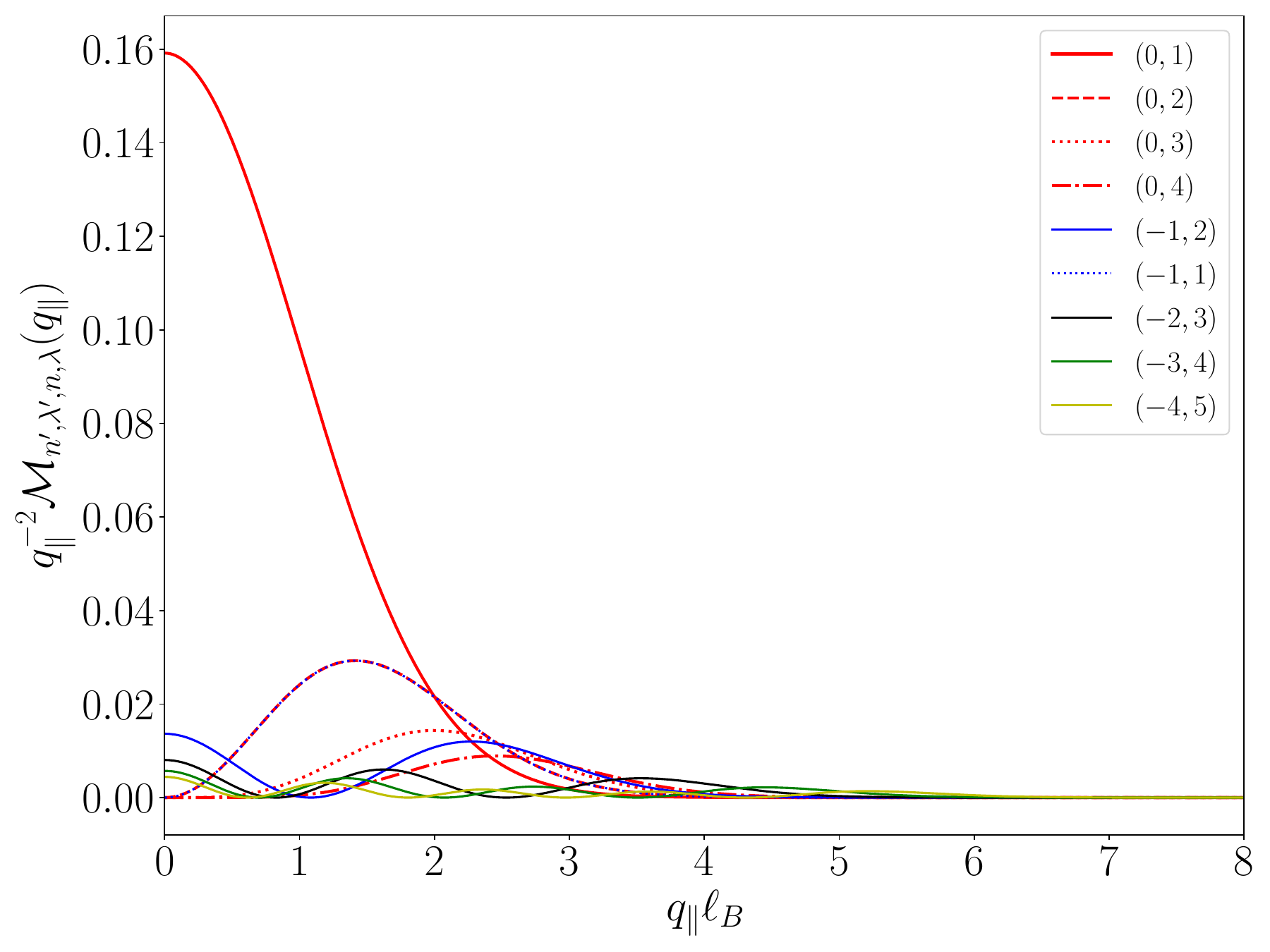}\put(0,80){\normalsize (c)}\end{overpic} 
 \caption{ Panel (a) shows the energy differences $\varepsilon_{\lambda,n} - \varepsilon_{\lambda^\prime,n^\prime}$ associated with the LL transitions as a function of the ratio $\hbar v_{\rm F}/\ell_B$. Panel (b) displays the squared modulus of the matrix element of the density, $\ell^2_{B}\mathcal{M}_{n^\prime,\lambda^\prime,n,\lambda}(q_\parallel)$, as a function of $q_\parallel \ell_{B}$. Panel (c) displays the quantity $q_\parallel^{-2}\mathcal{M}_{n^\prime,\lambda^\prime,n,\lambda}(q_\parallel)$ as a function of $q_\parallel \ell_{B}$. In all panels, each transition is labeled in the legend by the quantum numbers $( \lambda^\prime n^\prime,\lambda n)$, and solid (dashed) lines indicate dipole-allowed (dipole-forbidden) transitions.
 \label{figS2}}
\end{figure*}

This Section reviews the LLs in graphene under a uniform magnetic field $B$ and their coupling to a sub-wavelength cavity. We derive the Hamiltonian governing LLs, emphasizing the eigenstates and associated energies \(\varepsilon_{\lambda, n}\), which depend on the magnetic field strength via the magnetic length \(\ell_B\). Subsequently, we use eigenstates and eigenenergies to calculate the density--density response function \(\chi^{(0)}_{nn}(q_\parallel,\omega)\). These findings will be useful later to calculate the polaritonic spectrum.

\subsection{Energies and wavefunctions of LLs}
\label{subsec:4a}

The Hamiltonian describing graphene electrons in the presence of a uniform magnetic field $B$ reads \cite{castroneto_rmp_2009,Goerbig11}:

\begin{equation}
\label{eq:graphene_LL}
\hat{\mathcal{H}}_{\rm LLs} = v_{\rm F}\hat{\bm \pi}_\parallel\cdot \hat{\bm\sigma}_\parallel~,
\end{equation}
where $v_{\rm F}$ is the Fermi velocity, $\hat{\bm\sigma}_\parallel$ are the Pauli operators acting on the valley degrees of freedom and the kinetic momentum operators $\hat{\pi}_x,\hat{\pi}_y$ are
\begin{equation}
\begin{split}
\hat{\pi}_x &= \hat{p}_x - \frac{eB}{c}\hat{y}~, \\
\hat{\pi}_y &= \hat{p}_y~. \\
\end{split}
\end{equation}
Here, we have chosen the Landau gauge where the vector potential reads $\bm{A}(\bm{r}) = A_x(\bm{r})\hat{\bm{e}}_x+ A_y(\bm{r})\hat{\bm{e}}_y = -By\hat{\bm{e}}_x$. With this gauge choice, $\hat{x}$ does not appear in the Hamiltonian, which allows us to use the plane wave states $\ket{k_x}$, such that $\braket{x | k_x} = e^{-\I k_x x} / \sqrt{L_x}$ and $\hat{p}_x \ket{k_x} = \hbar k_x \ket{k_x}$. In terms of these operators, the LLs Hamiltonian $\hat{\mathcal{H}}_{\rm LLs}$ reads:

\begin{equation}
\begin{split}
\label{eq:graphene_LL_1}
\hat{\mathcal{H}}_{\rm LLs} &= v_{\rm F}\left[\hbar\left( k_x  - \hat{y}/{{\ell}^2_B} \right)\hat{\sigma}_x+\hat{\pi}_y\hat{\sigma}_y\right]~,
\end{split}
\end{equation}
where $ {\ell}_B=\sqrt{\left({\hbar c}/{eB}\right)}$ is the magnetic length.
By defining the displaced position $\hat{{y}}_{k_x}$

\begin{equation}
\begin{split}
\hat{{y}}_{k_x} &= \hat{y} - k_x \ell_B^2 ~,
\end{split}
\end{equation}
we can define the ladder operators $\hat{a}_{k_x}, \hat{a}_{k_x}^\dagger$ associated with $\hat{{y}}_{k_x}$ such that

\begin{equation}
\begin{split}
\hat{{y}}_{k_x} &= \frac{\ell_B}{\sqrt{2}} \left( \hat{a}_{k_x} + \hat{a}_{k_x}^\dagger \right)~, \\
\hat{p}_y &= \frac{\hbar\I}{\sqrt{2 }\ell_B} \left( \hat{a}_{k_x}^\dagger - \hat{a}_{k_x}  \right)~.
\end{split}
\end{equation}
In terms of these ladder operators the LLs Hamiltonian $\hat{\mathcal{H}}_{\rm LLs}$ reads:

\begin{equation}
\begin{split}
\label{eq:graphene_LL_2}
\hat{\mathcal{H}}_{\rm LLs} &=- \frac{\hbar v_{\rm F}\sqrt{2}}{\ell_B}
\begin{pmatrix}
0 &  \hat{a}_{k_x}  \\
\hat{a}^\dagger_{k_x} & 0
\end{pmatrix}
\end{split}~.
\end{equation}
It can be seen by inspection that the eigenstates of the LLs Hamiltonian $\hat{\mathcal{H}}_{\rm LLs}$ are
$ \ket{k_x,n,\lambda}$:

\begin{equation}
    \ket{k_x,n,\lambda} = \begin{pmatrix}
C_n^- \ket{ k_x}\ket{(n-1)_{k_x}} \\
-\lambda C_n^+ \ket{ k_x}\ket{n_{k_x}}
\end{pmatrix}~,
\end{equation}
with $\lambda=\pm$ and 

\begin{align}
\ket{ n_{k_x}}= \frac{(\hat{a}_{k_x}^\dagger)^n}{\sqrt{n!}}\ket{0_{k_x}}
\end{align} 
and $\ket{0_{k_x}}$ is the $k_x$-dependent vacuum, such that $\hat{a}_{k_x}\ket{0_{k_x}}=0$, and
\begin{equation}
C^\pm_n=\sqrt{\frac{1\pm\delta_{0n}}{2}}~.
\end{equation}
In real space, these LL states combine a plane-wave component along the $x$-direction with harmonic-oscillator eigenfunctions along the $y$-direction. More explicitly, the spatial representation of the single-particle wavefunction $\Psi_{k_x,n,\lambda}(\bm{r}_\parallel)\equiv\braket{x,y|k_x,n,\lambda}$ is expressed as:

\begin{equation}
\label{eq:wavefunction_definition}
\Psi_{k_x,n,\lambda}(\bm{r}_\parallel)
=\frac{e^{-\I k_x x}}{\sqrt{L_x}}\,
\begin{pmatrix}
C_n^-\varphi_{n-1}\!\bigl(y_{k_x}\bigr)\\[4pt]
-\lambda\,C_n^+\varphi_{n}\!\bigl(y_{k_x}\bigr)
\end{pmatrix},
\end{equation}
where the harmonic oscillator wavefunctions along the $y$-direction are defined by:

\begin{equation}
\varphi_n(y_{k_x}) = 
\frac{1}{\sqrt{2^n n! \, \sqrt{\pi} \, \ell_B}} \,
\exp\left[-\frac{1}{2} \left( \frac{y_{k_x}}{\ell_B} \right)^2 \right] 
\, H_n\left( \frac{y_{k_x}}{\ell_B} \right).
\end{equation}
for $n\geq0$ while $\varphi_{-1}(y_{k_x})\equiv 0$ and $y_{k_x}\equiv y - k_x\ell_B^2$ are shifted coordinates. The characteristic spatial scale of these Landau-level wavefunctions along $\hat{\bm{e}}_x$ is given by the magnetic length $\ell_B$. The associated energies are 

\begin{equation}
    \varepsilon_{\lambda, n}\equiv\lambda(\hbar v_{\rm F}/\ell_{B})\sqrt{2n}~.
\end{equation}
Note that these energies do not depend on the quantum number $k_x$, which reflects the degeneracy of the Landau levels in the $x$-direction. 
Notice that $n=0$ corresponds to only one state, without the degeneracy $\lambda = \pm$. Conventionally, we associate the quantum number $\lambda = -$ with $n=0$.
Each LL has a degeneracy of
\begin{equation}
N_{\text{L}} = \frac{gS}{2\pi\ell_B^2}~,
\end{equation}
where $g=g_{\rm s}g_{\rm v}=4$ takes into account the spin ($g_{\rm s}=2$) and valley 
($g_{\rm v}=2$) degeneracies in graphene.
 In second quantized formalism, the electronic field operator $\hat{\Psi}(\bm{r})$ can be expanded in the basis of LL single-particle wavefunction as
\begin{equation}
\hat{\Psi}(\bm{r})=\sum_{ k_x,n,\lambda}\Psi_{k_x,n,\lambda}(\bm{r}_\parallel)\hat{c}_{k_x,n,\lambda}~,
\end{equation}

where $\hat{c}_{k_x,n,\lambda}$ annihilates an electron in the $\ket{k_x,n,\lambda}$ state. 

Realistic values of the parameters for the graphene Hamiltonian in a magnetic field \cite{castroneto_rmp_2009,Goerbig11} are provided in Table~\ref{tab:2}, while the ranges of the magnetic field $B$, magnetic length $\ell_B$, and cyclotron frequency $\hbar v_{\rm F}/\ell_B$ used in the paper are summarized in Table~\ref{tab:3}.

\begin{table}[h]
    \centering
     \vspace{0.2cm}
    \begin{tabular}{|l|c|}
        \hline
        $v_{\rm F}$ & $10^6~{\rm m/s}$  \\
        $\ell_B$ & $25.6/\sqrt{B[{\rm T}]}~{\rm nm}$ \\
        $\hbar v_{\rm F}/\ell_B$ &$ 25.7\sqrt{B[{\rm T}]}~{\rm meV}$ \\
        \hline
    \end{tabular}
    \caption{Realistic values of the parameters for the graphene Hamiltonian coupled with a magnetic field $B$ \cite{castroneto_rmp_2009,Goerbig11}.}
    \label{tab:2}
\end{table}

\begin{table}[h]
    \centering
    \vspace{0.2cm}
    \begin{tabular}{|l|c|}
        \hline
        $B$& $5.4-49.1 {\rm T}$\\
        $\ell_B$ & $11.0-3.7~{\rm nm}$ \\
        $\hbar v_{\rm F}/\ell_B$ &$ 0.04-0.18~{\rm eV}$ \\
        \hline
    \end{tabular}
    \caption{Values of parameters relative to LLs spanned in this article.}
    \label{tab:3}
\end{table}
Fig.~\ref{figS2}(a) shows the energy differences $\varepsilon_{\lambda,n} - \varepsilon_{\lambda^\prime,n^\prime}$ associated with the LL transitions as a function of the ratio $\hbar v_{\rm F}/\ell_B$. Since $\ell_B$ depends on the magnetic field, it is experimentally feasible to adjust $\ell_B$ by varying the field strength. This allows for precise control over the LL transition energies, enabling them to be tuned into resonance with the cavity modes.

\subsection{Density-density response function and dipole-allowed and forbidden transitions}
\label{subsec:4b}

In Appendix~\ref{app1} we calculate the bare density-density response function, $\chi^{(0)}_{nn}( q_\parallel,\omega)$. This is relevant as the coupling with the cavity is through the density operator as in Eq.~\eqref{eq:H_int}. The density--density response function reads \cite{roldan2009,roldan2010}

\begin{align}
\label{eq:density--densityresponse01}
\chi^{(0)}_{nn}( q_\parallel,\omega) =
\sum_{n,\lambda,n',\lambda'}
\frac{(f_{n\lambda}-f_{n'\lambda'})\mathcal{M}_{n^\prime,\lambda^\prime,n,\lambda}({q}_\parallel)}{
\hbar\omega+\varepsilon_{\lambda,n}-\varepsilon_{\lambda',n'}+\I\eta}~.
\end{align}
where $f_{n\lambda}$ are the Fermi occupation numbers and $\mathcal{M}_{n^\prime,\lambda^\prime,n,\lambda}({q}_\parallel)$ represents the squared modulus of the matrix element of the density,

\begin{equation}
\begin{split}
  &\mathcal{M}_{n^\prime,\lambda^\prime,n,\lambda}({q}_\parallel)\equiv \sum_{k^\prime_x,k_x}
\lvert\braket{k_x^\prime,n',\lambda'|\hat n_{\bm q_\parallel}|k_x,n,\lambda}\rvert^2~,\\
&=\frac{g}{2\pi\ell_B^2}e^{-q_\parallel^2\ell_{B}^2/2} |\mathcal{F}_{n^\prime,\lambda^\prime,n,\lambda}( {q}_\parallel)|^2~, 
\end{split}
\end{equation}
where $\mathcal{F}_{n^\prime,\lambda^\prime,n,\lambda}( q_\parallel)$ is defined in Appendix~\ref{app1}. 
Fig.~\ref{figS2}(b) shows the squared modulus of the matrix element of the density, $\mathcal{M}_{n^\prime,\lambda^\prime,n,\lambda}(q_\parallel)$, as a function of $q_\parallel \ell_{B}$. The maximum of $\mathcal{M}_{n^\prime,\lambda^\prime,n,\lambda}(q_\parallel)$ is reached when $q_\parallel \ell_{B}$ is on the order of unity for all the considered LL transitions.
Fig.~\ref{figS2} (c) shows  $q_\parallel^{-2}\mathcal{M}_{n^\prime,\lambda^\prime,n,\lambda}(q_\parallel)$, which is particularly relevant because the bare longitudinal conductivity scales as \( \sigma^{(0)}_{\rm L}(q_\parallel,\omega) \sim q_\parallel^{-2} \chi^{(0)}_{nn}(q_\parallel,\omega) \). Hence, this quantity directly identifies which transitions are optically active.  
In the long-wavelength limit \(q_\parallel\ell_B \ll 1\), the density operator can be expanded as:

\begin{equation}
    \hat n_{\bm q_\parallel}\equiv \exp(-\I \bm{q}_\parallel\cdot \bm{r}_\parallel)\approx 1-\I\bm{q}_{\parallel}\cdot\bm{r}_\parallel\;+\;\cdots~,
\end{equation}
and the operator \(\bm{r}_\parallel\) couples LLs differing by exactly one quantum in the harmonic-oscillator index, \(\Delta n = \pm 1\). These transitions are the {\it dipole-allowed} transitions. Indeed, in Fig.~\ref{figS2}(c) we see that in the long-wavelength limit \(q_\parallel\ell_B \ll 1\) only \(\Delta n = \pm 1\) transitions are non-zero.
However, at larger wavevectors, \(q_\parallel \ell_B \sim 1\), the higher-order terms in the exponential expansion $e^{\I\bm{q}_{\parallel}\cdot \bm{r}_\parallel} $ allow additional couplings (\(\Delta n = \pm2, \pm3,\dots\)). In other words, transitions that are {\it dipole-forbidden} in the strict long-wavelength limit can acquire finite matrix elements at sufficiently large \(q_\parallel\). As we set ourselves in the latter regime \(q_\parallel \ell_B \sim 1\), all these transitions are optically active and contribute to the overall light-matter coupling in the cavity. As shown in Fig.~\ref{figS2}(b), when \(q_\parallel \ell_{B}\) is on the order of unity, the magnitude \(\mathcal{M}_{n^\prime,\lambda^\prime,n,\lambda}(q_\parallel)\) of otherwise forbidden processes are comparable to that of the \(\Delta n = \pm 1\) transitions. As an illustrative example, the matrix element associated with the transition \( (0,2) \) exhibits an enhancement by a factor of \( 2 \times 10^4 \) when the wavevector increases from \( q_\parallel= 0.1 \ell_B^{-1} \) to \( q_\parallel = 2\ell_B^{-1} \).

At zero temperature, $T=0$, the density--density response function is
\begin{widetext}
\begin{align}
\label{eq:density--densityresponse1}
\chi^{(0)}_{nn}(  q_\parallel,\omega) =
\sum_{n,\lambda\in \mathcal{N}_{\rm occ}}\sum_{n',\lambda'\in\mathcal{N}_{\rm unocc}}\mathcal{M}_{n^\prime,\lambda^\prime,n,\lambda}({q}_\parallel)\left( \frac{1}{\hbar\omega+\varepsilon_{\lambda,n}-\varepsilon_{\lambda',n'}+\I\eta}-\frac{1}{{\hbar\omega+\varepsilon_{\lambda',n'}-\varepsilon_{\lambda,n}+\I\eta}}\right)~,
\end{align}
\end{widetext}
where the sums run over occupied ($\mathcal{N}_{\rm occ}$) and unoccupied ($\mathcal{N}_{\rm unocc}$) LLs, respectively.

\section{Details on ``The polaritonic spectrum and the vacuum Rabi frequency"}
\label{sec:Polaritons}

In this Section, we derive and diagonalize the Hopfield Hamiltonian to describe the polaritonic spectrum of our system. 
By introducing bright modes, the LLs Hamiltonian is expressed in terms of bosonic operators and coupled to the cavity Hamiltonian via the interaction Hamiltonian. 
Once the Hopfield Hamiltonian is derived, the vacuum Rabi frequencies can be identified as the coupling constant associated with light-matter interaction. This approach establishes a direct connection between the vacuum Rabi frequencies, the vacuum fluctuation coefficients of the cavity, and the matrix elements of the electronic density operator. Additionally, we discuss the depolarization shift, linking it to the non-resonant contribution of the scalar Green's function. Finally, we identify a regime where only a single cavity mode is resonant with a single LL transition, allowing for an effective description within a simplified two-transition model.

\subsection{Diagonalizing the Hopfield Hamiltonian}

We now proceed to derive and diagonalize the Hopfield polaritonic Hamiltonian \cite{hopfield1958,Kavokin11} for our system, which is given by:

\begin{equation}
\label{eq:HHopfield}
\hat{\mathcal{H}} = \hat{\mathcal{H}}_{\rm LLs} + \hat{\mathcal{H}}_{\rm C} + \hat{\mathcal{H}}_{\rm cav} + \hat{\mathcal{H}}_{\rm int}~,
\end{equation}
where $\hat{\mathcal{H}}_{\rm LLs}$ represents the Hamiltonian for the LLs, $\hat{\mathcal{H}}_{\rm C}$ describes the Coulomb interactions among the electrons, $\hat{\mathcal{H}}_{\rm cav}$ is the cavity Hamiltonian (see Eq.~\eqref{eq:H_cav}), and $\hat{\mathcal{H}}_{\rm int}$ accounts for the interaction between the LLs and the cavity mode (see Eq.~\eqref{eq:H_int}).
LL states with energy $\varepsilon_{\lambda,n}$ are created (destroyed) by the fermionic operators $\hat{c}^\dagger_{k_x,\lambda,n}$ ($\hat{c}_{k_x,\lambda,n}$). Thus the matter LL Hamiltonian reads

\begin{equation}
\hat{\mathcal{H}}_{\rm LLs}=\sum_{k_x,\lambda,n}\varepsilon_{\lambda,n}\hat{c}^\dagger_{k_x,\lambda,n}\hat{c}_{k_x,\lambda,n}~.
\end{equation}
We now want to describe a system with many electrons occupying the Fermi Sea.
For convenience, we decompose this Hamiltonian into contributions from electrons and holes, such that $\hat{\mathcal{H}}_{\rm LLs} = \hat{\mathcal{H}}^{(+)}_{\rm LLs} + \hat{\mathcal{H}}^{(-)}_{\rm LLs}$ where

\begin{equation}
\hat{\mathcal{H}}^{(+)}_{\rm LLs}=\sum_{n',\lambda'\in\mathcal{N}_{\rm unocc}}\sum_{k_x'}\varepsilon_{\lambda',n'}\hat{c}^\dagger_{k_x',\lambda',n'}\hat{c}_{k_x',\lambda',n'}~,
\end{equation}
and
\begin{equation}
\hat{\mathcal{H}}^{(-)}_{\rm LLs}=\sum_{n,\lambda\in \mathcal{N}_{\rm occ}}\sum_{k_x}\varepsilon_{\lambda,n}\hat{c}^\dagger_{k_x,\lambda,n}\hat{c}_{k_x,\lambda,n}~.
\end{equation}
We now introduce the bright modes $\hat{b}^\dagger_{n,\lambda,n',\lambda'}(q_x)$ defined as
\begin{equation}
    \begin{split}
        \hat{b}^\dagger_{n,\lambda,n',\lambda'}(q_x)&\equiv \frac{1}{\sqrt{N_{\rm L}}}\sum_{k_x}\hat{c}^\dagger_{k_x-q_x,\lambda',n'}\hat{c}_{k_x,\lambda,n}~,\\
    \end{split}
\end{equation}
where ${n',\lambda'\in\mathcal{N}_{\rm unocc}}$ and ${n,\lambda\in \mathcal{N}_{\rm occ}}$. It is worth noting the symmetry relation $\hat{b}^\dagger_{n,\lambda,n',\lambda'}(q_x)=\hat{b}_{n',\lambda',n,\lambda}(-q_x)$.
Bright modes are approximately bosonic in the dilute limit \cite{hagenmuller2010}. Indeed, the commutator reads 

\begin{equation}
    \begin{split}
       & [\hat{b}_{n'',\lambda'',n''',\lambda'''}(q_x'),\hat{b}^\dagger_{n,\lambda,n',\lambda'}(q_x)]=\\
       & = \frac{1}{N_{\rm L}} \sum_{k_x'} \big[ \delta_{\lambda''', \lambda'} \delta_{n''', n'} \hat{c}^\dagger_{k_x', \lambda'', n''} \hat{c}_{k_x' - q_x' + q_x, \lambda, n}+\\&  -  \delta_{\lambda, \lambda''} \delta_{n, n''}\hat{c}^\dagger_{k_x' - q_x, \lambda', n'} \hat{c}_{k_x' - q_x', \lambda''', n'''} \big]~.
    \end{split}
\end{equation}
We now approximate the remaining operators by their expectation values over the Fermi sea,

\begin{equation}
|{\rm FS}\rangle \equiv \prod_{n,\lambda\in \mathcal{N}_{\rm occ}}\prod_{k_x}\hat{c}^\dagger_{k_x,\lambda,n} |0\rangle~,
\end{equation}

which represents the ground state of the system at zero temperature ($T=0$). Specifically, we consider the expectation value of the product of creation and annihilation operators as follows:

\begin{equation}
    \begin{split}
        &\hat{c}^\dagger_{k_x', \lambda'', n''} \hat{c}_{k_x' - q_x' + q_x, \lambda, n}\approx\\
        &\approx \tensor[_{}]{\langle}{} {\rm FS}|  \hat{c}^\dagger_{k_x', \lambda'', n''} \hat{c}_{k_x' - q_x' + q_x, \lambda, n}|{\rm FS} \tensor[]{\rangle}{}=\delta_{q_x', q_x}\delta_{\lambda'',\lambda}\delta_{n'',n}~,
    \end{split}
\end{equation}
since $n, \lambda, n'', \lambda'' \in \mathcal{N}_{\rm occ}$ at $T=0$. Similarly, for transitions between unoccupied states, we have:

\begin{equation}
    \begin{split}
        &\hat{c}^\dagger_{k_x' - q_x, \lambda', n'} \hat{c}_{k_x' - q_x', \lambda''', n'''} \approx\\
        &\approx \tensor[_{}]{\langle}{} {\rm FS}|\hat{c}^\dagger_{k_x' - q_x, \lambda', n'} \hat{c}_{k_x' - q_x', \lambda''', n'''} |{\rm FS} \tensor[]{\rangle}{}=0~,
    \end{split}
\end{equation}

for $n', \lambda', n''', \lambda''' \in \mathcal{N}_{\rm unocc}$ at $T=0$. Additionally, noting that $\sum_{k_x'} / N_{\rm L} = 1$, the commutators behave as bosonic:

\begin{equation}
    \begin{split}
       & [\hat{b}_{n'',\lambda'',n''',\lambda'''}(q_x'),\hat{b}^\dagger_{n,\lambda,n',\lambda'}(q_x)]=\\&= \delta_{\lambda''', \lambda'} \delta_{n''', n'} \delta_{\lambda'', \lambda} \delta_{n'', n}  \delta_{q_x' ,q_x}~.
    \end{split}
\end{equation}
Thus, we can write an effective Hamiltonian for the electron-hole excitations above the Fermi sea as:

\begin{equation}
\begin{split}
\hat{\mathcal{H}}^{\rm eff}_{\rm LLs} &= \sum_{n', \lambda' \in \mathcal{N}_{\rm unocc}} \sum_{n, \lambda \in \mathcal{N}_{\rm occ}} \sum_{q_x} \big(  \varepsilon_{\lambda', n'}-\varepsilon_{\lambda, n}\big) \\
& \times \hat{b}^\dagger_{n, \lambda, n', \lambda'}(q_x) \hat{b}_{n, \lambda, n', \lambda'}(q_x)~,
\end{split}
\end{equation}
which is quadratic and diagonal in terms of the bright modes.
By inserting the mode expansion in Eq.~\eqref{eqn:potential}, the light-matter interaction Hamiltonian $\hat{\mathcal{H}}_{\rm int}$ in Eq.~\eqref{eq:H_int} can be recasted as

\begin{equation}
\hat{\mathcal{H}}_{\rm int}=-e\sum_{\bm{q}_{\parallel},n_z,s} \frac{ \phi_{\bm{q}_\parallel,n_z,s}(0)}{\sqrt{S}}\left(\hat{a}^\dagger_{\bm{q}_\parallel,n_z,s}+  \hat{a}_{-\bm{q}_\parallel,n_z,s}\right)\hat{n}_{\bm{q}_\parallel}~,
\end{equation}
where we used the property $ \phi_{-\bm{q}_\parallel, n_z, s}(0) = \phi_{\bm{q}_\parallel, n_z, s}(0)$.

We project the density operator $\hat{n}_{\bm{q}_\parallel}$ onto the particle-hole excitation sector. This approximation retains only the terms responsible for the creation and annihilation of electron-hole pairs across the Fermi surface. Scattering processes entirely within the occupied or unoccupied manifolds are neglected; at $T=0$, the former are forbidden by the Pauli exclusion principle (or contribute only to the static background canceled by the Jellium), while the latter annihilate the Fermi sea $|{\rm FS}\rangle$. Formally, this corresponds to keeping only transitions where the difference in thermal occupation is non-zero ($f_{n\lambda k_x} - f_{n'\lambda' k_x'} \neq 0$). The projected density operator thus reads:
\begin{equation}
    \begin{split}
        \hat{n}_{\bm{q}_\parallel}&\approx \sum_{n', \lambda' \in \mathcal{N}_{\rm unocc}} \sum_{n, \lambda \in \mathcal{N}_{\rm occ}}\sum_{k_x,k_x'}\times \\&\times \big(\braket{k_x',n',\lambda'|\hat{n}_{\bm{q}_\parallel}|k_x,n,\lambda} \hat{c}^\dagger_{k_x',n',\lambda'}\hat{c}_{k_x,n,\lambda}+\\&+\braket{k_x,n,\lambda|\hat{n}_{\bm{q}_\parallel}|k_x',n',\lambda'} \hat{c}^\dagger_{k_x,n,\lambda}\hat{c}_{k_x',n',\lambda'}\big)~.
    \end{split}
\end{equation}
All other contributions vanish at $T=0$.

Next, we incorporate the explicit form of the matrix element of the density operator, as obtained in Appendix~\ref{app1}:
\begin{equation}
\begin{split}
&\braket{k_x^\prime,n',\lambda'|\hat n_{\bm q_\parallel}|k_x,n,\lambda}=\\
&=e^{- q_\parallel^2\ell_{ B}^2/4}\delta_{k_x^\prime,k_x-q_x} \mathcal{F}_{n^\prime,\lambda^\prime,n,\lambda}(\bm{q}_\parallel)~. 
\end{split}
\end{equation}
Substituting this result into the expression for $\hat{n}_{\bm{q}_\parallel}$, the density operator can be expressed in terms of the bright modes as

\begin{equation}
    \begin{split}
    \label{eq:bright_density}
        &\hat{n}_{\bm{q}_\parallel}= \sum_{n', \lambda' \in \mathcal{N}_{\rm unocc}} \sum_{n, \lambda \in \mathcal{N}_{\rm occ}}e^{- q_\parallel^2\ell_{ B}^2/4}\sqrt{N_{\rm L}}\times \\&\times \Big[\mathcal{F}_{n^\prime,\lambda^\prime,n,\lambda}(\bm{q}_\parallel)\hat{b}^\dagger_{n,\lambda,n',\lambda'}(q_x)+ \\&+\mathcal{F}_{n,\lambda,n',\lambda'}(\bm{q}_\parallel) \hat{b}_{n,\lambda,n',\lambda'}(-q_x)\Big]~.
    \end{split}
\end{equation}
In this form, the density operator is expressed as a sum of creation and annihilation operators for the bright modes. By substituting the expression from Eq.~\eqref{eq:bright_density} into the interaction Hamiltonian we obtain:

\begin{equation} \begin{split} \label{eq:Hintboson} \hat{\mathcal{H}}_{\rm int}&=-\sum_{\bm{q}_{\parallel},n_z,s} \sum_{n', \lambda' \in \mathcal{N}_{\rm unocc}} \sum_{n, \lambda \in \mathcal{N}_{\rm occ}}\Big(\hat{a}^\dagger_{\bm{q}_\parallel,n_z,s} +\hat{a}_{-\bm{q}_\parallel,n_z,s} \Big)\times\\ &\times\Big[\hbar g_{\bm{q}_{\parallel},n_z,s,n',\lambda',n,\lambda}  \hat{b}^\dagger_{n,\lambda,n',\lambda'}(q_x)+\\ &+\hbar g^*_{-\bm{q}_{\parallel},n_z,s,n',\lambda',n,\lambda}   \hat{b}_{n,\lambda,n',\lambda'}(-q_x)\Big]~, \end{split} \end{equation}
where the couplings are defined as
\begin{equation}
g_{\bm{q}_{\parallel},n_z,s,n',\lambda',n,\lambda} \equiv   \frac{e}{\hbar}  \phi_{{q}_\parallel,n_z,s}(0) \sqrt{\frac{N_{\rm L}}{S}}e^{- q_\parallel^2\ell_{ B}^2/4}\mathcal{F}_{n',\lambda',n,\lambda}(\bm{q}_\parallel)~,
\end{equation}
and the Rabi frequencies are defined as the modulus of the coupling terms,

\begin{equation}
\Omega_{\bm{q}_{\parallel},n_z,s,n,\lambda,n',\lambda'}=|g_{\bm{q}_{\parallel},n_z,s,n,\lambda,n',\lambda'}|~.
\end{equation}
Notice that the interaction Hamiltonian in Eq.~\eqref{eq:Hintboson} is quadratic in the bosonic operators $\hat{a}^\dagger_{\bm{q}_\parallel, n_z}, \hat{a}_{-\bm{q}_\parallel, n_z}, \hat{b}^\dagger_{n, \lambda, n', \lambda'}(q_x), \hat{b}_{n, \lambda, n', \lambda'}(-q_x)$. 

We now focus on the Coulomb term

\begin{equation}
\hat{\mathcal{H}}_{\rm C} = \frac{1}{2S}\sum_{\bm{q}_\parallel} \hat{n}_{-\bm{q}_\parallel}e^2g_{\infty}^{(0)}( q_{\parallel},0,0)\hat{n}_{\bm{q}_\parallel}~,
\end{equation}
where $g_{\infty}^{(0)}(\bm q_{\parallel},0,0)$ is given in Eq.~\eqref{eqn:green_function_hBN_classical_inf}. As the density operator $\hat{n}_{\bm{q}_\parallel}$, as expressed in Eq.~\eqref{eq:bright_density}, is linear in the bright modes, the Coulomb term is thus quadratic in the bright modes.

At this point, all the Hamiltonian terms in Eq.~\eqref{eq:HHopfield} are quadratic in terms of the bright and cavity modes $\hat{b}^\dagger_{n,\lambda,n',\lambda'}(q_x),\hat{b}_{n,\lambda,n',\lambda'}(q_x),\hat{a}^\dagger_{\bm{q}_\parallel,n_z,s},\hat{a}_{\bm{q}_\parallel,n_z,s}$. Hence, the Hamiltonian $\hat{\mathcal{H}}$ in Eq.~\eqref{eq:HHopfield} can be diagonalized by a Bogoliubov transformation. The resulting eigenfrequencies, denoted as ${\omega}_\nu$, describe the energies of the excitation modes in the system. Here, the index $\nu$ encapsulates all relevant quantum numbers. Such eigenfrequencies are given by

\begin{equation}
\begin{split}
\label{eqn:epsilonSM}
1 - e^2 g^{(0)}({q}_\parallel, 0, 0, \omega_\nu) \chi^{(0)}_{\rm nn}({q}_{\parallel}, \omega_\nu)=0~,
\end{split}
\end{equation}
corresponding to setting to zero the RPA dielectric function in Eq.~\eqref{eqn:epsilon}, $\epsilon_{\rm RPA}({q}_{\parallel},\omega_\nu)=0$. \\
\begin{figure*}[t!]
\centering
  \vspace{1.em}
  \begin{overpic}[width=0.9\columnwidth]{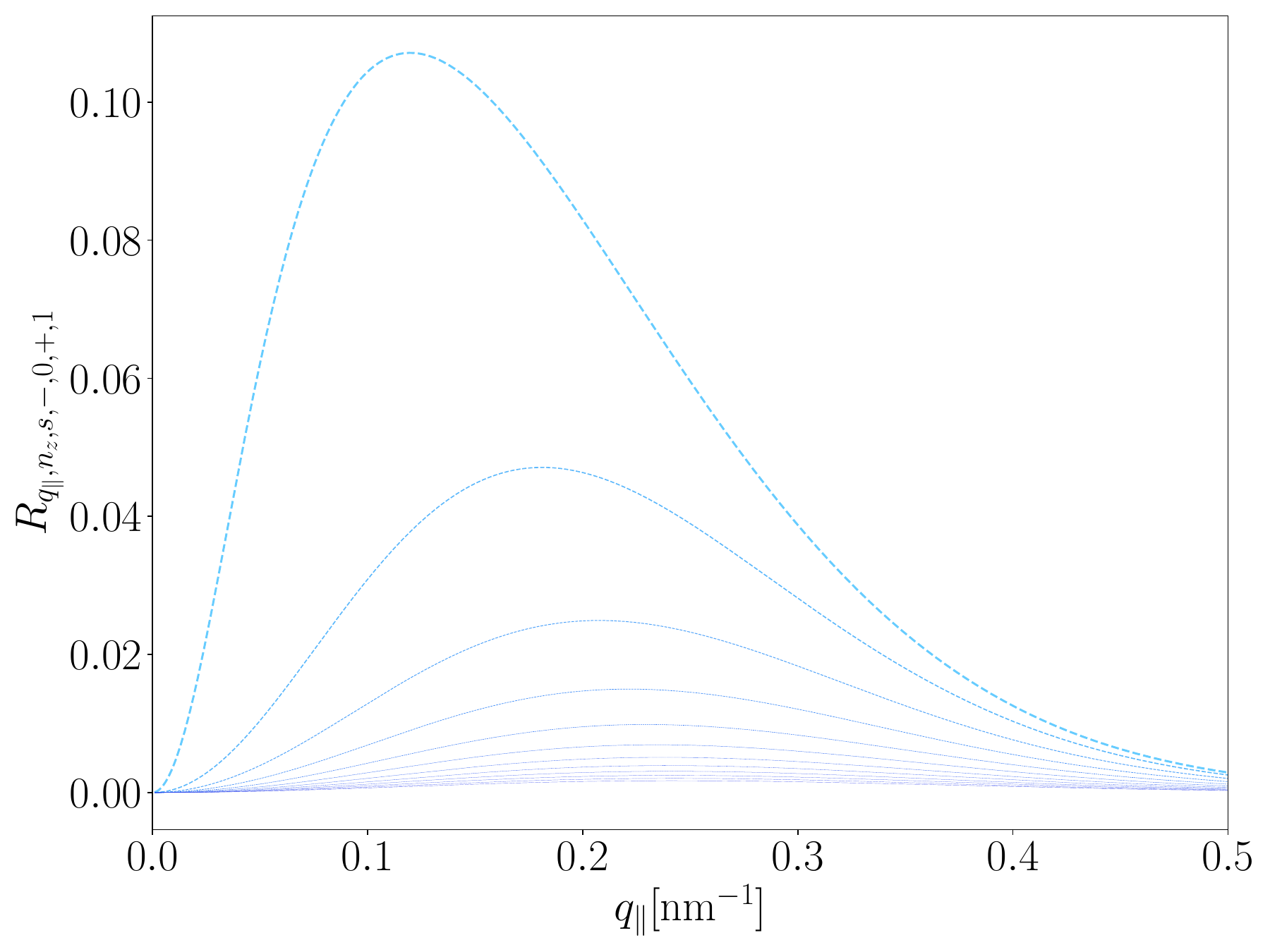}\put(0,80){\normalsize (a)}\end{overpic}  
  \begin{overpic}[width=0.9\columnwidth]{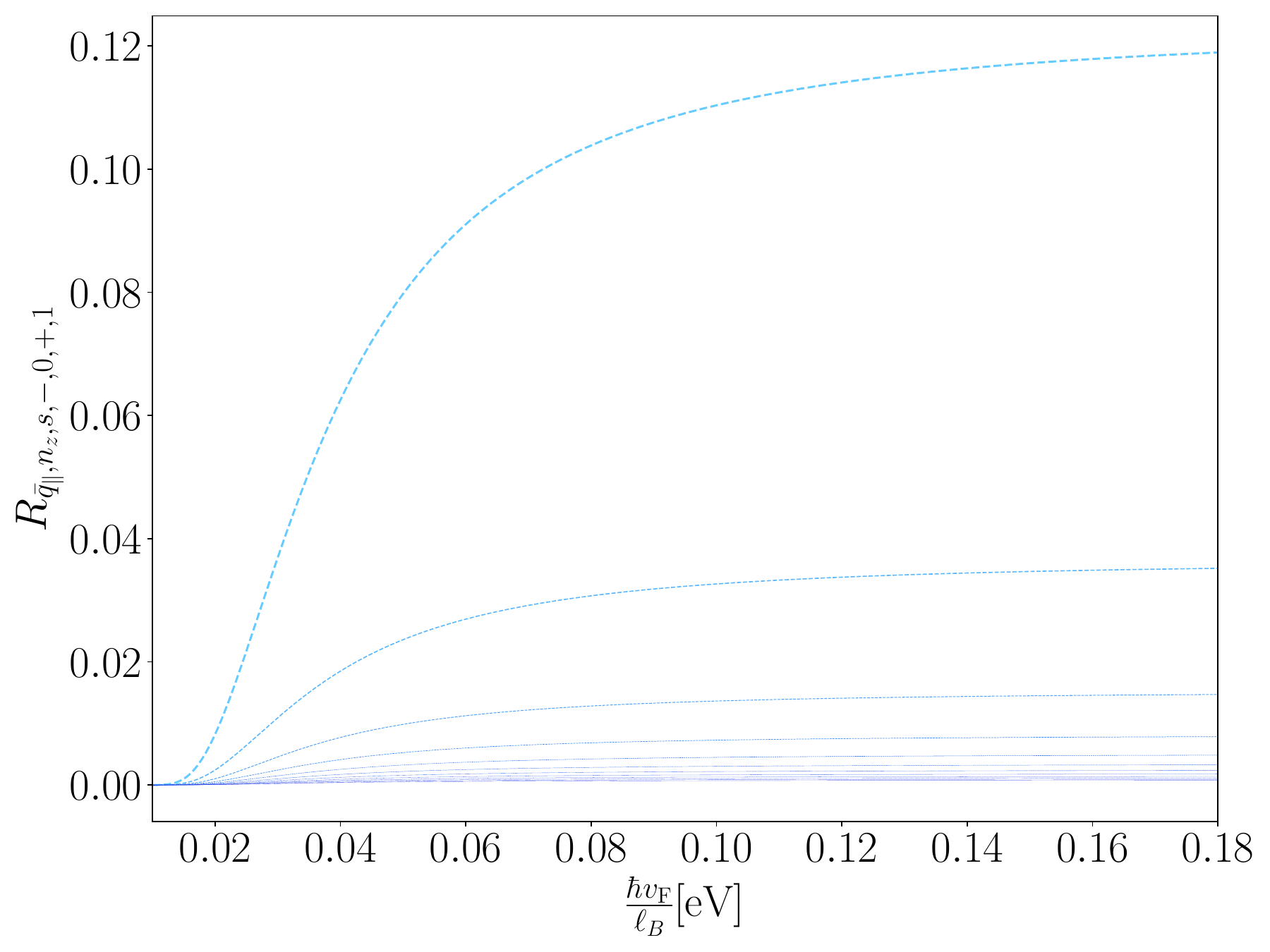}\put(0,80){\normalsize (b)}\end{overpic} 
    \begin{overpic}[width=0.9\columnwidth]{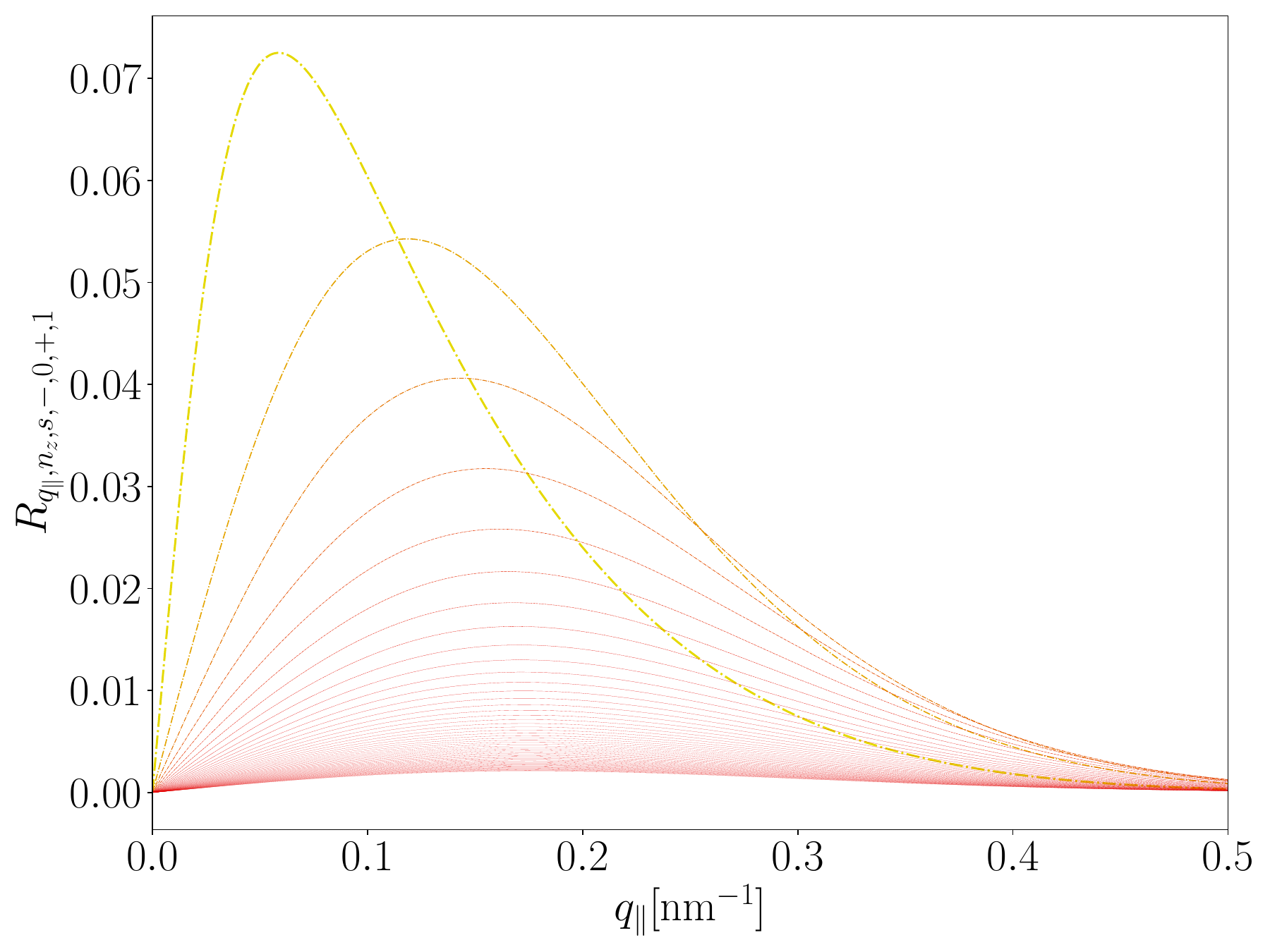}\put(0,80){\normalsize (c)}\end{overpic}  
  \begin{overpic}[width=0.9\columnwidth]{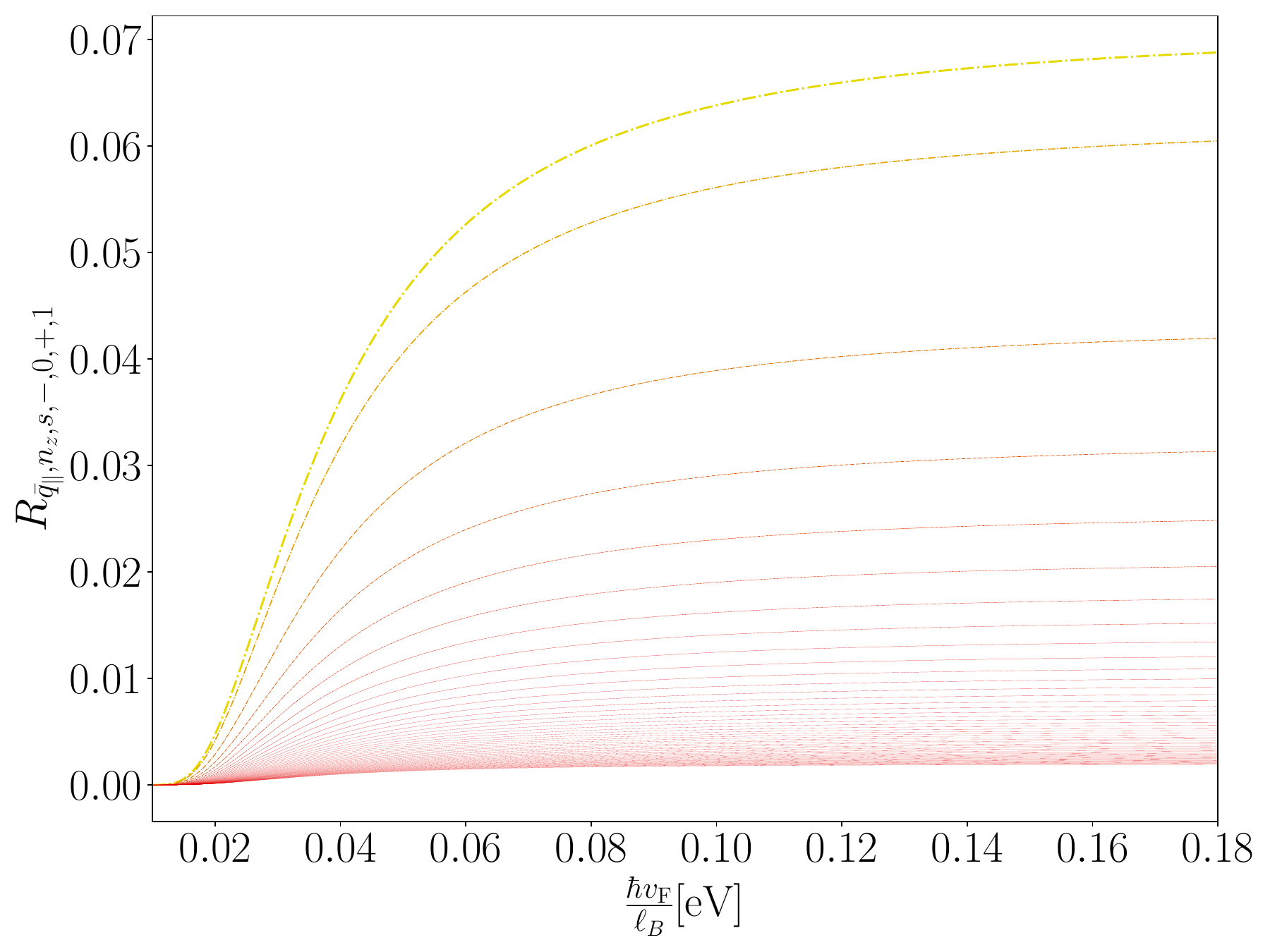}\put(0,80){\normalsize (d)}\end{overpic} 
 \caption{(Color online) Panels (a,c) display the ratio $R_{q_\parallel,n_z,s,-,0,+, 1} = \Omega_{q_\parallel,n_z,s,-,0,+, 1} / \omega_{q_\parallel,n_z,s}$, where $\Omega_{q_\parallel,n_z,s,-,0,+, 1}$ denotes the Rabi frequency for the Landau level (LL) transition $(0, +1)$, and $\omega_{q_\parallel,n_z,s}$ is the cavity mode frequency. This ratio is shown as a function of $q_\parallel$ (in ${\rm nm}^{-1}$) for different values of $n_z$, with $L_z = 30 {\rm nm}$ and $B = 10 {\rm T}$.
 Panels (b,d) present the same ratio as a function of $\hbar v_{\rm F}/\ell_B$ (in ${\rm eV}$) for different $n_z$. Panels (a,b) show these quantities for the upper Reststrahlen band using blue dotted curves, while panels (c,d) present the lower Reststrahlen band with red dashed curves. The line thickness decreases and the color of the curves darkens as the quantum number $n_z$ increases.
 \label{figS3}}
\end{figure*}

\subsection{The vacuum Rabi frequencies}
We now focus on the real part of the dielectric function. The real part of the scalar Green function in Eq.~\eqref{eqn:green_function_hBN_expansion_quantum}, for $z=z^\prime=0$, reads
\begin{equation}
{\rm Re}~g_{\rm q}^{(0)}({q}_\parallel,0,0,\omega)=\sum_{n_z,s} \phi^2_{{q}_\parallel,n_z,s}({0})\left(\frac{2\hbar\omega_{q_\parallel,n_z,s}}{(\hbar\omega)^2-(\hbar\omega_{q_\parallel,n_z,s})^2}\right)~. 
\end{equation}

The real part of the density--density response function, given in Eq.~\eqref{eq:density--densityresponse1} is
\begin{equation}
    \begin{split}
        {\rm Re}\chi^{(0)}_{nn}( q_\parallel,\omega) =
\sum_{n,\lambda\in \mathcal{N}_{\rm occ}}\sum_{n',\lambda'\in\mathcal{N}_{\rm unocc}}  \mathcal{M}_{n^\prime,\lambda^\prime,n,\lambda}({q}_\parallel)\times\\
\times\left( \frac{2(\varepsilon_{\lambda',n'}-\varepsilon_{\lambda,n})}{(\hbar\omega)^2-(\varepsilon_{\lambda,n}-\varepsilon_{\lambda',n'})^2}\right)~.
    \end{split}
\end{equation}
Thus, the real part of the dielectric function can be conveniently written as  

\begin{widetext}

\begin{align}
\label{eq:epsilon_oscillators}
{\rm Re}~\epsilon_{\rm RPA}( q_\parallel,\omega) =1-\sum_{n_z,s}\sum_{n,\lambda\in \mathcal{N}_{\rm occ}}\sum_{n',\lambda'\in\mathcal{N}_{\rm unocc}} {(\hbar\Omega_{q_\parallel,{n}_z,{s},{\lambda},{n},{\lambda}',{n}'})^2}
\left( \frac{2\hbar\omega_{q_\parallel,n_z,s}}{\left[(\hbar\omega)^2-(\hbar\omega_{q_\parallel,n_z,s})^2\right]}\frac{{2(\varepsilon_{\lambda',n'}-\varepsilon_{\lambda,n})}}{\left[(\hbar\omega)^2-(\varepsilon_{\lambda,n}-\varepsilon_{\lambda',n'})^2\right]}\right) ~,
\end{align}

\end{widetext}

where the Rabi frequencies $\Omega_{q_\parallel,{n}_z,{s},{\lambda},{n},{\lambda}',{n}'}$ are defined as
\begin{align}
\label{eq:Rabi}
& \hbar\Omega_{q_\parallel,{n}_z,{s},{\lambda},{n},{\lambda}',{n}'} =e \phi_{{q}_\parallel,n_z,s}(0)\sqrt{{ \mathcal{M}_{n^\prime,\lambda^\prime,n,\lambda}({q}_\parallel)}} ~. 
\end{align}
Eq.~\eqref{eq:epsilon_oscillators} can be interpreted as describing the response of a system composed of harmonic modes with frequencies $\omega_{q_\parallel,n_z,s}$, which are coupled to other harmonic modes with frequencies $(\varepsilon_{\lambda',n'} - \varepsilon_{\lambda,n})/\hbar$. The coupling strength between these modes is characterized by the Rabi frequency $\hbar\Omega_{q_\parallel,{n}_z,{s},{\lambda},{n},{\lambda}',{n}'}$. 
Eq.~\eqref{eq:Rabi} demonstrates how the vacuum Rabi frequency is determined by the fluctuations of both the electronic density and the electric field, extending the standard concept of C-QED \cite{dalibard1982,DDB_arxiv_2023,ciuti2005} to non-local (\( q_\parallel \neq 0 \)) sub-wavelength cavities interacting with a 2D material.

In Fig.~\ref{figS3}, we present the ratio  $R_{q_\parallel,n_z,s,-,0,+, 1}\equiv(\Omega_{q_\parallel,n_z,s,-,0,+, 1}/\omega_{q_\parallel,n_z,s})$, where $\Omega_{q_\parallel,n_z,s,-,0,+, 1}$ is the Rabi frequency associated with the LL transition $(0,+1)$ and $\omega_{q_\parallel,n_z,s}$ is the cavity frequency as a function of different parameters. 
In Fig.~\ref{figS3}(a,c),  the ratio  $R_{q_\parallel,n_z,s,-,0,+, 1}$ is plotted as a function of $q_\parallel$ (in ${\rm nm}^{-1}$) for various $n_z$. This plot reveals a non-monotonic behavior in $q_\parallel$, where each mode exhibits a specific $q_\parallel$ that maximizes the Rabi frequency. Fig.~\ref{figS3}(b,d) shows the Rabi frequency as a function of $\hbar v_{\rm F}/\ell_B$, which can be controlled experimentally by adjusting the external magnetic field. At low values, specifically around $\hbar v_{\rm F}/\ell_B \approx 0.01~{\rm eV}$, increasing the magnetic field—and thus this ratio—leads to a stronger coupling. However, the coupling saturates when $\hbar v_{\rm F}/\ell_B$ reaches approximately $0.15~{\rm eV}$ or higher.

In Fig.~\ref{fig_Rabi} (a-b) and Fig.~\ref{figS3}(a-d), we observe that the maximum ratio $R_{q_\parallel,n_z,s,-,0,+, 1}$, achieved for $n_z = 1$, exceeds $10\%$ for the upper Reststrahlen band and approaches $10\%$ for the lower band, placing our system in the ultra-strong coupling regime, where the coupling strength is a significant fraction of the cavity frequency \cite{Kockum19}.

\subsection{The depolarization shift}

Fig.~\ref{fig2}(a) shows $-{\rm Im}~{\chi}_{\rm nn} ({q}_{\parallel}, \omega)$ as a function of the wavevector $q_\parallel$ for a fixed magnetic field, $B=5~{\rm T}$. This plot reveals that, outside the Reststrahlen bands, LL excitations exhibit a $q_\parallel$-dependent positive frequency shift, known as the {\it depolarization shift}  \cite{Giuliani_and_Vignale,shtrichman2001}. Here, we investigate the Coulomb origin of this effect.
First, we recall that the dressed density--density response function is defined as ${\chi}_{\rm nn}({q}_{\parallel},\omega) = {\chi}^{(0)}_{\rm nn}({q}_{\parallel},\omega)/\epsilon_{\rm RPA}({q}_{\parallel},\omega)$. We further define the frequency-independent dressed density--density response function ${\chi}^{\infty}_{\rm nn}({q}_{\parallel},\omega)$ as the response function dressed by $\epsilon^{\infty}_{\rm RPA}({q}_{\parallel},\omega)$, which considers only the high-frequency limit of the Green function $g_{\infty}^{(0)}(\bm{r},\bm{r}^\prime)$, excluding contributions from quantum modes:
\begin{equation}
    \begin{split}
        {\chi}^{\infty}_{\rm nn}({q}_{\parallel},\omega)&=\frac{{\chi}^{(0)}_{\rm nn}({q}_{\parallel},\omega)}{\epsilon^{\infty}_{\rm RPA}({q}_{\parallel},\omega)}~,\\
        {\epsilon^{\infty}_{\rm RPA}({q}_{\parallel},\omega)}&=1-e^2 g_{\infty}^{(0)}({q}_\parallel, 0, 0) \chi^{(0)}_{\rm nn}({q}_{\parallel}, \omega)~.\\
    \end{split}
\end{equation}

In these equations, we isolate a single dominant LL transition
$\alpha\equiv(\bar{n},\bar{\lambda})\to(\bar{n}',\bar{\lambda}')$ with $\bar{n},\bar{\lambda}\in\mathcal{N}_{\rm occ}$ and $\bar{n}',\bar{\lambda}'\in\mathcal{N}_{\rm unocc}$,
bare frequency $\omega_{\alpha}=(\varepsilon_{\bar{\lambda}',\bar{n}'}-\varepsilon_{\bar{\lambda},\bar{n}})/\hbar>0$,
and oscillator strength $\mathcal{M}_{\bar{n}',\bar{\lambda}',\bar{n},\bar{\lambda}}(q_\parallel)$ defined in Eq.~\eqref{eq:density--densityresponse01}.
\begin{equation}
{\rm Re}~\chi^{(0)}_{\rm nn}(q_\parallel,\omega)=
\frac{2\hbar\omega_{\alpha}\mathcal{M}_{\bar{n}',\bar{\lambda}',\bar{n},\bar{\lambda}}(q_\parallel)}{(\hbar\omega)^2-(\hbar\omega_{\alpha})^2}
+{\rm Re}~\chi^{(0)}_{\rm reg}(q_\parallel,\omega)\,,
\label{eq:chi0_single_pole}
\end{equation}
where $\chi^{(0)}_{\rm reg}$ is regular at $\omega=\omega_{\alpha}$.

Near $\omega\simeq \omega_{\alpha}$, the regular part can be evaluated at the bare LL frequency $\chi^{(0)}_{\rm reg}(q_\parallel,\omega)\approx \chi^{(0)}_{\rm reg}(q_\parallel,\omega_{\alpha})$ . The collective resonance is then determined by the zero of
${\rm Re}~\epsilon^{\infty}_{\rm RPA}(q_\parallel,\omega)$, i.e.

\begin{equation}
\label{eq:epsilonRPAeff}
\epsilon_{\rm bg}(q_\parallel)-{e^2 g^{(0)}_\infty(q_\parallel,0,0)} \mathcal{M}_{\bar{n}',\bar{\lambda}',\bar{n},\bar{\lambda}}(q_\parallel)\,
\frac{2\hbar\omega_{\alpha}}{(\hbar\omega)^2-(\hbar\omega_{\alpha})^2}\simeq 0~,
\end{equation}
where we defined $\epsilon_{\rm bg}(q_\parallel)\equiv1-{e^2 g^{(0)}_\infty(q_\parallel,0,0)}{\rm Re}\chi^{(0)}_{\rm reg}(q_\parallel,\omega_{\alpha})$, which represents the background dielectric function arising from all other LL transitions.
Eq.~\eqref{eq:epsilonRPAeff} is solved by
\begin{equation}
\hbar\omega_{\alpha}^{(\infty)}(q_\parallel)\simeq\hbar\omega_{\alpha}\sqrt{1+2\frac{\Delta_{\alpha}^{\rm dep}(q_\parallel)}{{\hbar\omega_\alpha}}}~,
\label{eq:pole_condition_single_transition}
\end{equation}
with the Hartree depolarization shift
\begin{equation}
\Delta_{\alpha}^{\rm dep}(q_\parallel)=\frac{e^2}{\epsilon_{\rm bg}(q_\parallel)}
~g^{(0)}_\infty(q_\parallel,0,0)\,\mathcal{M}_{\bar{n}',\bar{\lambda}',\bar{n},\bar{\lambda}}(q_\parallel)~.
\label{eq:depol_shift_single_transition}
\end{equation}
Since $g^{(0)}_\infty(q_\parallel,0,0)>0$ being a repulsive Coulomb interaction and
$\mathcal{M}_{\bar{n}',\bar{\lambda}',\bar{n},\bar{\lambda}}(q_\parallel)>0$ by definition, one has
$\Delta\omega_{\alpha}^{\rm dep}>0$, i.e. a blue shift. 

\begin{figure*}[t]
\centering
  \vspace{1.em}
  \begin{overpic}[width=0.68\columnwidth]{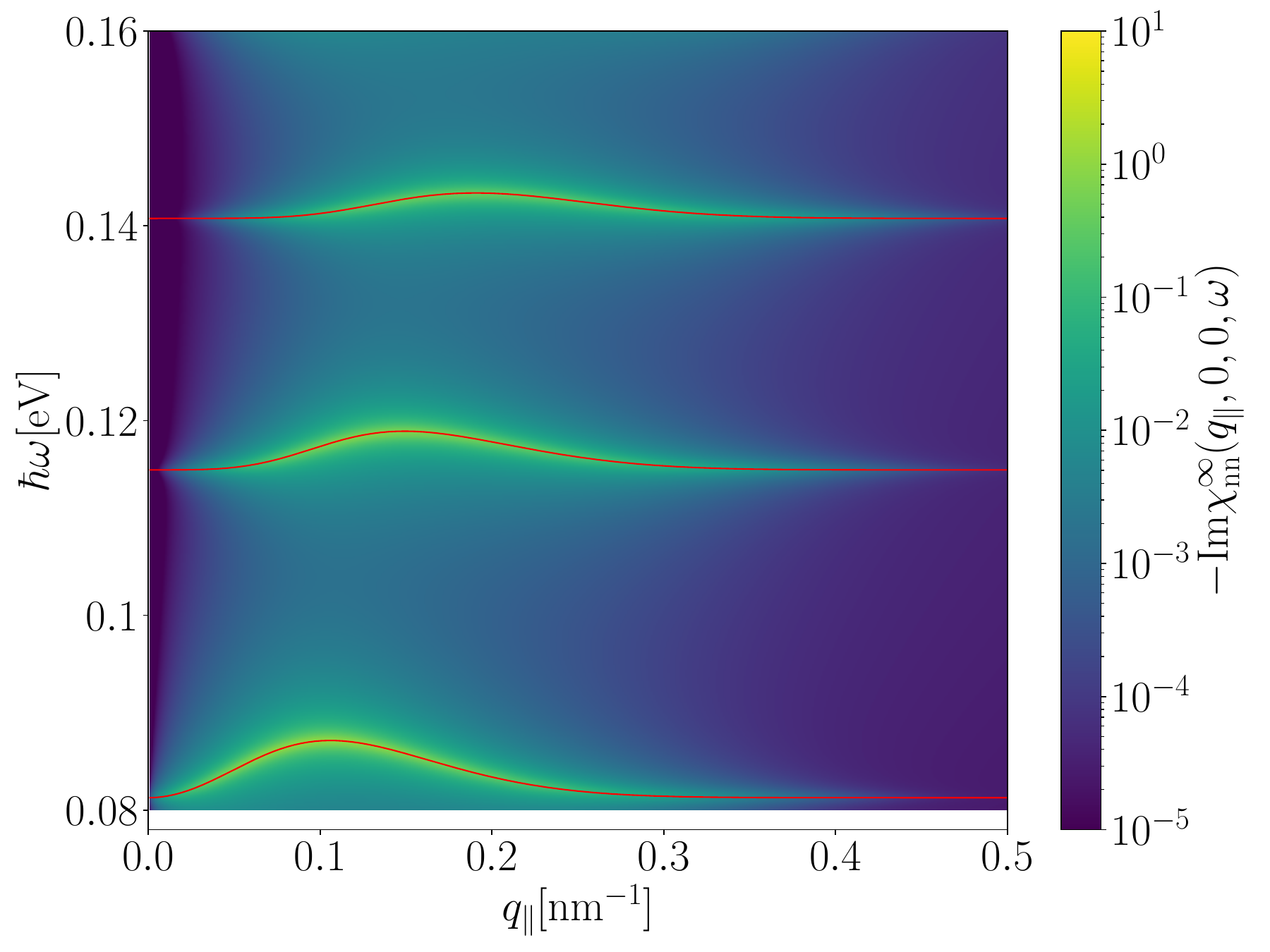}\put(0,80){\normalsize (a)}\end{overpic} 
  \begin{overpic}[width=0.68\columnwidth]{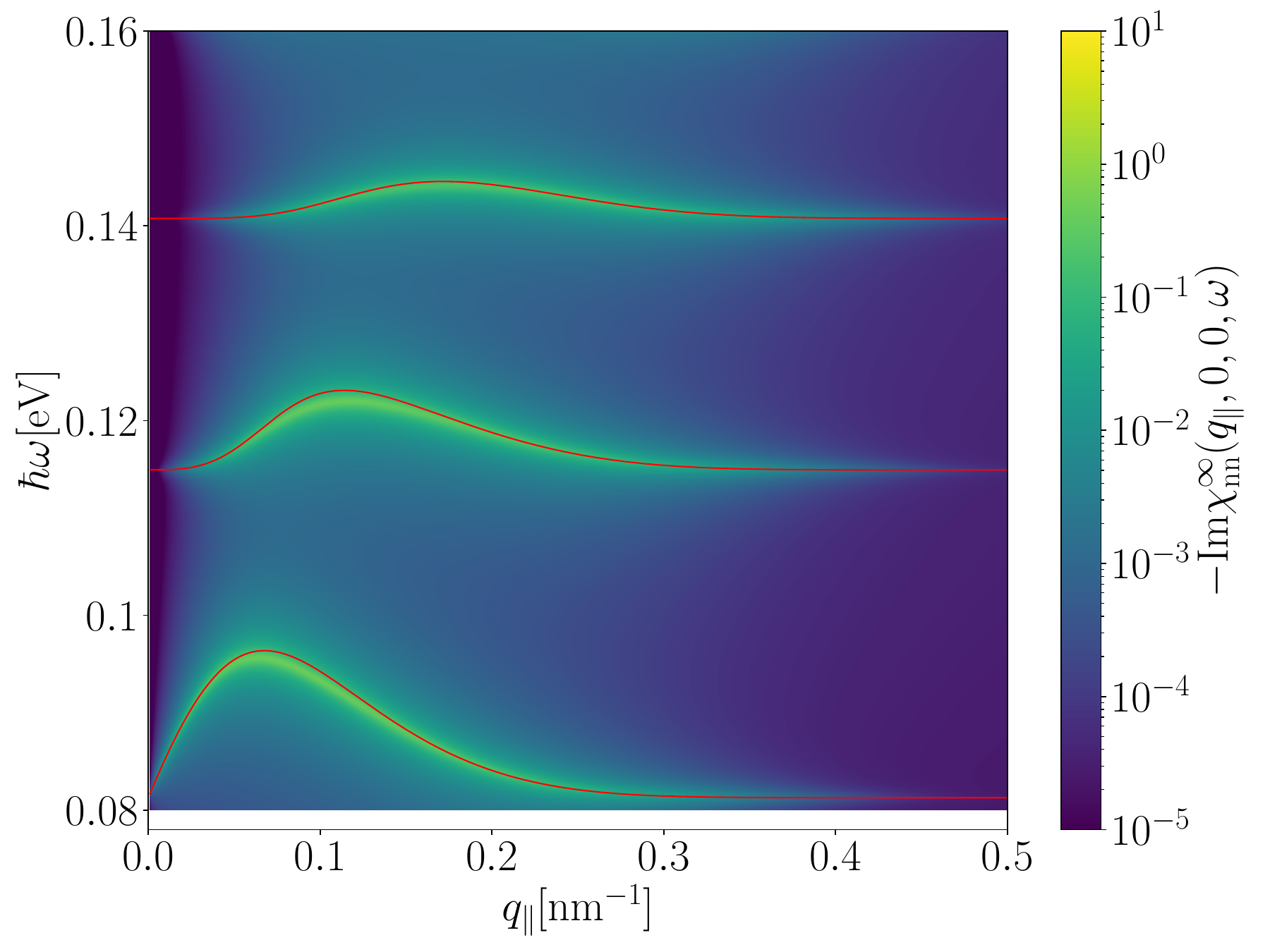}\put(0,80){\normalsize (b)}\end{overpic} 
  \begin{overpic}[width=0.68\columnwidth]{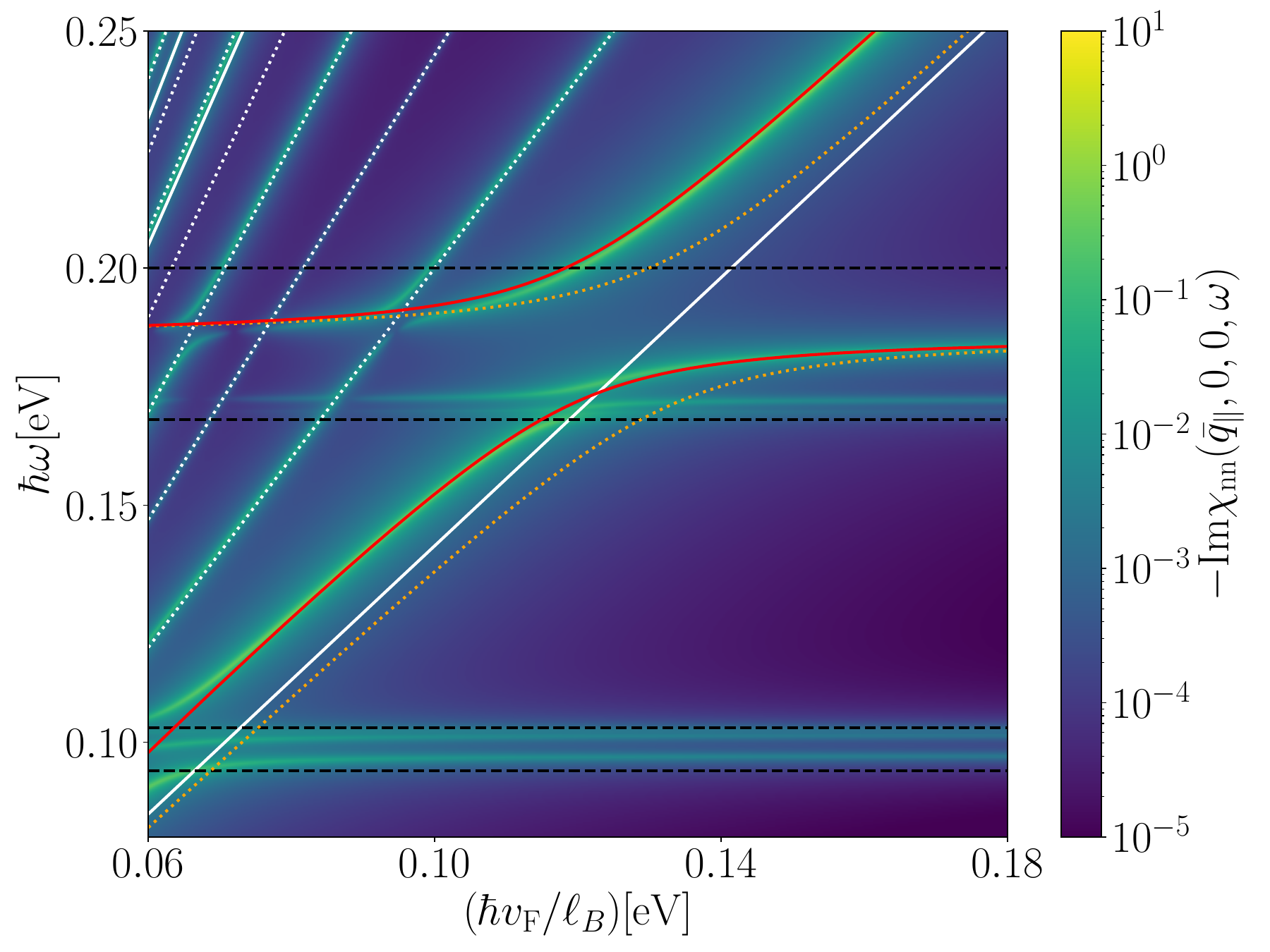}\put(0,80){\normalsize (c)}\end{overpic} 
 \caption{(Color online)  Panel (a) shows $-{\rm Im}~{\chi}^{\infty}_{\rm nn} ({q}_{\parallel}, \omega)$ as a function of the wavevector $q_\parallel$ for a given magnetic field, $B=5~{\rm T}$, and for $L_z=5~{\rm nm}$. 
 Panel (b) shows the same quantity in the absence of the cavity confinement $L_z\to \infty$. 
 Panel (c) shows $-{\rm Im}~{\chi}_{\rm nn} ({q}_{\parallel}, \omega)$ as a function of $\hbar v_{\rm F}/\ell_B$, for fixed momentum $\bar{q}_\parallel=0.05~{\rm nm}^{-1}$ and for $L_z=50~{\rm nm}$.
 Other parameters are $N_{\rm c}=30$, $\hbar \eta= 5\times 10^{-4}~{\rm eV}$. For the scalar Green function, the expression resummed over all modes was employed. \label{figS4} }
\end{figure*}

Fig.~\ref{figS4}(a) displays $-{\rm Im}~\chi^{\infty}_{\rm nn}(q_{\parallel},\omega)$ as a function of the wavevector $q_\parallel$ for $B=5~{\rm T}$ for strong confinement $L_z=5 {\rm nm}$. Fig.~\ref{figS4}(b), shows the same quantity in the absence of cavity confinement ($L_z\to\infty$). In both figures the $q_\parallel$ dependence of the depolarization shift is evident. By comparing the two, it is clear that reducing the cavity thickness $L_z$ suppresses the depolarization shift.
The energies calculated from Eq.~\eqref{eq:depol_shift_single_transition} are plotted as continuous red lines in Fig.~\ref{figS4}(a,b), showing good agreement between the analytical expression and the numerical results.
Eq.~\eqref{eq:depol_shift_single_transition} further shows that the shift is controlled by the static Coulomb propagator $g^{(0)}_\infty(q_\parallel,0,0)$. As discussed below Eq.~\eqref{eqn:green_function_hBN_classical_inf}, cavity confinement reduces the effective strength of the Coulomb repulsion when $q_\parallel L_z \lesssim 1$, in agreement with the trend observed by comparing Fig.~\ref{figS4}(a) and Fig.~\ref{figS4}(b).

\subsection{Two-transition model}

Here, we identify a regime where a given LL transition is coupled only to a single cavity mode, allowing for an effective simplified description.
Indeed, Fig.~\ref{figS4}(c) shows $-{\rm Im}~{\chi}_{\rm nn} (q_{\parallel}, \omega)$ as a function of $\hbar v_{\rm F}/\ell_B$, for a small wavevector $\bar{q}_\parallel=0.05~{\rm nm}^{-1}$ and $L_z=50~{\rm nm}$. In this regime, where $q_\parallel L_z \gg 1$, only the Rabi frequency with $n_z = 1$ remains significant, as the contributions from other $n_z$ modes rapidly decay to zero, as evident in Fig.~\ref{figS3}(a). This behavior allows us to simplify the analysis by focusing exclusively on the interaction between a specific cavity mode $(q_\parallel, \bar{n}_z, \bar{s})$ and a single LL transition$(\bar{n}, \bar{\lambda}) \rightarrow (\bar{n}', \bar{\lambda}')$. 
Under this approximation, the dielectric function ${\rm Re}~\epsilon^{(2)}_{\rm RPA}(q_\parallel, \omega)$ can be expressed as:

\begin{equation} \label{eq:ttmodel}
    \begin{split}
     &{\rm Re}~\epsilon^{(2)}_{\rm RPA}( q_\parallel,\omega) \equiv \\
&=1 -\frac{4(\hbar\Omega_{q_\parallel,\bar{n}_z,\bar{s},\bar{\lambda},\bar{n},\bar{\lambda}',\bar{n}'})^2{(\varepsilon_{\bar{\lambda}',\bar{n}'}-\varepsilon_{\bar{\lambda},\bar{n}})}\hbar\omega_{q_\parallel,\bar{n}_z,\bar{s}}}{\left[(\hbar\omega)^2-(\hbar\omega_{q_\parallel,\bar{n}_z,\bar{s}})^2\right]\left[(\hbar\omega)^2-(\varepsilon_{\bar{\lambda},\bar{n}}-\varepsilon_{\bar{\lambda}',\bar{n}'})^2\right]}~.   
    \end{split}
\end{equation}

where all quantum numbers $x$ are fixed and denoted by $\bar{x}$. The corresponding frequencies of the coupled modes can be derived by solving the quadratic equation, obtained by setting ${\rm Re}~\epsilon^{(2)}_{\rm RPA}(q_\parallel, \omega_\nu)=0$:

\begin{equation}
        \begin{split}
          &(\hbar\omega_\nu )^2= \frac{(\hbar\omega_{q_\parallel,\bar{n}_z,\bar{s}})^2 + (\varepsilon_{\bar{\lambda},\bar{n}} - \varepsilon_{\bar{\lambda}',\bar{n}'})^2}{{2}}+\\
&\pm \frac{1}{2}\Big\{\left[(\hbar\omega_{q_\parallel,\bar{n}_z,\bar{s}})^2 - (\varepsilon_{\bar{\lambda},\bar{n}} - \varepsilon_{\bar{\lambda}',\bar{n}'})^2\right]^2 + \\
&+16(\hbar\Omega_{q_\parallel,\bar{n}_z,\bar{s},\bar{\lambda},\bar{n},\bar{\lambda}',\bar{n}'})^2\hbar\omega_{q_\parallel,\bar{n}_z,\bar{s}} (\varepsilon_{\bar{\lambda},\bar{n}} - \varepsilon_{\bar{\lambda}',\bar{n}'})\Big\}^{\frac{1}{2}}~.  
        \end{split}
    \end{equation}

These frequencies are plotted as orange dotted lines in Fig.~\ref{figS4}(c) for the $(0,1)$ LL transition coupled to the HPP mode with $n_z=1$ and $s=+$. The bare LL transitions are shown as white lines. As evident from the figure, this simplified model fails to capture the correct excitation spectrum of the system accurately. Even when the two bare excitations are detuned, the model only correctly predicts the cavity mode, while it fails to reproduce the shifted LL transition, which experiences a positive depolarization shift. This shift arises due to the influence of the static Coulomb potential $g_{\infty}^{(0)}({q}_\parallel,0,0)$.

Therefore, to achieve a more accurate description of the spectrum, it is essential to include the static Coulomb potential $g_{\infty}^{(0)}({q}_\parallel,0,0)$ in the analysis, which modifies Eq.~\eqref{eq:ttmodel} as follows

\begin{equation}
    \begin{split}
        &{\rm Re}~\epsilon^{(2+{\rm C})}_{\rm RPA}( q_\parallel,\omega) = \nonumber\\
&=1 -\frac{4(\hbar\Omega_{q_\parallel,\bar{n}_z,\bar{s},\bar{\lambda},\bar{n},\bar{\lambda}',\bar{n}'})^2{(\varepsilon_{\bar{\lambda}^\prime,\bar{n}^\prime}-\varepsilon_{\bar{\lambda},\bar{n}})}\hbar\omega_{q_\parallel,\bar{n}_z,\bar{s}}}{\left[(\hbar\omega)^2-(\hbar\omega_{q_\parallel,\bar{n}_z,\bar{s}})^2\right]\left[(\hbar\omega)^2-(\varepsilon_{\bar{\lambda},\bar{n}}-\varepsilon_{\bar{\lambda}',\bar{n}'})^2\right]}+ \nonumber\\
&-e^2g_{\infty}^{(0)}({q}_\parallel,0,0)\left( \frac{2(\varepsilon_{\bar{\lambda}',\bar{n}'}-\varepsilon_{\bar{\lambda},\bar{n}}) \mathcal{M}_{\bar{n}',\bar{\lambda}',\bar{n},\bar{\lambda}}({q}_\parallel)}{(\hbar\omega)^2-(\varepsilon_{\bar{\lambda},\bar{n}}-\varepsilon_{\bar{\lambda}',\bar{n}'})^2}\right)~.
    \end{split}
\end{equation}
The solutions of ${\rm Re}~\epsilon^{(2+{\rm C})}_{\rm RPA}( q_\parallel, \omega)$ are shown as red solid lines in Fig.~\ref{figS4}(c) for the $(0,1)$ LL transition coupled to the hPP mode with $n_z=1$ and $s=+1$. These solutions accurately reproduce the observed spectrum, demonstrating both the critical role of the coupling to the resonant mode with $n_z=1$ and $s=+$ and the importance of incorporating the static Coulomb potential $g_{\infty}^{(0)}({q}_\parallel, 0, 0)$ in the model. In this case, the Rabi frequency is approximately $\Omega_{q_\parallel,0,+1,-,0,+,1} \approx 0.014~{\rm eV}$, which corresponds to about $7\%$ of the bare cavity frequency.  This value indicates that the system, for low wavevectors, is near the threshold of the ultra-strong coupling regime.

\section{Conclusions}

Sub-wavelength cavities are described by a scalar potential $\phi$ solving the electrostatic Poisson equation rather than a vector potential\cite{andolina2024}. Starting from this insight, we developed a new theoretical framework to study C-QED of sub-wavelength cavities, by quantizing the scalar potential $\hat{\phi}$, assessing quantum vacuum fluctuations, and providing a way to calculate vacuum Rabi frequencies, without resorting to phenomenological compression factors. Our theory provides a quantitative tool to directly assess light-matter interactions in the strong and ultra-strong coupling regimes, offering a predictive tool for cavity design.

Motivated by recent experimental studies \cite{dapolito2023, wehmeier2024} that investigated the coupling between hBN phonon-polaritons and LL transitions in graphene in a SNOM setup,  we applied our framework to investigate the polaritonic spectrum of LLs in graphene embedded in an HPP cavity. Crucially, we incorporated non-local response beyond the {\it dipole approximation}. 
Our analysis demonstrated that the non-local response introduces spectral features absent in purely local theories, such as coupling to LL transitions that are optically forbidden in the dipole approximation.
Furthermore, our non-local theory shows that the vacuum Rabi frequency is maximized when the light's momentum, $q_\parallel$, is comparable to the inverse of the magnetic length, $q_\parallel \ell_B \sim 1$. This insight underscores the importance of considering non-local effects for achieving optimal light-matter coupling in such systems.
These results indicate that {\it ultra-strong} coupling between LL transitions and cavity modes can be effectively engineered under high magnetic fields and low temperatures.

Beyond this specific case, our approach paves the way for engineering the electronic properties of materials embedded in cavities. Given the numerous predictions in this field \cite{gao2020,schlawin2019,wang2019,nguyen2023,zezyulin2022,lenk2022,arwas2023,pavosevic2022}, it would be important to revisit their conclusion in light of our new theoretical framework for sub-wavelength cavities. 
Previous studies largely relied on Fabry–Pérot geometries or phenomenological models of sub-wavelength confinement. In contrast, our scalar Green's function theory provides a rigorous foundation to analyze how cavities modify quantum materials.
For example, it could enable the design of superconductivity in moiré materials through tailored cavity structures \cite{BlochReview}, the modulation of topological phenomena such as the quantum Hall effect \cite{appugliese2022,enkner2024}, or the enhancement and control of interaction effects, including fractional quantum Hall states \cite{enkner2024b,bacciconi2023} or correlated states in twisted bilayer graphene \cite{Cao} in a cavity. Although the present analysis is restricted to the linear-response regime and the RPA, the description of HPP cavities via the scalar Green’s function provides a natural pathway for extending the framework to a more general diagrammatic formulation that can incorporate non-linear many-body effects.
Indeed, our microscopic theory of sub-wavelength cavities provides a rigorous framework to explore how cavity environments can enhance \cite{Ebbesen19,Sentef2018} or induce \cite{schlawin2019} superconductivity. To achieve this, our theory could be interfaced with Eliashberg theory, allowing for a systematic treatment of the strong non-retarded electron-electron interactions induced by the cavity.

Another promising direction for future research would be the study of polaritonic lattices \cite{Jacqmin2014}, wherein HPP cavities are constructed from a lattice of different materials or incorporate a grating grown on the cavity, thereby breaking strict translational symmetry. In such systems, energy bands naturally emerge and can be described using Bloch theory. This additional degree of control could significantly enhance the tunability of polaritonic systems.  Our theoretical framework could be readily generalized to describe these more complex lattice-based systems. 

Altogether, our work establishes a new theoretical paradigm for the microscopic description of C-QED in sub-wavelength cavities. It not only advances the fundamental understanding of light–matter interactions in confined geometries but also offers a powerful framework for designing next-generation cavity-embedded quantum materials.
The simulation scripts are available from Zenodo \cite{Zenodo}, and all the figures are available from Figshare \cite{Figshare}.

\subsection{ Acknowledgments } G.M.A acknowledges useful discussion with I. Torre, F.M.D Pellegrino, L. Orsini, J. Sueiro, G. Mazza, Z. Bacciconi, D. De Bernardis, G. Chiriacò, M. Michael, A. Rubio, M. Ruggenthaler  and F. Schlawin.  G.M.A acknowledge funding from the European Union’s Horizon 2020 research and innovation programme under the Marie Sklodowska-Curie
(Grant Agreement No. 101146870 --  COMPASS). G.M.A and M.S. acknowledge funding from the European Research Council (ERC) under the European Union's Horizon 2020 research and innovation programme (Grant agreement No. 101002955 -- CONQUER). B.T. acknowledges funding from “la Caixa” Foundation (ID 100010434). This research was supported in part by grant NSF PHY-2309135 to the Kavli Institute for Theoretical Physics (KITP). 
R.R. and M.P. were supported by the MUR - Italian Ministry of University and Research under the ``Research projects of relevant national interest  - PRIN 2020''  - Project No.~2020JLZ52N (``Light-matter interactions and the collective behavior of quantum 2D materials, q-LIMA'') and by the European Union under grant agreement No. 101131579 - Exqiral and No.~873028 - Hydrotronics. 
F.H.L.K. acknowledges support from the Gordon and Betty Moore Foundation through Grant GBMF12212, and the Government of Spain (FIS2016-81044; PID2019-106875GB-100; Severo Ochoa CEX2019-000910-S [MCIN/ AEI/10.13039/501100011033], PCI2021-122020-2A, and PDC2022-133844-100 funded by MCIN/AEI/10.13039/501100011033). This work was also supported by the European Union NextGenerationEU/PRTR (PRTR-C17.I1) and EXQIRAL 101131579, Fundació Cellex, Fundació Mir-Puig, and Generalitat de Catalunya (CERCA, AGAUR, 2021 SGR 014431656). Views and opinions expressed are those of the author(s) only and do not necessarily reflect those of the European Union Research Executive Agency. Neither the European Union nor the granting authority can be held responsible for them. This material is based upon work supported by the Air Force Office of Scientific Research under award number FA8655-23-1-7047. Any opinions, findings, conclusions, or recommendations expressed in this material are those of the author(s) and do not necessarily reflect the views of the United States Air Force.

\appendix 

\section{The dielectric tensor in hBN}
\label{app0}

\begin{figure}[t]
  \vspace{1.em}
  \begin{overpic}[width=0.9\columnwidth]{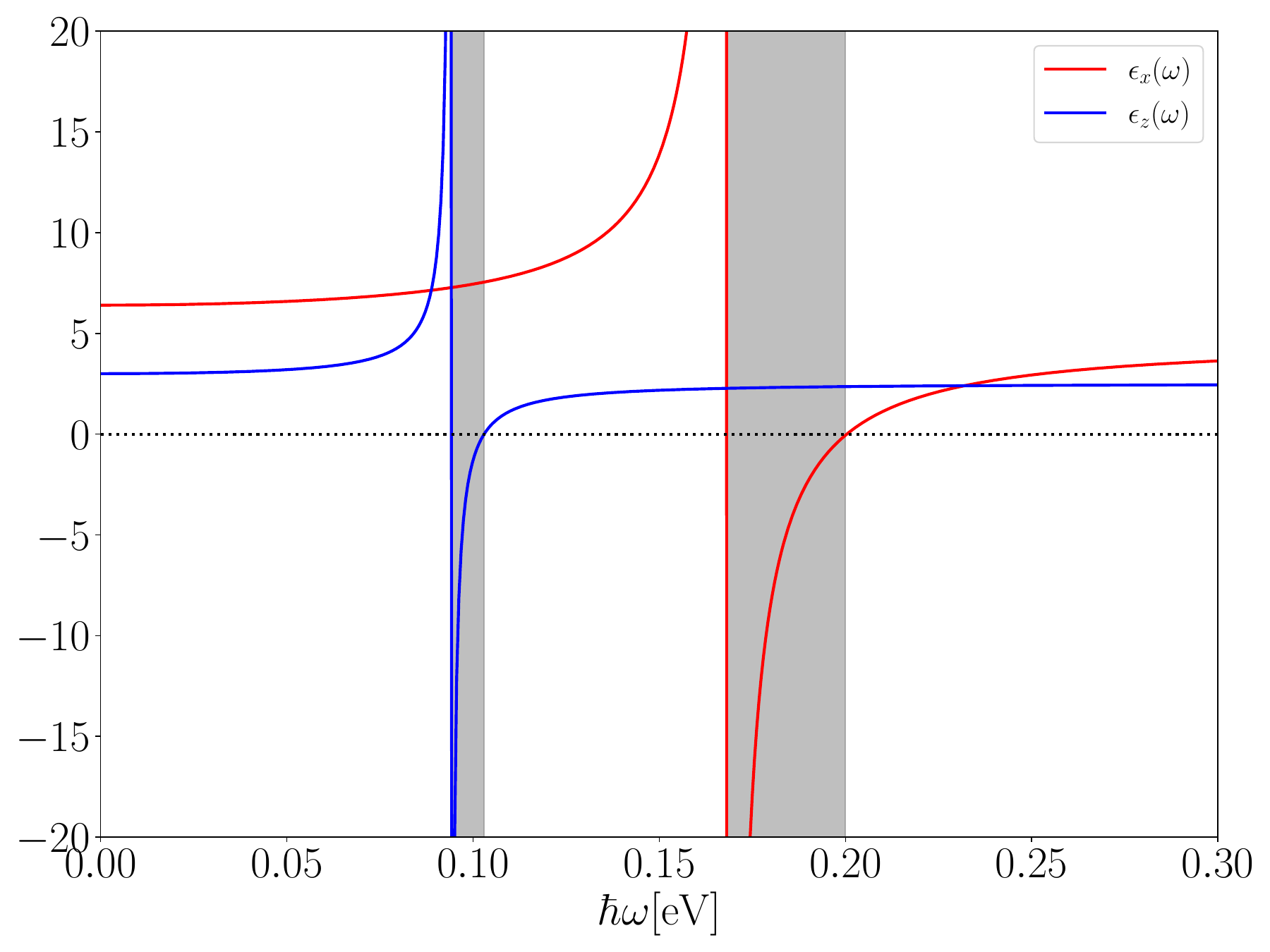}\put(0,80){\normalsize}\end{overpic} \\
  
 \caption{(Color online) The in-plane (out-of-plane) dielectric function $\epsilon_x(\omega)$ ($\epsilon_z(\omega)$) is plotted as a function of frequency $\omega$, using the realistic values reported in Table~\ref{tab:1}. The two Reststrahlen bands, where $\epsilon_x(\omega)\epsilon_z(\omega)<0$, are highlighted in gray. \label{figS0}}
\end{figure}
hBN is a two-dimensional material composed of boron and nitrogen atoms arranged in a hexagonal lattice structure, akin to graphene. Here, we consider stacked layers of hBN, forming a three-dimensional structure with weak van der Waals interactions between the layers. This material exhibits hyperbolic behavior due to its anisotropic dielectric tensor $\bm{\epsilon}(\bm{r},\omega)={\rm diag}(\epsilon_x(\bm{r},\omega),\epsilon_x(\bm{r},\omega),\epsilon_z(\bm{r},\omega))$, where the real parts of its principal components have opposite signs in the Reststrahlen bands frequency ranges. The components of the dielectric tensor are given by the Drude-Lorentz formula \cite{tomadin2015}:

\begin{align}
\label{eq:Drude-Lorentz}
 \epsilon_i(\omega)=\epsilon_{i,\infty}+\frac{\epsilon_{i,0}-\epsilon_{i,\infty}}{1-(\omega/\omega_{{\rm T},i})^2}~,
\end{align}

where $i=x,z$ and $\omega_{{\rm T},i}$ are the transverse phonon frequencies, $\epsilon_{i,0}$ is the static dielectric constant ($\epsilon_i(0)=\epsilon_{i,0}$) while $\epsilon_{i,\infty}$ is the high-frequency limit, $\epsilon_i(\omega\to\infty)=\epsilon_{i,\infty}$.
The longitudinal phonon frequencies, defined by $ \epsilon_i(\omega)$ can be obtained by the relation $\omega_{{\rm L},i}=\omega_{{\rm T},i}\sqrt{(\epsilon_{i,0}/\epsilon_{i,\infty})}$. Realistic values for hBN of these quantities \cite{Geick66} are reported in Tab.~\ref{tab:1}.

\begin{table}[h]
    \centering
    \begin{tabular}{|l|c|c|}
        \hline
        & $i = x$ & $i = z$ \\
        \hline
        $\epsilon_{i,\infty}$ & 4.5 & 2.5 \\
        $\epsilon_{i,0}$ & 6.4 & 3.0 \\
        $\omega_{{\rm T},i}$ (meV) & 168 & 94 \\
        $\omega_{{\rm L},i}$ (meV) & 200 & 103 \\
        \hline
    \end{tabular}
    \caption{Realistic values of the parameters for the hBN dielectric tensor \cite{Geick66} entering in the Drude-Lorentz formula in Eq.~\eqref{eq:Drude-Lorentz}.}
    \label{tab:1}
\end{table}

Small dissipation can be included in the dielectric tensor in Eq.~\eqref{eq:Drude-Lorentz-Dissipative} by performing the substitution $\omega\to\omega+\I\eta$ and getting:

\begin{align}
\label{eq:Drude-Lorentz-Dissipative}
 \epsilon_i(\omega)\approx\epsilon_{i,\infty}+\frac{\epsilon_{i,0}-\epsilon_{i,\infty}}{1-(\omega/\omega_{{\rm T},i})^2-\I2\omega\eta/(\omega_{{\rm T},i})^2}~,
\end{align}

valid for $\eta\to 0$.

\section{Alternative quantization via a Hamiltonian approach}
\label{app_alt}

Here we provide an alternative derivation of the quantization of the scalar field $\hat{\phi}(\bm{r})$.
Instead of relying on the scalar Green’s function, we derive the equations of motion for the polarization field and from them construct a Hamiltonian for the normal modes, which is then quantized.

For an infrared-active optical mode, let $u_i(\bm{r},t)$ denote the relative ionic displacement along the axis $i$. With reduced mass $M_i$, elastic constant $K_i = M_i \omega_{T,i}^2$, and effective charge $e^*$, the equation of motion reads

\begin{equation}
M_i \ddot{u}_i(\bm{r},t) + M_i \omega_{T,i}^2 u_i(\bm{r},t) = e^* E_i(\bm{r},t)~,
\end{equation}

where $E_i(\bm{r},t)$ is the electric field.
The polarization along the axis $i$ is given by the dipole moment per unit volume,

\begin{equation}
P_i(\bm{r},t) = n_{\rm I}e^* u_i(\bm{r},t)~,
\end{equation}

where $n_{\rm I}$ is the ionic density. Dividing the equation of motion by $M_i$ and defining the {\it ionic plasma frequency}

\begin{equation}
\Omega_{{\rm p},i} = e^*\sqrt{\frac{4\pi n_{\rm I} }{M_i}}~,
\end{equation}

one obtains the driven polarization equation

\begin{equation} \label{eq:EOM}
\ddot{P}_i(\bm{r},t) + \omega_{{\rm T},i}^2 P_i(\bm{r},t)
= \frac{\Omega_{{\rm p},i}^2}{4\pi} E_i(\bm{r},t)~.
\end{equation}

Eq.~\eqref{eq:EOM} describes the harmonic response of the lattice polarization to an external electric field, where the coupling strength is governed by $\Omega_{{\rm p},i}$. This forms the starting point of a macroscopic quantization scheme in which the polarization plays the role of the field coordinate. To close the system, the electric field must be expressed in terms of the polarization. The scalar potential satisfies the Poisson equation (equivalent to Eq.~\eqref{eqn:scalar_Poisson}),

\begin{equation}
\sum_i\partial^2_i\epsilon_{i,\infty}\phi(\bm{r},t)=4\pi\left[\sum_i\partial_i P_i(\bm{r},t)-\rho_{\rm ext}(\bm{r},t)\right]~,
\end{equation}

where the ionic charge density $\rho_{\rm I}(\bm{r},t)$ has been expressed as $\rho_{\rm I}(\bm{r},t) = -\partial_i P_i(\bm{r},t)$. As we focus on the hBN cavity alone, we assume no external charges, $\rho_{\rm ext}(\bm{r},t)=0$. This formulation explicitly connects the microscopic lattice motion to the macroscopic electrostatic potential. In Eq.~\eqref{eq:EOM} $E_i(\bm{r},t)$ represents the electrostatic field generated by the polarization and can be expressed in terms of the high-frequency Green’s function
$g^{(0)}_\infty (\bm r_\parallel, \bm r'_\parallel,z,z') \equiv \lim_{\omega \to \infty} g^{(0)}(\bm r_\parallel, \bm r'_\parallel,z,z',\omega)$, that is the Green's function of Eq.~\eqref{eqn:green_function_hBN_expansion}. Using the electrostatic relation

\begin{equation}\label{eq:pot_E}
    E_i(\bm{r},t) = -\partial_i \phi(\bm{r},t)~,
\end{equation}

the scalar potential is obtained as

\begin{equation}\label{eq:pot_ginfty}
    \phi(\bm{r},t)=-\sum_j\int d \bm{r}^\prime g_\infty^{(0)}( \bm{r},  \bm{r}^\prime)  \partial^\prime_jP_j(\bm{r}^\prime,t)~,
\end{equation}

which, upon substitution into Eq.~\eqref{eq:EOM}, yields the closed equation of motion for the polarization field,

\begin{equation}\begin{split}
  \label{eq:EOM1}
  &\ddot{P}_i(\bm{r},t)+\omega_{T,i}^2P_i(\bm{r},t)+\\&+ \frac{\Omega_{{\rm p},i}^2}{4\pi} \sum_j\int d \bm{r}^\prime \partial_i\partial^\prime_j g_\infty^{(0)}( \bm{r},  \bm{r}^\prime) P_j(\bm{r}^\prime,t) =0~.  
\end{split}
\end{equation}

Eq.~\eqref{eq:EOM1} provides a self-contained description of the polarization dynamics, in which the non-local electrostatic coupling between different spatial regions is mediated by $g_\infty^{(0)}$.
For practical calculations, it is convenient to expand $g_\infty^{(0)}$ in the orthonormal electrostatic eigenfunctions,

\begin{equation}
\begin{split}
\label{eqn:green_function_hBN_expansion_S}
&g^{(0)}_\infty (\bm r_\parallel, \bm r'_\parallel,z,z')=
4\pi \sum_{\bm{q}_{\parallel},n_z}\frac{f_{\bm{q}_{\parallel},n_z}(\bm{r})f^*_{\bm{q}_{\parallel},n_z}(\bm{r}^\prime)}{\lambda_{\bm{q}_{\parallel},n_z}^{(\infty)}}~.
\end{split}
\end{equation}

where the eigenvalues and normalized mode functions are defined as

\begin{align}
&\lambda_{\bm{q}_{\parallel},n_z}^{(\infty)}\equiv\epsilon_{z,\infty}q_{z,n_z}^2+\epsilon_{x,\infty}q_\parallel^2~,\\
& f_{\bm{q}_{\parallel},n_z}(\bm{r})=\sqrt{\frac{2}{L_zS}}e^{\I\bm{q}_\parallel\cdot\bm{r}_\parallel}\sin\left(q_{z,n_z}\big[z+\frac{L_z}{2}\big]\right)~,
\end{align}

as introduced in Eq.~\eqref{eqn:mode_functions_hBN1}.
These mode functions form a complete and orthonormal basis (see Eq.~\eqref{eqn:completeness}), satisfying

\begin{align}
&\sum_{\bm{q}_{\parallel},n_z}f_{\bm{q}_{\parallel},n_z}(\bm{r})f^*_{\bm{q}_{\parallel},n_z}(\bm{r}')=\delta(\bm{r}-\bm{r}')~,\\
& \int d \bm{ r}f^*_{\bm{q}_{\parallel},n_z}(\bm{r})f_{\bm{q}_{\parallel}',n_z'}(\bm{r})=\delta_{\bm{q}_{\parallel},\bm{q}_{\parallel}'}\delta_{n_z,n_z'}~.
\label{eqn:ortonormal}
\end{align}

The equation of motion for the polarization field in Eq.~\eqref{eq:EOM1} follows from the Euler–Lagrange equations, $\frac{d}{dt} [\delta L_{\rm cav}/\delta\dot{P}_i(\bm{r},t)]=[\delta L_{\rm cav}/\delta P_i(\bm{r},t)]$, applied to the Lagrangian

\begin{equation}
\label{eq:Lagrangian}
\begin{split}
      &L_{\rm cav}=\int d\bm{r}  
\sum_i \frac{2\pi}{\Omega_{{\rm p},i}^{2}}\Big[\dot P^2_i(\bm{r},t)-\omega_{T,i}^2\,P_i^2(\bm{r},t)\Big]
+\\&-\frac{1}{2}\sum_{i,j}\int d\bm{r}d\bm{r}'\;
P_i(\bm{r},t)~
\partial_i\partial'_j\,g_\infty^{(0)}(\bm{r},\bm{r}')~
P_j(\bm{ r}',t)~.\\
\end{split}
\end{equation}

Since the polarization field is driven by the local electric field, and due to Eq.~\eqref{eq:pot_E} its spatial dependence must follow the gradients of the electrostatic eigenmodes that solve the Poisson equation. The scalar potential $\phi(\bm{r},t)$ can be expanded in the eigenfunctions $f_{\bm q_\parallel,n_z}(\bm r)$ of the Laplace operator, which already encode the boundary conditions of the system. Consequently, the polarization inherits the same spatial structure through their gradients.

We therefore expand the polarization field in the basis of these mode gradients,

\begin{equation}
\begin{split}
\label{eq:polarization_modes}
&\bm{P}(\bm{r},t)=\sum_{\bm{q}_{\parallel},n_z,s} \big[c^{(\parallel)}_{\bm{q}_{\parallel},n_z,s}(t)\partial_{\bm{r}_\parallel} f_{\bm{q}_{\parallel},n_z}(\bm{r})
\bm{e}_{\bm{q}_\parallel}+\\&+c^{(z)}_{\bm{q}_{\parallel},n_z,s}(t)\partial_{z} f_{\bm{q}_{\parallel},n_z}(\bm{r})\bm{e}_{z}\big]~,
\end{split}
\end{equation}

where the reality of the polarization imposes the constraint $[c^{(i)}_{\bm{q}_{\parallel},n_z,s}(t)]^*=c^{(i)}_{-\bm{q}_{\parallel},n_z,s}(t)$.  The derivatives of the mode functions $f_{\bm{q}_{\parallel},n_z}(\bm{r})$ are given by

\begin{align}
&\partial_{\bm{r}_\parallel} f_{\bm{q}_{\parallel},n_z}(\bm{r})\bm{e}_{\bm{q}_\parallel}=\I\bm{q}_\parallel f_{\bm{q}_{\parallel},n_z}(\bm{r})~,\\
& \partial_{z} f_{\bm{q}_{\parallel},n_z}(\bm{r})=q_{z,n_z} \tilde{f}_{\bm{q}_{\parallel},n_z}(\bm{r})~,
\end{align}
where $\tilde{f}_{\bm{q}_{\parallel},n_z}(\bm{r})=\sqrt{({2}/{L_zS})}e^{\I\bm{q}_\parallel\cdot\bm{r}_\parallel}\cos\left(q_{z,n_z}\big[z+\frac{L_z}{2}\big]\right)$. Note that the mode functions $\tilde{f}_{\bm{q}_{\parallel},n_z}(\bm{r})$ form an orthonormal set:

\begin{align}
& \int d \bm{ r}\tilde f^*_{\bm{q}_{\parallel},n_z}(\bm{r})\tilde f_{\bm{q}_{\parallel}',n_z'}(\bm{r})=\delta_{\bm{q}_{\parallel},\bm{q}_{\parallel}'}\delta_{n_z,n_z'}~\label{eqn:ortonormal1}~.
\end{align}
Inserting Eq.~\eqref{eq:polarization_modes} into the Lagrangian, Eq.~\eqref{eq:Lagrangian}, and expanding $g_\infty^{(0)}$ in terms of the mode, see Eq.~\eqref{eqn:green_function_hBN_expansion_S},
and employing the orthonormality relations in Eqs.~(\ref{eqn:ortonormal}, \ref{eqn:ortonormal1}) to perform the spatial integrals, we obtain

\begin{equation}
\label{eq:Lagrangian11}
\begin{split}
L_{\rm cav}=
&{2\pi}\sum_{\bm{q}_{\parallel},n_z,s}\Big\{\sum_{i=\parallel,z} \frac{\mathfrak{q}_i^2}{\Omega^2_{{\rm p},i}}\Big[|\dot c^{(i)}_{\bm{q}_{\parallel},n_z,s}(t)|^2- \omega^2_{{\rm T},i}| c^{(i)}_{\bm{q}_{\parallel},n_z,s}(t)|^2\Big]+\\
&-\sum_{i,j}c^{(i)*}_{\bm{q}_{\parallel},n_z,s}(t)c^{(j)}_{\bm{q}_{\parallel},n_z,s}(t) \frac{\mathfrak{q}_i^2\mathfrak{q}_j^2}{\lambda_{\bm{q}_{\parallel},n_z}^{(\infty)}}\Big\},
\end{split}
\end{equation}

where ${ \mathfrak{q}}=(\bm{q}_\parallel,q_{z,n_z})^{\rm T}$. This equation determines the equations of motion for  $c^{(i)}_{\bm{q}_{\parallel},n_z,s}(t)$, which in turn yield the eigenfrequencies

\begin{widetext}
\begin{equation}
\begin{split}
&\omega_{\bm{q}_{\parallel},n_z,s}^2
=\frac{
{\lambda_{\bm q_\parallel,n_z}^{(\infty)}(\omega_{{\rm T},x}^2+\omega_{{\rm T},z}^2)
+q_\parallel^2\Omega_{{\rm p},x}^2+q_{z,n_z}^2\Omega_{{\rm p},z}^2} +s \sqrt{\left[\lambda_{\bm q_\parallel,n_z}^{(\infty)}(\omega_{{\rm T},x}^2-\omega_{{\rm T},z}^2)
+q_\parallel^2\Omega_{{\rm p},x}^2-q_{z,n_z}^2\Omega_{{\rm p},z}^2\right]^2
+4 q_\parallel^2 q_{z,n_z}^2\Omega_{{\rm p},x}^2\Omega_{{\rm p},z}^2}}{{2\lambda_{\bm q_\parallel,n_z}^{(\infty)}}}~,
\end{split}
\end{equation}
\end{widetext}
where $s=\pm1$. 
Compared with the solutions of Eq.~\eqref{eqn:mode_functions_hBN}, the two expressions coincide when identifying $\Omega_{p,i} = \omega_{{\rm T},i} \sqrt{\epsilon_{i,0} - \epsilon_{i,\infty}}$.

In the 2D plane $\bm{e}_{\bm{q}_\parallel}-\bm{e}_{z}$, the eigenvectors corresponding to the eigenfrequencies $\omega_{\bm{q}_{\parallel},n_z,s}$ are given by
\begin{equation}\label{eq:Q}
    \begin{pmatrix}
    {c^{(\parallel)}_{\bm{q}_{\parallel},n_z,s}}(t)\\
        {c^{(z)}_{\bm{q}_{\parallel},n_z,s}}(t)
    \end{pmatrix}=\begin{pmatrix}
         \cos(\theta_{\bm{q}_{\parallel},n_z,s})\\
         \sin(\theta_{\bm{q}_{\parallel},n_z,s})
    \end{pmatrix} Q_{\bm{q}_{\parallel},n_z,s}(t)~,
\end{equation}
where $Q_{\bm{q}_{\parallel},n_z,s}(t)$ are the amplitudes of normal modes and the angle  $\theta_{\bm{q}_{\parallel},n_z,s}$ is determined by
\begin{equation}
\tan(\theta_{\bm{q}_{\parallel},n_z,s})
=-\frac{\lambda_{\bm q_\parallel,n_z}^{(\infty)}(\omega_{{\rm T},x}^2-\omega_{\bm{q}_{\parallel},n_z,s}^2)/\Omega_{{\rm p},x}^2+q_\parallel^{2}}{q_{z,n_z}^{2}}~.
\end{equation}

The normal-mode amplitudes $Q_{\bm{q}_{\parallel},n_z,s}(t)$ satisfy the condition $Q^*_{\bm{q}_{\parallel},n_z,s}(t)=Q_{-\bm{q}_{\parallel},n_z,s}(t)$  which ensures the reality of the solution.
The Lagrangian in Eq.~\eqref{eq:Lagrangian11} can be written in terms of the normal modes as $Q_{\bm{q}_{\parallel},n_z,s}(t)$:

\begin{equation}
\begin{split}
L_{\rm cav} &= \sum_{\bm q_\parallel,n_z,s} \frac{M_{\bm q_\parallel,n_z,s}}{2}\times  \\
&\quad \times \Big(|\dot Q_{\bm q_\parallel,n_z,s}(t)|^2-\omega_{\bm q_\parallel,n_z,s}^2|Q_{\bm q_\parallel,n_z,s}(t)|^2\Big)~,
\end{split}
\end{equation}

where $M_{\bm q_\parallel,n_z,s}$ plays the role of a mass and reads

\begin{equation}
M_{\bm q_\parallel,n_z,s}
\equiv 4\pi\left(\frac{q_\parallel^{2}}{\Omega_{{\rm p},x}^{2}}\cos^2\theta_{\bm q_\parallel,n_z,s}
+\frac{q_{z,n_z}^{2}}{\Omega_{{\rm p},z}^{2}}\sin^2\theta_{\bm q_\parallel,n_z,s}\right)~.
\end{equation}

By combining Eqs. (\ref{eq:pot_ginfty},\ref{eqn:green_function_hBN_expansion_S},\ref{eq:polarization_modes},\ref{eq:Q}), it is possible to express the scalar field in terms of the normal modes $Q_{\bm{q}_{\parallel},n_z,s}(t)$ as

\begin{equation}
\label{eq:scalar11}
\begin{split}
     &\phi(\bm{r},t)=4\pi\sum_{\bm{q}_{\parallel},n_z,s}\frac{f_{\bm{q}_{\parallel},n_z}(\bm{r})}{{\lambda_{\bm{q}_{\parallel},n_z}^{(\infty)}}}\times\\& \times\left(q_\parallel^2 \cos(\theta_{\bm{q}_{\parallel},n_z,s})+q^2_{z,n_z}\sin(\theta_{\bm{q}_{\parallel},n_z,s})\right) Q_{\bm{q}_{\parallel},n_z,s}(t)~.
\end{split}
\end{equation}

We now compute the cavity Hamiltonian as the Legendre transform of the Lagrangian, $\mathcal H_{\rm cav}
=\sum_{\bm q_\parallel,n_z,s}\Pi_{\bm q_\parallel,n_z,s}\dot Q_{\bm q_\parallel,n_z,s}-L_{\rm cav}$, where

\begin{equation}
\Pi_{\bm q_\parallel,n_z,s}\equiv \frac{\partial L}{\partial \dot Q_{\bm q_\parallel,n_z,s}}
= M_{\bm q_\parallel,n_z,s}\dot Q^*_{\bm q_\parallel,n_z,s}~.
\end{equation}

The cavity Hamiltonian explicitly reads

\begin{equation}
\label{eq:H_cav_alt}
\mathcal H_{\rm cav}
=\sum_{\bm q_\parallel,n_z,s}\left[
\frac{|\Pi_{\bm q_\parallel,n_z,s}|^2}{2 M_{\bm q_\parallel,n_z,s}}
+\frac{ M_{\bm q_\parallel,n_z,s}\omega_{\bm q_\parallel,n_z,s}^2}{2}
|Q_{\bm q_\parallel,n_z,s}|^2\right]~.
\end{equation}

The Hamiltonian in Eq.~\eqref{eq:H_cav_alt} can be quantized by promoting $ Q_{\bm q_\parallel,n_z,s}$ and $\Pi_{\bm q'_\parallel,n'_z,s'}$ to conjugated quantum variables fulfilling the commutator

\begin{equation}
  \big[\hat Q_{\bm q_\parallel,n_z,s},\hat \Pi_{\bm q'_\parallel,n'_z,s'}\big]
=i\hbar\delta_{\bm q_\parallel,\bm q'_\parallel}\delta_{n_z,n'_z}\delta_{s,s'} ~.
\end{equation}

This allows us to introduce the ladder operators $ \hat a_{\bm q_\parallel,n_z,s}, \hat a^\dagger_{\bm q_\parallel,n_z,s}$ defined by
\begin{equation}
\label{eq:a_adagger}
\begin{split}
    \hat a_{\bm q_\parallel,n_z,s}
=\sqrt{\frac{M_{\bm q_\parallel,n_z,s}\omega_{\bm q_\parallel,n_z,s}}{2\hbar}}\hat Q_{\bm q_\parallel,n_z,s}+
\\+\frac{i}{\sqrt{2\hbar M_{\bm q_\parallel,n_z,s}\omega_{\bm q_\parallel,n_z,s}}}\hat \Pi^\dagger_{\bm q_\parallel,n_z,s}~,\\
\hat a^\dagger_{\bm q_\parallel,n_z,s}
=\sqrt{\frac{{M}_{\bm q_\parallel,n_z,s}\omega_{\bm q_\parallel,n_z,s}}{2\hbar}}\hat Q^\dagger_{\bm q_\parallel,n_z,s}+
\\-\frac{i}{\sqrt{2\hbar {M}_{\bm q_\parallel,n_z,s}\omega_{\bm q_\parallel,n_z,s}}}\hat \Pi_{\bm q_\parallel,n_z,s}~,
\end{split}
\end{equation}

and fulfilling $\big[\hat a_{\bm q_\parallel,n_z,s},\hat a^\dagger_{\bm q'_\parallel,n'_z,s'}\big]
=\delta_{\bm q_\parallel,\bm q'_\parallel}\delta_{n_z,n'_z}\delta_{s,s'}$.
The cavity Hamiltonian in Eq.~\eqref{eq:H_cav_alt} becomes diagonal when expressed in terms of the modes $ \hat a_{\bm q_\parallel,n_z,s}, \hat a^\dagger_{\bm q_\parallel,n_z,s}$:

\begin{equation}
\hat{\mathcal H}_{\rm cav}
=\sum_{\bm q_\parallel,n_z,s}
\hbar \omega_{\bm q_\parallel,n_z,s}\left(\hat a^\dagger_{\bm q_\parallel,n_z,s}\hat a_{\bm q_\parallel,n_z,s}+\frac{1}{2}\right)~.
\end{equation}

Inverting Eq.~\eqref{eq:a_adagger} we can express the quantized normal modes $\hat Q_{\bm{q}_{\parallel},n_z,s}$ in terms of the ladder operators:

\begin{equation}
\begin{split}
\hat Q_{\bm q_\parallel,n_z,s}= \sqrt{\frac{\hbar}{2M_{\bm q_\parallel,n_z,s} \omega_{\bm q_\parallel,n_z,s}}}\left(\hat a_{\bm q_\parallel,n_z,s}+\hat a^\dagger_{-\bm q_\parallel,n_z,s}\right)~.
\end{split}
\end{equation}

Combining the above expression with Eq.~\eqref{eq:scalar11}, we express the
quantized scalar field $\hat{\phi}(\bm r)$ in terms of ladder operators as:

\begin{equation}
\begin{split}\label{eq:phi_in_modes_SM}
    \hat \phi(\bm{r})=&\sum_{\bm{q}_{\parallel},n_z,s}\left(\frac{1}{\sqrt{N'_{{\bm q_\parallel,n_z,s}}}}  f_{\bm{q}_{\parallel},n_z}(\bm{r})\hat a^\dagger_{\bm q_\parallel,n_z,s}+{\rm h.c}\right)~,
\end{split}
\end{equation}

where the mode–normalization factor obtained within the polarization–field quantization scheme reads

\begin{equation}
N'_{\bm q_\parallel,n_z,s} \equiv \frac{M_{\bm q_\parallel,n_z,s}\omega_{\bm q_\parallel,n_z,s}[{\lambda_{\bm{q}_{\parallel},n_z}^{(\infty)}}]^2}{8\pi^2\hbar \left[q_\parallel^2 \cos(\theta_{\bm{q}_{\parallel},n_z,s})+q^2_{z,n_z}\sin(\theta_{\bm{q}_{\parallel},n_z,s})\right]^2}~.
\end{equation}

Using the explicit expressions for the mode mass $M_{\bm q_\parallel,n_z,s}$, and the eigenvector angle $\theta_{\bm{q}_{\parallel},n_z,s}$ this expression reduces to the compact form

\begin{equation}
\begin{split}
    N'_{\bm q_\parallel,n_z,s} =&\frac{\omega_{\bm{q}_{\parallel},n_z,s}}{2\pi \hbar}\times\\&\times \left[ \frac{q_{z,n_z}^2 \Omega_{p,z}^2}{(\omega_{T,z}^2 - \omega_{\bm{q}_{\parallel},n_z,s}^2)^2} + \frac{q_{\parallel}^2 \Omega_{p,x}^2}{(\omega_{T,x}^2 - \omega_{\bm{q}_{\parallel},n_z,s}^2)^2} \right]~.
\end{split}
\end{equation}

The expression above is obtained from the {polarization--field quantization},
where the fundamental dynamical variable is the lattice polarization 
$P_i(\bm{r},t)$. In the main text we also introduced the {scalar--potential quantization}, 
where the starting point is the electrostatic scalar field 
$\phi(\bm{r})$, which is quantized by resorting to the scalar Green’s 
functions.
The two procedures lead to distinct intermediate expressions for the mode 
normalization, denoted respectively by $N'_{\bm{q}_\parallel,n_z,s}$ and $N_{\bm{q}_\parallel,n_z,s}$.

The two quantization schemes are fully consistent if the corresponding 
normalization constants coincide,
\begin{equation}
N'_{\bm{q}_\parallel,n_z,s}=N_{\bm{q}_\parallel,n_z,s}~.
\end{equation}
By inserting the explicit expression of $N_{\bm{q}_\parallel,n_z,s}$ from 
Eq.~\eqref{eq:N_q}, and using the dielectric response of hBN, one verifies that the identity above holds exactly.  
This confirms that both approaches lead to the same quantized scalar field 
$\hat{\phi}(\bm{r})$.

\section{The density--density response function for graphene in a magnetic field}\label{app1}

Here, we calculate the bare density-density response function, $\chi^{(0)}_{nn}( q_\parallel,\omega)$. First of all, it is useful to define the ladder operators $\hat{a}$ and $\hat{a}^\dagger$ for the case when the displacement $k_x\ell^2_B$ is zero, as:

\begin{equation}
    \begin{split}
\hat{{y}}&= \frac{\ell_B}{\sqrt{2}} \left( \hat{a} + \hat{a}^\dagger \right)~, \\
\hat{p}_y &= \frac{\hbar\I}{\sqrt{2 }\ell_B} \left( \hat{a}^\dagger - \hat{a}  \right)~.
   \end{split}
\end{equation}
It is now useful to introduce the displacement operators defined as:

\begin{equation}
    \begin{split}
         \hat{\mathcal{D}}\left(\alpha\right)&\equiv \exp\left( \alpha\hat{a}^\dagger -\alpha^*\hat{a} \right)~
    \end{split}
\end{equation}

where $\alpha$ is a complex number.
The displaced ladder operators $\hat{a}_{k_x}$ and $\hat{a}_{k_x}^\dagger$ can be related to the ladder operators $\hat{a}$ and $\hat{a}^\dagger$ by introducing the displacement operator $\hat{\mathcal{D}}\left(\frac{k_x\ell_B}{\sqrt{2}}\right)$ as:

\begin{equation}
    \begin{split}
         \hat{\mathcal{D}}\left(\frac{k_x\ell_B}{\sqrt{2}}\right)&\equiv \exp\left(\frac{k_x\ell_B}{\sqrt{2}}\left( \hat{a}^\dagger -\hat{a} \right)\right)~,\\
    &=\exp\left(-\I k_x \ell_B^2 \hat{p}_y/\hbar\right)~.
    \end{split}
\end{equation}
   
The operator $\hat{\mathcal{D}}\left(\frac{k_x\ell_B}{\sqrt{2}}\right)$ links the ladder operators $\hat{a}_{k_x}$ with $\hat{a}$:

\begin{equation}
    \begin{split}
\hat{a}_{k_x}&=\hat{\mathcal{D}}\left(\frac{k_x\ell_B}{\sqrt{2}}\right)~\hat{{a}}~\hat{\mathcal{D}}^\dagger\left(\frac{k_x\ell_B}{\sqrt{2}}\right)~,\\
&=\hat{a}-\frac{k_x\ell_B}{\sqrt{2}}~.
    \end{split}
\end{equation}

Hence, the operator $ \hat{\mathcal{D}}\left(\frac{k_x\ell_B}{\sqrt{2}}\right)$ allows us to connect $\hat{y}_{k_x}$ and $\hat{y}$:

\begin{equation}
    \begin{split}
\hat{{y}}_{k_x}&=\hat{\mathcal{D}}\left(\frac{k_x\ell_B}{\sqrt{2}}\right)~\hat{{y}}~\hat{\mathcal{D}}^\dagger\left(\frac{k_x\ell_B}{\sqrt{2}}\right)~,\\
&=\hat{{y}}-{k_x}{\ell_B^2}~.
    \end{split}
\end{equation}

Hence, by using the displacement operator $\hat{\mathcal{D}}\left(\frac{k_x\ell_B}{\sqrt{2}}\right)$, we can relate the displaced vacuum $\ket{ 0_{k_x}}$ (defined by $\hat{a}_{k_x} \ket{0_{k_x}} = 0$) with the vacuum $\ket{0}$, such that $\hat{a} \ket{0} = 0$:

\begin{align}
\ket{0_{k_x}}&=\hat{\mathcal{D}}\left(\frac{k_x\ell_B}{\sqrt{2}}\right)\ket{0}~.
\end{align} 

Similarly, we can link the excited states:

\begin{align}
\label{eq:displaced_Fock}
\ket{n_{k_x}}&= \hat{\mathcal{D}}\left(\frac{k_x\ell_B}{\sqrt{2}}\right)\ket{n},
\end{align} 

where $\ket{n}$ is the Fock state with respect to $\hat{a}$, $\ket{n}=\left[{(\hat{a}^\dagger)^n}/{\sqrt{n!}}\right]\ket{0}$ .

We now calculate the matrix elements of the density operator $\hat n_{\bm q_\parallel}\equiv \exp(-\I \bm{q}_\parallel\cdot \bm{r}_\parallel)$.

\begin{equation}
\label{eq:expqx}
\braket{{k_x}^\prime|e^{-\I q_x \hat{x}}|{k_x}}= \delta_{k_x^\prime,k_x+q_x}~,
\end{equation}

\begin{equation}
\begin{split}
 \label{eq:expqy}
&\braket{n^\prime_{{k_x}^\prime}|e^{-\I q_y \hat{y}}|n_{k_x}}=\\&=\braket{n^\prime|\hat{\mathcal{D}}^\dagger\left(\frac{k_x^\prime\ell_B}{\sqrt{2}}\right)\hat{\mathcal{D}}^\dagger\left(\frac{\I q_y\ell_B}{\sqrt{2}}\right)\hat{\mathcal{D}}\left(\frac{k_x\ell_B}{\sqrt{2}}\right)|n}~,     
\end{split}  
\end{equation}

where we used that $e^{-\I q_y \hat{y}} = \hat{\mathcal{D}}^\dagger(\I q_y \ell_B / \sqrt{2})$ and applied Eq.~\eqref{eq:displaced_Fock} to connect the states $\ket{n^\prime_{k_x^\prime}}$ with $\ket{n^\prime}$ and $\ket{n_{k_x}}$ with $\ket{n}$. Using the braiding relation $\hat{\mathcal{D}}^\dagger(\alpha)\hat{\mathcal{D}}^\dagger(\beta)=\exp(\alpha^*\beta-\alpha\beta^*)\hat{\mathcal{D}}^\dagger(\beta)\hat{\mathcal{D}}^\dagger(\alpha)$ we get:
\begin{equation}
\begin{split}
& \braket{n^\prime_{{k_x}^\prime}|e^{-\I q_y \hat{y}}|n_{k_x}}=\\
 &=\exp(-\I k^\prime_xq_y\ell_B^2)\braket{n^\prime|\hat{\mathcal{D}}^\dagger\left(\frac{[(k_x^\prime-k_x)+\I q_y]\ell_B}{\sqrt{2}}\right)|n}~,    
\end{split}
\label{eq:expqy1}   
\end{equation}

Employing Eqs. (\ref{eq:expqx},\ref{eq:expqy1}) we can express the matrix element of the density operator $\hat n_{\bm q_\parallel}$ as:

\begin{equation}
    \begin{split}
       &\braket{{k_x}^\prime,n^\prime|\hat n_{\bm q_\parallel}|{k_x},n}=\\&= \exp(-\I k^\prime_xq_y\ell_B^2)\delta_{k_x^\prime,k_x+q_x}\braket{n^\prime|\hat{\mathcal{D}}^\dagger\left(\frac{q_z\ell_B}{\sqrt{2}}\right)|n}~. 
    \end{split}
\end{equation}

where we defined $q_z\equiv q_x+\I q_y$.
By using the Baker--Campbell--Hausdorff formula for
$[\hat{A},[\hat A,\hat B]]=[\hat B,[\hat A,\hat B]]=0$,
\begin{equation}
e^{\hat A+\hat B}=e^{\hat A}e^{\hat B}e^{-[\hat A,\hat B]/2},
\end{equation}

we get:
\begin{equation}
    \begin{split}
     & \braket{n^\prime|\hat{\mathcal{D}}^\dagger\left(\frac{q_z\ell_B}{\sqrt{2}}\right)|n}=\\& =e^{-|q_z|^2\ell_B^2/4}\braket{n^\prime|\exp{\left(-\frac{q_z\ell_B \hat{a}^\dagger}{\sqrt{2}}\right)}\exp{\left(\frac{{q}^*_z\ell_B \hat{a}}{\sqrt{2}}\right)}|n}~.  
    \end{split}
\end{equation}

We further define:

\begin{align}
{G}_{n^\prime,n}({q}_z\ell_B)\equiv\braket{n^\prime|\exp{\left(-\frac{q_z\ell_B \hat{a}^\dagger}{\sqrt{2}}\right)}\exp{\left(\frac{{q}^*_z\ell_B \hat{a}}{\sqrt{2}}\right)}|n}~.
\end{align} 

In the case where $n'>n$, ${G}_{n^\prime,n}({q}_z\ell_B)$ can be calculated as:

\begin{equation}
G_{n'n}(q_z\ell_B) =
\sqrt{\frac{n!}{n'!}}
\bigg(\frac{\I q_z\ell_B }{\sqrt2}\bigg)^{\!n'-n}
L^{n'-n}_n\bigg(\frac{\lvert q_z\rvert^2\ell_B^2}{2}\bigg)~,
\end{equation}
 where $L^{m}_n\left(X\right)$ is a Laguerre polynomial defined as:
\begin{equation}
L^{m}_n\left(X\right) =\sum_{k=0}^n\binom{n+m}{n-k}\left(-X\right)^{\!k}~.
\end{equation}

The case $n'<n$ can be obtained by noticing that 

\begin{equation}
G^*_{n'n}(X) =G_{nn'}(X)~.
\end{equation}

The matrix element of the density operator reads
\begin{equation}
\begin{split}
&\braket{k_x^\prime,n',\lambda'|\hat n_{\bm q_\parallel}|k_x,n,\lambda}=\\
&=e^{-\lvert q_\parallel\rvert^2\ell_{ B}^2/4}\delta_{k_x^\prime,k_x-q_x} \mathcal{F}_{n^\prime,\lambda^\prime,n,\lambda}(\bm{q}_\parallel)~, 
\end{split}
\end{equation}

where 
\begin{equation}
\begin{split}
 & \mathcal{F}_{n^\prime,\lambda^\prime,n,\lambda}(\bm{q}_\parallel)\equiv\\
  &\equiv C^-_{n'}C^-_nG_{n'-1,n-1}({q}^*_z\ell_B)+\lambda\lambda'C^+_{n'}C^+_nG_{n'n}({q}^*_z\ell_B)~.
\end{split}
\end{equation}

We can thus calculate the density--density response function as
\begin{align}
\label{eq:density--densityresponse0}
\chi^{(0)}_{nn}( q_\parallel,\omega) =
\sum_{n,\lambda,n',\lambda'}
\frac{(f_{n\lambda}-f_{n'\lambda'})\mathcal{M}_{n^\prime,\lambda^\prime,n,\lambda}({q}_\parallel)}{
\hbar\omega+\varepsilon_{\lambda,n}-\varepsilon_{\lambda',n'}+\I\eta}~.
\end{align}
where $\varepsilon_{\lambda,n}\equiv\lambda(\hbar v_{\rm F}/\ell_{B})\sqrt{2n}$ and $\mathcal{M}_{n^\prime,\lambda^\prime,n,\lambda}({q}_\parallel)$ represents the squared modulus of the matrix element of the density,

\begin{equation}
\begin{split}
  &\mathcal{M}_{n^\prime,\lambda^\prime,n,\lambda}({q}_\parallel)\equiv \sum_{k^\prime_x,k_x}
\lvert\braket{k_x^\prime,n',\lambda'|\hat n_{\bm q_\parallel}|k_x,n,\lambda}\rvert^2~, \\
&=\frac{g}{2\pi\ell_B^2}e^{-q_\parallel^2\ell_{B}^2/2} |\mathcal{F}_{n^\prime,\lambda^\prime,n,\lambda}( {q}_\parallel)|^2~, 
\end{split}
\end{equation}

where $|\mathcal{F}_{n^\prime,\lambda^\prime,n,\lambda}( q_\parallel)|^2$ depends only on the modulus $q_\parallel$ of $\bm{q}_\parallel$ and reads

\begin{widetext}
\begin{equation}
 |\mathcal{F}_{n^\prime,\lambda^\prime,n,\lambda}( q_\parallel)|^2= \bigg(\frac{q_\parallel^2\ell_B^2 }{2}\bigg)^{\!n_>-n_<}\Big|C^-_{n'-1}C^-_{n-1}\sqrt{\frac{(n_<-1)!}{(n_>-1)!}}
L^{n_>-n_<}_{n_<-1}\bigg(\frac{q_\parallel^2\ell_B^2}{2}\bigg)
+\lambda\lambda'C^+_{n^\prime}C^+_{n}\sqrt{\frac{n_<!}{n_>!}}
L^{n_>-n_<}_{n_<}\bigg(\frac{q_\parallel^2\ell_B^2}{2}\bigg)\Big|^2~,
\end{equation}
and $n_>={\rm max}(n^\prime,n)$, $n_<={\rm min}(n^\prime,n)$.

At $T=0$, the density--density response function is
\begin{align}
\label{eq:density--densityresponse_app}
\chi^{(0)}_{nn}(  q_\parallel,\omega) =
\sum_{n,\lambda\in \mathcal{N}_{\rm occ}}\sum_{n',\lambda'\in\mathcal{N}_{\rm unocc}}\mathcal{M}_{n^\prime,\lambda^\prime,n,\lambda}({q}_\parallel)\left( \frac{1}{\hbar\omega+\varepsilon_{\lambda,n}-\varepsilon_{\lambda',n'}+\I\eta}-\frac{1}{{\hbar\omega+\varepsilon_{\lambda',n'}-\varepsilon_{\lambda,n}+\I\eta}}\right)~.
\end{align}

\end{widetext}

This concludes the calculation of the density--density response function $\chi^{(0)}_{nn}(  q_\parallel,\omega)$ for Landau levels in graphene.

\end{document}